\documentclass[aps,prd,draft,nofootinbib,groupedaddress,preprintnumbers,%
              showpacs,showkeys]{revtex4}       
\usepackage{psfig}
\usepackage{colordvi}
\usepackage{graphics}
\usepackage{lscape}
\usepackage{rotfloat}
\usepackage{rotating}
\usepackage{amsmath}
\usepackage{amssymb}

\def\lsim{\mathrel{\rlap{\lower4pt\hbox{\hskip1pt$\sim$}}
    \raise1pt\hbox{$<$}}}                
\def\gsim{\mathrel{\rlap{\lower4pt\hbox{\hskip1pt$\sim$}}
    \raise1pt\hbox{$>$}}}                

\def\slashed{{/}\mskip-10.0mu}
\def\pcircslash{\slashed {\buildrel \circ\over p}}
\def\kc{{{\buildrel \circ\over k}}}

\def\smn{{\sigma_{\mu\nu}}}

\def\openone{\leavevmode\hbox{\small1\kern-3.3pt\normalsize1}}

\def\Zq{Z_{\rm q}}
\def\ZS{Z_{\rm S}}
\def\ZP{Z_{\rm P}}
\def\ZV{Z_{\rm V}}
\def\ZA{Z_{\rm A}}
\def\ZT{Z_{\rm T}}

\newcommand{\be}{\begin{equation}}
\newcommand{\ee}{\end{equation}}
\newcommand{\bea}{\begin{eqnarray}} 
\newcommand{\eea}{\end{eqnarray}}

\newcommand{\pslash}{{\not{\hspace{-0.001cm}p}}}  
\newcommand{\ve}{\varepsilon}  
\newcommand{\csw}{\, c_{\rm SW}}  
\newcommand{\ggcf}{\frac{g^2 C_F}{16 \, \pi^2}\; }  

\newcommand{\gttwo}{\Green{\tilde{g}^2}}
\newcommand{\MB}{\MidnightBlue}

\newcommand{\eins}{\openone} 

\newcommand{\s}{{\mathcal S}} 
\newcommand{\Op}{\mathcal{O}} 

\newcommand{\J}{\mathcal{J}} 

\usepackage{flexisym}
\usepackage{breqn}
\newcommand{\bdm}{\begin{dmath}}
\newcommand{\edm}{\end{dmath}}
\newcommand{\bdms}{\begin{dmath*}}
\newcommand{\edms}{\end{dmath*}}
\newcommand{\bdg}{\begin{dgroup*}}
\newcommand{\edg}{\end{dgroup*}}

\begin{document}

\title{Renormalization constants of local operators for Wilson type improved fermions\\
\vspace*{1cm}
\centerline{\psfig{figure=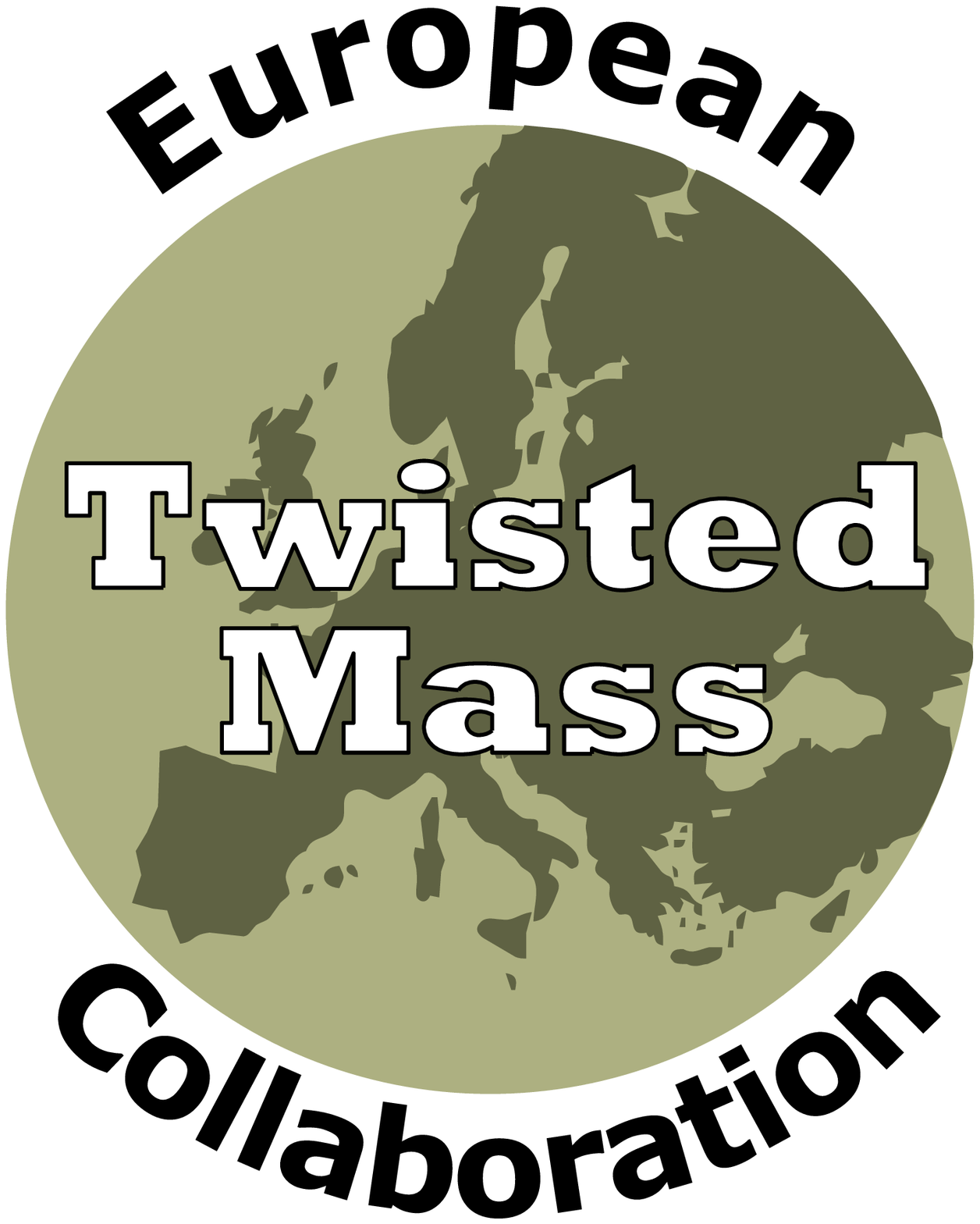,height=2.2truecm}}
\vspace*{1cm}
}

\author{C. Alexandrou $^{a, b}$, M.~Constantinou~$^a$, T. Korzec~$^c$,
H.~Panagopoulos~$^a$, F. Stylianou~$^a$ }
\email{alexand@ucy.ac.cy, marthac@ucy.ac.cy,
korzec@physik.hu-berlin.de, haris@ucy.ac.cy, fstyl01@ucy.ac.cy}

\affiliation{\\ \vskip 0.35cm
$a\,\, Department\,\, of\,\, Physics,\,\, University\,\, of\,\, Cyprus,$\\ 
$PoB\,\, 20537,\,\, 1678\,\, Nicosia,\,\, Cyprus$   \\ \vskip 0.25cm
$b\,\, Computation\,\, based\,\, Science\,\, and\,\, Technology\,\, Research\,\, Center,\,\, 
The\,\, Cyprus\,\, Institute,$ \\
$15\,\, Kypranoros\,\, Str.,\,\, 1645\,\, Nicosia,\,\, Cyprus$\\\vskip 0.25cm
$c\,\, Institut\,\, fur\,\, Physik,\,\, 
Humboldt\,\, Universitat\,\, zu\,\, Berlin,$\\ 
$Newtonstrasse\,\, 15,\,\, 12489\,\, Berlin,\,\, Germany $\\ \vskip 0.25cm
}
\date{\today}
\vskip 0.25cm

\begin{abstract} 
 Perturbative and non-perturbative results are presented on
the renormalization constants of the quark field and the vector, axial-vector,
pseudoscalar, scalar and tensor currents.
The perturbative computation, carried out at one-loop level and up to second
order in the lattice spacing, is performed for
a fermion action, which includes the clover term and the twisted mass
parameter yielding results that are applicable for unimproved Wilson
fermions, as well as for improved clover and twisted mass fermions.
We consider ten variants of the Symanzik improved gauge action
corresponding to ten different  values of the
plaquette coefficients.
Non-perturbative results are
obtained using the twisted mass Wilson 
fermion formulation employing two degenerate dynamical quarks and the
tree-level Symanzik improved gluon action. The simulations are
performed for pion masses in the range of 480~MeV to 260~MeV and at
three values of the lattice spacing, $a$, corresponding to
$\beta=3.9,\,4.05,\,4.20$. 
 For each renormalization factor computed non-perturbatively we subtract its perturbative ${\cal
O}(a^2)$ terms so that we eliminate part of the cut-off artifacts.
The renormalization constants are converted to  ${\overline{\rm MS}}$ at 
a scale of $\mu=2$~GeV. The perturbative results depend on a large number of parameters and are
made easily accessible to the reader by including them in
the distribution package of this paper, as a Mathematica input file.


\end{abstract}

\maketitle

\newpage

\section{Introduction}
Simulations of Quantum Chromodynamics (QCD) are nowadays being carried out 
at almost physical parameters. Therefore studies of 
hadron structure within lattice QCD are beginning to yield results that can
be connected to experiment more reliably than ever before.
In these lattice QCD studies one calculates matrix elements of local
operators between hadron states. Unless these operators correspond to
a conserved current they have to be renormalized. Calculation of
renormalization factors can be carried out using lattice perturbation
theory. Although perturbation theory on the lattice is computationally more 
complex than in the continuum these calculations can be extended beyond 
one-loop order~\cite{Bode:2001uz,Skouroupathis:2007jd,Skouroupathis:2008mf}. Various
methods to improve the convergence of lattice perturbation theory have
been introduced~\cite{Lepage:1992xa,Constantinou:2006hz}
yielding valuable first input to the values of the renormalization constants.
In this work we will use the perturbative results
to improve the non-perturbative evaluation of 
the renormalization constants.
We use the Rome-Southampton method (also known as the RI-MOM
scheme)~\cite{Martinelli:1994ty} to compute renormalization
coefficients of arbitrary quark- antiquark operators non-perturbatively.
 In this approach the procedure is similar to that used in continuum 
perturbation theory. In particular, the renormalization conditions are defined 
similarly in perturbative and non-perturbative calculations. The
renormalization factors, 
obtained for different values of the renormalization scale,
are evolved perturbatively to a reference scale $\mu=2$ GeV. In
addition, they are translated to ${\overline{\rm MS}}$ at 2~GeV using
3-loop perturbative results for the conversion factors.
Since in the end one wants to make contact with phenomenological
studies, which almost exclusively refer to operators renormalized in
the ${\overline{\rm MS}}$ scheme of dimensional regularization, one
needs the renormalization factors leading from the bare operators on
the lattice to the ${\overline{\rm MS}}$ operators in the continuum. 

 A number of lattice groups are
producing results on nucleon form factors and first moments of
structure functions closer to the physical regime both in terms of
pion mass as well as in terms of the continuum limit~\cite{Hagler:2007xi,Syritsyn:2009mx,
Brommel:2007sb,Alexandrou:2008rp,Yamazaki:2009zq,Alexandrou:2009ng,
Alexandrou:tomasz}. In these lattice QCD computations one
calculates hadron matrix elements of bilocal operators.
In order to compare hadron matrix elements of local operators to
experiment one needs to renormalize them. The aim of this paper is to
calculate non-perturbatively the renormalization factors of the
vector, axial-vector, scalar, pseudoscalar and tensor currents within
the twisted mass formulation of Wilson lattice QCD~\cite{Frezzotti:2000nk}.
We show that, although the lattice spacings considered in this work are
smaller than $1$~fm, ${\cal O}(a^2\,p^2)$ terms are non-negligible and are
significantly larger than statistical errors. We therefore compute the
${\cal O}(a^2\,g^2)$-terms perturbatively and subtract them from the
non-perturbative results. This subtraction suppresses lattice
artifacts considerably depending on the operator under study and leads
to a more accurate determination of the renormalization constants.
This approach was applied to evaluate the renormalization constants
for one-derivative bilinear operators in Ref.~\cite{Alexandrou:2010me}.

The paper is organized as follows: in Section \ref{sec2} we give the
expressions for the fermion and gluon actions we employed, and
define the operators. Sections \ref{prop} and \ref{oper} concentrate on the
perturbative procedure, and the ${\cal O}(a^2)$-corrected
expressions for the renormalization constants $\Zq$ and $Z_{\cal O}$. 
In Section \ref{renorm} we provide the renormalization
prescription of the RI$'$-MOM scheme, and we discuss alternative ways
for its application, while in Section \ref{conversion} we provide all
necessary formulae for the conversion to $\overline{\rm MS}$ and the
evolution to a reference scale of 2 GeV. Section \ref{sec4} focuses on the
non-perturbative computation, where we explain the different steps of
the calculation.
The main results of this work are presented in Section \ref{sec5}: the
reader can find numerical values for the $Z$-factors of the fermion
field and fermion operators, which are computed non-perturbatively and 
corrected using the perturbative ${\cal O}(a^2)$ terms presented in
Sections~\ref{prop} and \ref{oper}. For comparison with
phenomenological and experimental results, we convert the $Z$-factors to
the ${\overline{\rm MS}}$ scheme at 2~GeV. In Section~\ref{sec6} we
give our conclusions. 

\section{Formulation}
\label{sec2}
\subsection{Lattice actions}

Our perturbative calculation makes use of the twisted mass fermion action
including the usual clover (SW) term with a clover parameter that is left free.
 For
$N_F$ flavor species and using standard notation, this action reads
\bea
S_F=\hspace{-0.2cm} &-&\hspace{-0.2cm} {a^3\over 2}\sum_{x,\,f,\,\mu}\bigg{[}\bar{
 \psi}_{f}(x) \left( r - \gamma_\mu\right) U_{x,\, x+a\,\mu}\psi_f(x+a\,\mu) 
+\bar{\psi}_f(x+a\,\mu)\left( r + \gamma_\mu\right)U_{x+a\,\mu,\,x}\psi_{f}(x)\bigg{]}\nonumber \\
&+&\hspace{-0.2cm} a^4 \sum_{x,\,f} (\frac{4r}{a}+m^f_0+ i \mu_0^f \gamma_5\tau^3 )\bar{\psi}_{f}(x)\psi_f(x) \nonumber \\
&-&\hspace{-0.2cm} {a^5\over 4}\,\sum_{x,\,f,\,\mu,\,\nu} r\,c_{\rm SW}\,\bar{\psi}_{f}(x)
\smn F_{\mu\nu}(x) \psi_f(x),
\label{clover}
\eea
where the Wilson parameter $r$ is henceforth set to $r=1$, $f$ is a flavor
index, $\smn =[\gamma_\mu,\,\gamma_\nu]/2$ and the clover
coefficient $c_{\rm SW}$ as well as the twisted mass parameter $\mu_0^f$
are kept as free parameters throughout. $F_{\mu\nu}$ is the standard
clover discretization of the gluon field tensor~\cite{Sheikholeslami:1985ij}.

\vspace{0.75cm}
For the non-perturbative calculation, we consider 
the purely twisted mass fermion action (no clover term), which for two
degenerate flavors of quarks is given by: 
\be
S_F= a^4\sum_x  \overline{\chi}(x)\bigl(\frac{1}{2}
\gamma_{\mu}(\overrightarrow\nabla_{\mu} +
\overrightarrow\nabla_{\mu}^{*})
-\frac{ar}{2} \overrightarrow\nabla_{\mu}
\overrightarrow\nabla^*_{\mu}
 + m_0 
+ i \mu_0 \gamma_5\tau^3  \bigr ) \chi(x) \, ,
\label{ch5_action}
\ee
with $\tau^3$ the Pauli matrix acting in the isospin space, and $\mu_0$
the bare twisted mass. Maximally twisted Wilson quarks are
obtained by setting the untwisted bare quark mass $m_0$ to its
critical value $m_{\rm cr}$, while the twisted quark mass parameter
$\mu_0$ is kept non-vanishing in order to give the light quarks their
mass~\cite{Frezzotti:2000nk,Frezzotti:2003ni}. The physical quantities
extracted from lattice simulations employing maximally twisted Wilson
quarks are automatically ${\cal O}(a)$ improved~\cite{Frezzotti:2003ni}. In
Eq.~(\ref{ch5_action}) the quark fields $\chi$ are in the 
so-called ``twisted basis''. The ``physical basis'' is obtained for
maximal twist by the simple transformation:
\be
\psi(x)=\exp\left(\frac {i\omega} 2\gamma_5\tau^3\right) \chi(x),\qquad
\overline\psi(x)=\overline\chi(x) \exp\left(\frac {i\omega}
2\gamma_5\tau^3\right),\qquad \omega=\frac{\pi}{2}.
\label{ch5_rotations}
\ee
In terms of the physical fields the action is given by:
\be
S_F^{\psi}= a^4\sum_x  \overline\psi(x)\left(\frac 12 \gamma_\mu 
[\overrightarrow\nabla_\mu+\overrightarrow\nabla^*_\mu]-i \gamma_5\tau^3 \left(- 
\frac{a}{2} \;\overrightarrow\nabla_\mu\overrightarrow\nabla^*_\mu+ m_{\rm cr}\right ) 
+ \mu_0 \right ) \psi(x).
\label{ch5_S_phy}
\ee
One can check that this action is equivalent to the action in the
twisted basis given by Eq.~(\ref{ch5_action}), by performing the rotations 
defined in Eq.~(\ref{ch5_rotations}) and identifying $m_0=m_{\rm cr}$.

\vspace{0.75cm}
For gluons we employ the Symanzik improved action, involving
Wilson loops with 4 and 6 links ($1\times 1$ {\em plaquette},
$1\times 2$ {\em rectangle}, $1\times 2$ {\em chair}, and $1\times
1\times 1$ {\em parallelogram} wrapped around an elementary 3-d
cube), which is given by 
\bea
\hspace{-1cm}
S_G=\frac{2}{g_0^2} \Bigl[ &c_0& \sum_{\rm plaq.} {\rm Re\,Tr\,}\{1-U_{\rm plaq.}\}
\,+\, c_1 \sum_{\rm rect.} {\rm Re \, Tr\,}\{1- U_{\rm rect.}\} 
\nonumber \\ 
+ &c_2& \sum_{\rm chair} {\rm Re\, Tr\,}\{1-U_{\rm chair}\} 
\,+\, c_3 \sum_{\rm paral.} {\rm Re \,Tr\,}\{1-U_{\rm paral.}\}\Bigr].
\label{Symanzik}
\eea
The coefficients $c_i$ can in principle be chosen arbitrarily, subject
to the following normalization condition, which ensures the correct
classical continuum limit of the action:
\be
c_0 + 8 c_1 + 16 c_2 + 8 c_3 = 1.
\label{norm}
\ee
Some popular choices of values for $c_i$ used in numerical simulations
will be considered in this work, and are itemized in Table~\ref{tab1}
(the acronym TILW represent the Tadpole Improved L\"uscher-Weisz action);
they are normally tuned in a way as to ensure ${\cal O}(a^2)$
improvement in the pure gluon sector. In our non-perturbative
computation presented here we employ the tree-level Symanzik action
($c_0=5/3$, $c_1=-1/12$, $c_2=c_3=0$).
Our 1-loop Feynman diagrams do not involve pure gluon vertices, and
the gluon propagator depends only on three combinations of the
Symanzik parameters: 
\bea
&&C_0 \equiv c_0 + 8 c_1 + 16 c_2 + 8 c_3 \,=1,\nonumber \\
&&C_1 \equiv c_2 + c_3, \\
&&C_2 \equiv c_1 - c_2 - c_3.\nonumber
\eea
Therefore, with no loss of generality we can set $c_2=0$.

\begin{table}
\begin{center}
\begin{minipage}{11cm}
\begin{tabular}{lr@{}lr@{}lr@{}l}
\hline
\hline
\multicolumn{1}{c}{Action}&
\multicolumn{2}{c}{$c_{0_{\Large{\phantom{A}}}}^{{\Large{\phantom{A}}}}$} &
\multicolumn{2}{c}{$c_1$} &
\multicolumn{2}{c}{$c_3$} \\
\hline
\hline
$\,\,$Plaquette               &  1&.0         &  0&         &  0&               \\
$\,\,$Symanzik                &  5&/3         & -1&/12      &  0&               \\
$\,\,$TILW, $\beta c_0=8.60$  &  2&.3168064   & -0&.151791  & -0&.0128098$\,\,$  \\
$\,\,$TILW, $\beta c_0=8.45$  &  2&.3460240   & -0&.154846  & -0&.0134070$\,\,$  \\
$\,\,$TILW, $\beta c_0=8.30$  &  2&.3869776   & -0&.159128  & -0&.0142442$\,\,$  \\
$\,\,$TILW, $\beta c_0=8.20$  &  2&.4127840   & -0&.161827  & -0&.0147710$\,\,$  \\
$\,\,$TILW, $\beta c_0=8.10$  &  2&.4465400   & -0&.165353  & -0&.0154645$\,\,$  \\
$\,\,$TILW, $\beta c_0=8.00$  &  2&.4891712   & -0&.169805  & -0&.0163414$\,\,$  \\
$\,\,$Iwasaki                 &  3&.648       & -0&.331     &  0&                \\
$\,\,$DBW2                    & 12&.2688      & -1&.4086    &  0&                \\
\hline
\hline
\end{tabular}
\end{minipage}
\end{center}
\vspace{-0.3cm}
\caption{Input parameters $c_0$, $c_1$, $c_3$.}
\label{tab1}
\end{table}

\subsection{Definition of operators and Renormalization condition}

The ultra-local bi-fermion operators considered in this work are the following:
\begin{eqnarray}
   \Op_S^a &= \bar \chi \tau^a \chi           &= \begin{cases} \bar \psi \tau^a          \psi   & a=1,2 \\
                                                             -i\bar \psi \gamma_5 \eins  \psi   & a=3 \end{cases} \\
   \Op_P^a &= \bar \chi \gamma_5\tau^a \chi   &= \begin{cases} \bar \psi \gamma_5 \tau^a \psi   & a=1,2 \\
                                                             -i\bar \psi          \eins  \psi   & a=3 \end{cases} \\
   \Op_V^a &= \bar \chi \gamma_\mu\tau^a \chi &= \begin{cases} \bar \psi  \gamma_5\gamma_\mu \tau^2 \psi   & a=1 \\
                                                              -\bar \psi  \gamma_5\gamma_\mu \tau^1 \psi   & a=2 \\
                                                               \bar \psi  \gamma_\mu         \tau^3 \psi   & a=3 \end{cases} \\
   \Op_A^a &= \bar \chi \gamma_5\gamma_\mu\tau^a \chi &= \begin{cases} \bar \psi  \gamma_\mu        \tau^2 \psi   & a=1 \\
                                                                      -\bar \psi  \gamma_\mu        \tau^1 \psi   & a=2 \\
                                                                       \bar \psi  \gamma_5\gamma_\mu\tau^3 \psi   & a=3 \end{cases}\\
   \Op_T^a &= \bar \chi \smn\tau^a \chi &= \begin{cases} \bar \psi  \smn      \tau^a \psi   & a=1,2 \\
                                                                  -i\bar \psi  \gamma_5\smn\eins\psi   & a=3 \end{cases}\\
   \Op_{Tp}^a &= \bar \chi \gamma_5\smn\tau^a \chi &= \begin{cases} \bar \psi \gamma_5 \smn      \tau^a \psi   & a=1,2 \\
                                                                  -i\bar \psi \smn\eins\psi   & a=3 \end{cases}
\end{eqnarray}
For convenience we have included $\Op_{Tp}^a$ even though its components
are related to those of $\Op_{T}^a$. 
We denote the corresponding $Z$-factors by $\ZS^a,\,\ZP^a,\,\ZV^a,\,\ZA^a,\ZT^a,Z_{\rm Tp}^a$.
In a massless renormalization scheme the renormalization constants are defined in the chiral limit,
where iso-spin symmetry is recovered. Hence $Z$-factors become independent of 
the isospin index $a=1,2,3$
and we drop the $a$ index on the $Z$-factors from here on. Still note that, for instance, the physical
$\bar\psi \gamma_\mu \tau^1 \psi$ is renormalized with $\ZA$ while $\bar\psi \gamma_\mu \tau^3 \psi$
needs $\ZV$, which differ from each other even in the chiral limit.

\vspace{0.75cm}

The renormalization constants are computed both perturbatively and
non-perturbatively in the RI$'$-MOM
scheme at different renormalization scales.
We translate them to the ${\overline{\rm MS}}$ scheme at $\mu=$2~GeV using a conversion
factor computed in perturbation theory to ${\cal O}(g^6)$ as described in Section~\ref{conversion}.
The $Z$-factors are determined by imposing the following conditions in
the massless theory, i.e., at critical mass and vanishing twisted mass
\begin{eqnarray}
   \Zq = \frac{1}{12} {\rm Tr} \left[(S^L(p))^{-1}\, S^{(0)}(p)\right] \Bigr|_{p^2=\mu^2}  \label{Zq_cond}\\[2ex]
   \Zq^{-1}\,Z_{\cal O}\,\frac{1}{12} {\rm Tr} \left[\Gamma^L(p) \,\Gamma^{(0)-1}(p)\right] \Bigr|_{p^2=\mu^2} &=& 1\, ,
\label{renormalization cond}
\end{eqnarray}
where the trace is taken over spin and color indices, $\mu$ is the
renormalization scale, while $S^L$ and $\Gamma^L$ correspond to the
perturbative or non-perturbative results, and
$S^{(0)}$ is the tree-level result for the propagator defined as:
\be
    S^{(0)}(p) = \frac{-i
     \sum_\rho \gamma_\rho \sin(p_\rho)}{\sum_\rho\sin(p_\rho)^2}\,,
\label{meth2a}
\ee
while $\Gamma^{(0)}$ is the tree-level value for the fermion operators S, P, V, A, T, T$'$, that is 
\be
\Gamma^{(0)}(p) = \openone,\,\,\gamma_5,\,\, \gamma_\mu,\,\,\gamma_5\,\gamma_\mu,
\,\, \gamma_5\,\smn,\,\,\smn\,,
\label{meth2}
\ee
respectively.
The trace is taken over spin and color indices. For alternative
renormalization prescriptions the reader can refer to
Ref.~\cite{Alexandrou:2010me}.

The choices for $S^{(0)}$ and $\Gamma^{(0)}$ given in
Eqs.~(\ref{meth2a}) - (\ref{meth2}) are optimal, since we obtain
$\Zq = 1$, $Z_{\cal O} = 1$ when the gauge field is set to unity.
Similarly, in the perturbative computation this condition leads to
$\Zq = 1$, and $Z_{\cal O} = 1$ at tree-level.


\section{Corrections to the Fermion Propagator}
\label{prop}
\begin{center}
\begin{figure}[!h]
\centerline{\psfig{figure=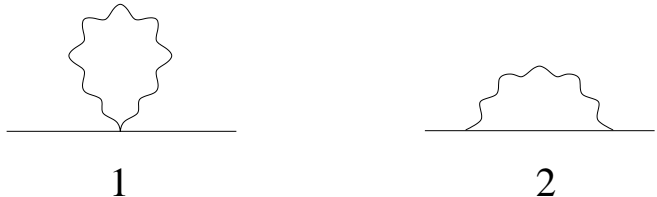,height=2.6truecm}}
\begin{minipage}{10cm}
\caption{One-loop diagrams contributing to the 
fermion propagator. Wavy (solid) lines represent gluons (fermions).}
\label{figprop1}
\end{minipage}
\end{figure}
\end{center}

The fermion propagator of the interacting theory is given by the
following 2-point correlation function (Green's function), with the
various quantities computed in perturbation theory: 
\be
\langle \chi_{\alpha}^{a,f}(x) \Bar\chi_{\beta}^{b,g}(y)\rangle =  
\int_{-\frac{\pi}{a}}^{\frac{\pi}{a}}\frac{d^4p}{(2\pi)^4} 
\delta^{a\,b}\,\,e^{i\,p (x-y)}
\bigg( S_{tree}\, \cdot \,\sum_{n=0}^\infty \left(-S^{-1}_{\rm amp}(p)\, \cdot \, S_{tree}\right)^n \bigg)_{\alpha \, \beta}^{f \, g},
\label{fermion_propagator_interacting}
\ee
with $S^{-1}_{\rm amp}(p)$ being the amputated, 1PI 2-point function in momentum
space, computed perturbatively up to a desired order.
$S_{tree}$ is the tree-level propagator for the twisted mass action and it is
given by 
{\small{
\be
 S_{tree}= \frac{1}{i\,\pcircslash + M(p) + i \mu_0 \gamma^5 \tau^3} ,\quad
{\scriptsize{
 M(p)\equiv m^f_0 + \frac{2}{a} \sum_\mu \sin^2(a\,p_\mu/2),\quad
\pcircslash \equiv \sum_\mu \gamma_{\mu}\frac{1}{a} \sin(a\,p_\mu)}}\,,
\ee
}}

\noindent where $\alpha,\,\beta$ are Dirac indices, $f_1,\,f_2$ are flavor
indices in the fundamental representation of $SU(N_F)$, and $a,\,b$
are color indices in the fundamental representation of $SU(N_c)$. The
dot product runs over flavor and Dirac indices. Due to the diagonal
form of the $\tau^3$ matrix, and since we are studying the case of
only two degenerate quarks (up/down) we can simplify the expression of
$S_{tree}$ and omit $\tau^3$ by giving a flavor index to the twisted
mass parameter, and at the same time we take $m^f_0\to m_0$: 
\be
S_{tree}= \frac{1}{i\,\pcircslash + M(p) + i \mu_0^{(f)} \gamma^5},
\ee
where now $\mu_0^{(1)}=+\mu_0$ for the up quark propagator and
$\mu_0^{(2)}=-\mu_0$ for the down quark propagator. 
The 1-loop Feynman diagrams that enter our 2-point amputated Green's function
calculation ($S^{-1}_{amp}$), are depicted in Fig.~\ref{figprop1}. 

\bigskip
For the algebraic operations involved in evaluating Feynman
diagrams, we make use of our symbolic package in Mathematica. In a
nutshell, the required steps for the computation of a Feynman diagram
are the following (the reader can find more details in Ref.~\cite{Constantinou:2009tr}):

$\bullet\,$ A preliminary expression for each diagram can be obtained by
contracting the appropriate vertices, which is performed automatically
once the algebraic expression of the vertices and the topology
(``incidence matrix'') of the diagram are specified. To limit the
proliferation of the algebraic expressions we exploit symmetries of
the theory, and we simplify the color dependence, Dirac matrices and
tensor structures.

{$\bullet\,$ The ${\cal O}(a^2)$ computation introduces several
complications, especially when isolating logarithms and Lorentz
non-invariant terms, which leads to a whole family of infrared
divergent integrals. These can be reduced to a minimal set of
approximately 250 'basis' integrals. This is achieved by converting
all propagator denominators to a standard form $(\hat{q}^2 + M^2)^{-1}$ using two
kinds of subtractions, one for the fermion propagator
\be
\frac{1}{\tilde q^2}=\frac{1}{\hat{q}^2}+
\Bigg{\{}\frac{4\sum_\mu \sin^4(q_\mu/2)-4\left(\sum_\mu
  \sin^2(q_\mu/2)\right)^2 -4\,m_0\, \sum_\mu \sin^2(q_\mu/2)}
{\tilde q^2 \,\hat{q}^2}\Bigg{\}}\label{sub1}
\ee
where the denominator of the fermion propagator, $\tilde q^2$, is defined as
\be
\tilde q^2 = \sum_\mu \sin^2(q_\mu)+
\left(m_0+\frac{1}{2} \hat{q}^2\right)^2 + (\mu_0^{(f)})^2,\quad
\hat{q}^2 =4\sum_\mu \sin^2(\frac{q_\mu}{2}),\quad M^2 = m_0^2 + \mu^2_0\,,
\ee
and one for the gluon propagator :
\bea
D(q)= D_{plaq}(q) + \Big{\{}D(q) - D_{plaq}(q) \Big{\}} =  D_{plaq}(q) + D_{plaq}(q) \Big{\{}D^{-1}_{plaq}(q) - D^{-1}(q) \Big{\}} D(q)\,.
\label{sub2}
\eea
$D$ is the $4\times4$ Symanzik gluon propagator; the
expression for the matrix $\left(D^{-1}_{plaq}(q) - D^{-1}(q)
\right)$, which is ${\cal O}(q^4)$, is independent of the gauge
parameter, $\lambda$, and it can be easily obtained in closed
form. Moreover, we have
\be
\bigl(D_{plaq}(q)\bigr)_{\mu\nu} = \frac{\delta_{\mu\nu}}{\hat{q}^2} -
(1-\lambda)\frac{\hat{q}_\mu\,\hat{q}_\nu}{(\hat{q}^2)^2}
\ee
Terms in curly brackets of Eqs.~(\ref{sub1}) and (\ref{sub2}) are less
IR divergent than their unsubtracted counterparts, by one or 
two powers in $a$. These subtractions are performed
iteratively until all divergent integrals (initially
depending on the fermion and the Symanzik propagator) are expressed in
terms of the gluon propagator, $(\hat{q}^2 + M^2)^{-1}$. 
The computation of the divergent integrals is performed in a
non-integer number of dimensions $D>4$. Ultraviolet divergences are
explicitly isolated \`a la Zimmermann and evaluated as in the
continuum. The remainders are $D$-dimensional, parameter-free, zero
external momentum lattice integrals which can be recast in terms of
Bessel functions, and finally expressed as sums of a pole part plus
numerical constants. 

A small subset of the infrared divergent integrals, shown in
Appendix~\ref{appA}, contains the most demanding ones in the list;
they must be evaluated to two further orders in $a$, beyond the order
at which an IR divergence initially sets in. As a consequence,
their evaluation requires going to $D>6$ dimensions. A correct way to
evaluate strong divergent integrals is given in detail in a previous
publication~\cite{Constantinou:2009tr}.

$\bullet\,$ The required numerical integrations of the algebraic
expressions for the loop integrands are performed by highly optimized
Fortran programs; these are generated by our Mathematica `integrator'
routine. Each integral is expressed as a sum over the discrete
Brillouin zone of finite lattices, with varying size $L$ ($4^4\leq L^4
\leq 128^4$), and evaluated for all values of the Symanzik
coefficients listed in Table~\ref{tab1}.

$\bullet\,$ The last part of the evaluation is the extrapolation of
the numerical results to infinite lattice size. This procedure entails
a systematic error, which is reliably estimated, using a complex
inference technique; for one-loop quantities we expect a relative
error smaller than $10^{-7}$.

\bigskip
Next, we provide the total expression for the inverse fermion
propagator $S^{-1}_{\rm pert.}(p)$, computed up to 1-loop in
perturbation theory. Here we should point out that for
dimensional reasons, there is a global prefactor $1/a$ multiplying our
expressions for the inverse propagator, and thus, the ${\cal O}(a^2)$
correction is achieved by considering all terms up to ${\cal O}(a^3)$. 
The most general expression for the inverse propagator appears in
the Mathematica file Zfactors.m (see Appendix~\ref{appD} for notation). 
In the main text we provide a more compact expression, for special
values of the various parameters, that is tree-level Symanzik improved
gluon action, $\csw=0$, Landau gauge ($\lambda$=0), but we keep the
Lagrangian mass and the twisted mass parameter general.

\begin{minipage}{17.5cm}
\bdg[style={\normalsize}, spread={8pt}]
\bdm[label={Sinv}]
\label{propagator}
S^{-1}_{\rm pert.} \hiderel{=} 
m
+i\,\mu\,  \gamma^5
+i\,\pslash
+\frac{\Red{a}\,p^2}{2}
-\frac{\Red{a^2} \, i \,\pslash^3\,}{6}
\edm
\bdms
+\gttwo \Green{\left\{}
               -13.0232725(2)\, i\,\pslash\,
               + m \left(0.5834586(2)
                                       -3 \ln[a^2 M^2+a^2 p^2]
                                       -\frac{3 M^2 \ln[1+\frac{p^2}{M^2}]}{p^2}
                                \right)
               +i\,\mu \gamma^5 \left(8.7100834(2)
                                             -3 \ln[a^2 M^2+a^2 p^2]
                                             -\frac{3 M^2 \ln[1+\frac{p^2}{M^2}]}{p^2}
                                      \right) 
               +\Red{a} \Red{\Bigg{[}}
                          \left(-10.69642965(5)\,p^2
                                               -0.8530378(3)\,m^2
                                               -1.842911859(4)\,M^2
                                               +\frac{6 M^2 m^2 \ln[1+\frac{p^2}{M^2}]}{p^2}
                                               +\left(\frac{3 p^2}{2}+3 m^2+\frac{3 M^2}{2}\right) \ln[a^2 M^2+a^2 p^2] 
                                            \right)
                         +i\,m\,\pslash\, \left(0.3393996(2)
                                                                   +\frac{3 M^2}{2 p^2}
                                                                   +\frac{3}{2} \ln[a^2 M^2+a^2 p^2]  
                                                                   -\frac{3 M^4 \ln[1+\frac{p^2}{M^2}]}{2 (p^2)^2}
                                                          \right) 
                         +i\,\mu\,m\, \gamma^5 \left(-6.68582372(4)
                                                        +3 \ln[a^2 M^2+a^2 p^2]
                                                        +\frac{6 M^2 \ln[1+\frac{p^2}{M^2}]}{p^2}
                                                      \right)
                   \Red{\Bigg{]}}
               +\Red{a^2} \Red{\Bigg{[}} m  \left(2.3547298(1)\,p^2 
                                                   +2.3562747(1)\,m^2
                                                   +3.46524146(4)\,M^2
                                                   -\frac{M^4}{6 p^2}
                                                   +\frac{M^6}{3 (p^2)^2}
                                                   -\frac{3 m^2 p^2}{2 (M^2+p^2)}
                                                   -\left(\frac{p^2}{4} +3 m^2 +\frac{11 M^2}{3}\right) \ln[a^2 M^2+a^2 p^2] 
                                                   +\left(\frac{p^2}{3} -9 m^2 -2 M^2 -\frac{M^6}{3 (p^2)^2}\right) \frac{M^2 \ln[1+\frac{p^2}{M^2}]}{p^2}
                                             \right)
                             +i\,\mu \gamma^5 \left(0.70640552(8)\,p^2 
                                                          +6.79538844(2)\,m^2
                                                          +1.16985307(3)\,M^2
                                                          -\frac{M^4}{6 p^2}
                                                          + \frac{M^6}{3 (p^2)^2}
                                                          -\frac{3 m^2 p^2}{2 (M^2+p^2)}
                                                          -\left(\frac{p^2}{4} +3 m^2 +\frac{2 M^2}{3}\right) \ln[a^2 M^2+a^2 p^2]
                                                          +\left(\frac{p^2}{3} -9 m^2 -\frac{M^2}{2} -\frac{M^6}{3 (p^2)^2}\right) \frac{M^2 \ln[1+\frac{p^2}{M^2}]}{p^2} 
                                                    \right) 
                             +\frac{\left(\sum_\rho\,p^4_\rho\right)\, (m  +i\,\mu \gamma^5)}{p^2} \left(\frac{1}{2}
                                                                                            -\frac{2 M^2}{9 p^2}
                                                                                            +\frac{M^4}{3 (p^2)^2}
                                                                                            -\frac{2 M^6}{3 (p^2)^3}
                                                                                            +\left(-\frac{1}{3} +\frac{2 M^6}{3 (p^2)^3}\right) \frac{M^2 \ln[1+\frac{p^2}{M^2}]}{p^2} 
                                                                                      \right)
                            + i\,\pslash\, \left(1.1471643(7)\,p^2 
                                                                       -0.2145514(2)\,m^2
                                                                       +1.15904388(6)\,M^2
                                                                       -\frac{9 m^2 M^2}{2 p^2}
                                                                       -\frac{209 M^4}{360 p^2}
                                                                       -\frac{M^6}{240 (p^2)^2}
                                                                       +\frac{7 M^8}{40 (p^2)^3}
                                                                       -\left(\frac{73 p^2}{360}+\frac{3 m^2}{2}+\frac{2 M^2}{3}\right) \ln[a^2 M^2+a^2 p^2] 
                                                                       + \left(\frac{1}{24}+\frac{9 m^2}{2 p^2}+\frac{43 M^2}{72 p^2}-\frac{M^4}{12 (p^2)^2}-\frac{7 M^6}{40 (p^2)^3}\right) \frac{M^4 \ln[1+\frac{p^2}{M^2}]}{p^2} 
                                                                   \right) 
                             +i \,\pslash^3\, \left(4.2478764(2)
                                                                        -\frac{67 M^2}{120 p^2}
                                                                        +\frac{M^4}{120 (p^2)^2}
                                                                        -\frac{8 M^6}{15 (p^2)^3}
                                                                        +\frac{7 M^8}{30 (p^2)^4}
                                                                        -\frac{157}{180} \ln[a^2 M^2+a^2 p^2]
                                                                        +\left(\frac{1}{2}+\frac{5 M^2}{18 p^2}+\frac{5 M^4}{12 (p^2)^2}-\frac{7 M^6}{30 (p^2)^3}\right) \frac{M^4 \ln[1+\frac{p^2}{M^2}]}{(p^2)^2} 
                                                                      \right)
                              +\frac{i\,\left(\sum_{\rho'}\,p^4_{\rho'}\right)\,\,\pslash\,}{p^2} \left(\frac{7}{240}
                                                                                    +\frac{M^2}{48 p^2}
                                                                                    +\frac{67 M^4}{72 (p^2)^2}
                                                                                    +\frac{13 M^6}{24 (p^2)^3}
                                                                                    -\frac{7 M^8}{12 (p^2)^4}
                                                                                    +\left(-\frac{5}{12}-\frac{5 M^2}{4 p^2}-\frac{M^4}{4 (p^2)^2}+\frac{7 M^6}{12 (p^2)^3}\right) \frac{M^4 \ln[1+\frac{p^2}{M^2}]}{(p^2)^2} 
                                                                       \right)
                        \Red{\Bigg{]}}
        \Green{\right\}} 
+ {\cal O}(\Red{a^3},\Green{g^4})
\edms
\edg
\end{minipage}
\vspace*{0.5cm}

\noindent
where $\pslash=\sum_\rho \gamma^\rho\,p_\rho$ and $\pslash^3=\sum_\rho \gamma^\rho\,p_\rho^3$.
To make the above expressions less complicated we defined $m\equiv
m_0$ and $M^2=m_0^2+ \mu_0^2$. We would like to
point out that the up quark propagator is obtained by the choice
$\mu^{(1)}=+\mu_0$, while for the down quark propagator one should choose 
$\mu^{(2)}=-\mu_0$. Moreover, $\tilde{g}^2 \equiv \frac{g^2C_F}{16\pi^2}$ and 
$C_F \equiv \frac{N_c^2-1}{2N_c}$. 

\vspace{0.5cm}
Another byproduct of this part of the computation is the additive
critical fermion mass; its general expression depends on $\csw$ and
the Symanzik parameters. These are terms proportional to $1/a$ that
have been left out of Eq.~(\ref{propagator}) for conciseness:

\be
\label{criticalmass}
m_{\rm cr} = -\frac{\tilde{g}^2}{a} \Big[
        \ve_m^{(1)} + \ve_m^{(2)}\csw +\ve_m^{(3)}\csw^2 \Big] + \frac{1}{a} {\cal O}(g^4)\,.
\ee
The quantities $\ve_m^{(i)}$ (listed in Table~\ref{tabcriticalmass}) are numerical
coefficients depending on the Symanzik parameters.

\begin{table}[!h]
\begin{center}
\begin{minipage}{12cm}
\begin{tabular}{lr@{}lr@{}lr@{}l}
\hline
\hline
\multicolumn{1}{c}{Action}&
\multicolumn{2}{c}{$\ve_{m_{\phantom{A}}}^{{(1)}^{\phantom{A}}}$} &
\multicolumn{2}{c}{$\ve_m^{(2)}$} &
\multicolumn{2}{c}{$\ve_m^{(3)}$} \\
\hline
\hline
$\,\,$Plaquette                 &-&51.4347118(2)$\,\,\,\,\,$    &&13.73313097(5)$\,\,\,$    &&5.71513853(1)  \\
$\,\,$Symanzik                  &-&40.44324019(7)               &&11.94821988(5)            &&4.662672112(4)   \\
$\,\,$TILW (8.60)$\,\,\,\,\,$   &-&34.17747288(3)               &&10.76516514(3)            &&3.998348778(3)   \\
$\,\,$TILW (8.45)               &-&33.9488671(1)                &&10.71947605(3)            &&3.97345187(1) \\
$\,\,$TILW (8.30)               &-&33.6344391(1)                &&10.65632621(4)            &&3.939135834(8)   \\
$\,\,$TILW (8.20)               &-&33.43979350(6)               &&10.61705314(7)            &&3.917851255(1)  \\
$\,\,$TILW (8.10)               &-&33.1892274(1)                &&10.56629305(3)            &&3.890401337(1)   \\
$\,\,$TILW (8.00)               &-&32.87904072(9)               &&10.50313393(3)            &&3.856345868(2)   \\ 
$\,\,$Iwasaki                   &-&26.07292275(7)               &&$\,\,\,$9.01533524(3)     &&3.1061330684(3)   \\
$\,\,$DBW2                      &-&11.5127475(2)                &&$\,\,\,$4.9953066(1)      &&1.351772367(3)$\,\,$   \\
\hline
\hline
\end{tabular}
\end{minipage}
\end{center}
\vspace{-0.3cm}
\caption{Numerical results for the coefficients $\ve_m^{(1)},\,\ve_m^{(2)},\,\ve_m^{(3)}$
  (Eq.~(\ref{criticalmass})) for different actions. The
systematic errors in parentheses come from the extrapolation over
finite lattice size, $L \to \infty$.}
\label{tabcriticalmass} 
\end{table}

\section{Corrections to Fermion Bilinear Operators}
\label{oper}

In the context of this work we also study the perturbative ${\cal O}(a^2)$
corrections to Green's functions of local fermion operators (currents)
that have the form:
\be
 O_\Gamma=\sum_z\sum_{\alpha''\, \beta''}\left(\bar\chi^{f''}_{\alpha''}(z)\,\Gamma_{\alpha''\,\beta''}\,\chi^{g''}_{\beta''}(z)\right).
\ee
We restrict ourselves to forward matrix elements (i.e. 2-point Green's
functions, zero momentum operator insertions). The symbol $\Gamma$
corresponds to the following set of products of the Euclidean Dirac matrices:
\be
\Gamma \in \{S,P,V,A,T,T'\}\equiv \{\openone,\, \gamma^5,\, \gamma_\mu,\,
\gamma^5\gamma_\mu,\, \gamma^5\smn,\,\smn\}; \qquad \smn=\frac{1}{2}[\gamma_\mu,\gamma_\nu],
\label{Gamma}
\ee
for the scalar $O_S$, pseudoscalar $O_P$, vector
$O_V$, axial $O_A$, tensor $O_T$ and tensor prime $O_{T'}$
operator, respectively. The matrix elements of $O_{T'}$ can be related
to those of $O_T$; this is a
nontrivial check for our calculational procedure~\cite{Constantinou:2009tr}.
The relationship between the amputated 2-point Green functions
$\Lambda_{T}$ and $\Lambda_{T'}$ is:
\be
\Lambda_{T}^{\mu\,\nu}=-\frac{1}{2}
\sum_{\mu'\,\nu'}\epsilon_{\mu\,\nu\,\mu'\,\nu'}\Lambda_{T'}^{\mu'\,\nu'}.
\ee

The matrix elements of the above set of fermion bilinear operators
can be obtained as:
\be
\langle \chi_{\alpha}^{a,f}(x) O_{\Gamma} \Bar\chi_{\beta}^{b,g}(y)\rangle = 
\int_{-\frac{\pi}{a}}^{\frac{\pi}{a}}\frac{d^4p}{(2\pi)^4} 
\delta^{a\,b}\,e^{i\,p (x-y)}
\bigg( S\, \cdot \, \Lambda_{\Gamma}^{\rm pert.}(p) \, \cdot \, S \bigg)_{\alpha \,\beta }^{f \, g},
\label{fermion_bilinear_interacting}
\ee
where $\Lambda_{\Gamma}^{\rm pert.}(p)$ is the amputated 1PI 2-point Green's function
of each operator $O_{\Gamma}$, in momentum space, which upon contraction of indices becomes:
\be
\Lambda_\Gamma^{\rm pert.}(p) = \sum_{\alpha'\, \beta'} \langle \chi_{\alpha}^f(p) 
\,\left(\bar\chi^f_{\alpha'}(p)\,\Gamma_{\alpha'\,\beta'}\,\chi^g_{\beta'}(p)\right)\,
\Bar\chi_{\beta}^g(p)\rangle^{amp.}.
\label{fermion_bilinears}
\ee
The only 1-particle irreducible Feynman diagram that enters the
calculation of the above Green's function is shown in Fig.~\ref{figbil2}.

\begin{figure}[!h]
\begin{center}
\centerline{\psfig{figure=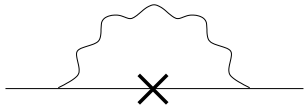,height=1.4truecm}}
\begin{minipage}{10cm}
\caption{One-loop diagram contributing to the 
bilinear operators. A wavy (solid) line represents gluons (fermions). A
cross denotes the Dirac matrix $\Gamma$.}
\label{figbil2}
\end{minipage}
\end{center}
\end{figure}

In this diagram there are two fermion propagators, for which we allowed
different $\mu$-values, in order to have more general results. In
other words, each of the two internal fermion lines on the left and on the right of the
operator insertion (see Fig.~\ref{figbil2}) can independently represent the up or
down propagator. For the evaluation of the $Z$-factors, we keep the two
flavors independent.
The amputated Green†¢s functions of the operators $O_{\Gamma}$ are
given in the Mathematica file Zfactors.m. 
As mentioned above, one may choose the two
fermion propagators of the diagram to correspond either to the up or
down quark, thus there are two twisted mass parameters $\mu^{(1)}$, and
$\mu^{(2)}$. These can have any sign and the only restriction is: $|\mu^{(1)}| = |\mu^{(2)}|$. 

\section{Quark Field and Quark Bilinear Renormalization Constants in
the RI$'$-MOM Scheme}
\label{renorm}
An operator renormalization constant (RC) can be thought of as the link
between its matrix element, regularized on the lattice, and its
renormalized continuum counterpart. The RCs of lattice operators are
necessary ingredients in the prediction
of physical probability amplitudes from lattice matrix elements.
In this section we present the multiplicative RCs, in the RI$'$-MOM scheme, of the quark field ($\Zq^{\rm pert.}$) and quark
bilinear operators ($Z_{\rm \Gamma}^{\rm pert.}$), obtained by using the perturbative
expressions of $S^{-1}(p)$ and $\Lambda_\Gamma(p)$.

The RI$'$-MOM renormalization scheme consists in imposing that the forward
amputated Green function $\Lambda_\Gamma(p)$, computed in the chiral
limit and at a given (large Euclidean) scale $p^2=\mu^2$, is equal to
its tree-level value. In practice, the renormalization condition is
implemented by requiring that in the chiral limit\footnote{A simpler
version of Eq.~(\ref{eq:rimom}) is given by the relation: 
  \be
   \Zq^{-1} Z_{\rm \Gamma} \frac{1}{4}{\rm Tr}\Big{[}\Lambda_\Gamma(p)\,\cdot\,\Lambda_\Gamma^{\rm
       tree}\Big{]}_{p_{\rho}=\mu_{\rho}} =  
   \frac{1}{4}{\rm Tr}\Big{[}\Lambda_\Gamma^{\rm tree} \,\cdot\,\Lambda_\Gamma^{\rm tree}\Big{]},
  \ee
where $\Lambda_\Gamma^{\rm tree}$ is the tree-level value of $\Lambda_\Gamma(p)$.}:
\be
\Zq^{-1} Z_{\rm \Gamma} \, {\cal V}_\Gamma(p)\vert_{p_{\rho}=\mu_{\rho}}=1 , 
\quad {\cal V}_\Gamma(p) \equiv \frac{1}{4} {\rm Tr} \Big{[}\Lambda_\Gamma(p)\,\cdot\, P_\Gamma\Big{]} ,
\label{eq:rimom}
\ee
where $P_\Gamma$ are the Dirac projectors defined as follows: 
\be
P_{\Gamma} \in \{P_S,P_P,P_V,P_A,P_T,P_{T'} \} \equiv 
\{ \openone,\, \gamma^5,\, \gamma_\mu,\,
-\gamma^5 \gamma_\mu,\, -\gamma^5\smn,\,-\smn \} ;
\ee
they are chosen to obey the relation ${\rm Tr}[\Gamma \cdot
P_{\Gamma}]\equiv 4$. The traces are always taken over the spin indices.
The quark field RC $\Zq$, which enters
Eq.~(\ref{eq:rimom}), is obtained by imposing, again in the chiral limit, the
condition\footnote{Strictly speaking, the renormalization condition of
Eq.~(\ref{eq:zqri}) defines the so called RI$'$ scheme. In the
original RI-MOM scheme the quark field renormalization condition reads:
  \be
  \Zq^{-1} \,\frac{-i}{16}\,{\rm Tr}\left[\gamma_\mu \frac{\partial S_q(p)^{-1}}
  {\partial p_\mu}\right] _{p^2=\mu^2}=1 \,.
  \ee
The two schemes differ in the Landau gauge at the N$^2$LO.}:

\be
\Zq^{-1} \, {\cal V}_q(p) \vert_{p_{\rho}=\mu_{\rho}}=1 , 
\quad {\cal V}_q(p) \equiv 
-\frac{i}{4}{\rm Tr}\Bigg{[} 
\frac{\frac{1}{a}\sum_{\rho}\gamma_\rho \sin(a\,p_\rho)}{\frac{1}{a^2}
  \sum_{\rho} \sin^2(a\,p_\rho)}\,\cdot\,S^{-1}(p)\Bigg{]}.
\label{eq:zqri}
\ee

We compute $\Zq$ in the RI$'$-MOM renormalization scheme, defined in
Eq.~(\ref{Zq_cond}), which can be Taylor expanded up to ${\cal
  O}(a^2)$ terms. This leads to: 

\bea
\Zq = &-& \frac{i}{4}{\rm Tr}\Bigg{[} \frac{\sum_{\rho} \gamma_\rho ( p_\rho -\frac{a^2}{6}
      p^3_\rho)}{\sum_{\rho} p^2_\rho}\Bigg(1+\frac{a^2}{3}\frac{\sum_{\rho}
    p_\rho^4}{\sum_{\rho} p^2_\rho} \Bigg)
  \,\cdot\, S^{-1}_{\rm
    1-loop}(p)\Bigg{]}_{p_{\rho}=\mu_{\rho}} \hspace{-0.15cm} + {\cal O}(a^4\,g^2,g^4)\nonumber \\[3ex]
= &-& \frac{i}{4}{\rm Tr}\Bigg{[}
\frac{\pslash}{p^2} \,\cdot\, S^{-1}_{\rm 1-loop}(p) -
\frac{a^2}{3} \Bigg(\frac{1}{2} \frac{\pslash\,^3}{p^2} -
\frac{\pslash\,p4}{(p^2)^2} \Bigg)
\,\cdot\, S^{-1}_{\rm 1-loop}(p)
\Bigg{]}_{p_{\rho}=\mu_{\rho}} \hspace{-0.15cm} + {\cal O}(a^4\,g^2,g^4)\,\,.
\label{RC2}
\eea

\noindent The trace is taken only over spin indices and
$S^{-1}_{\rm{1-loop}}$ is the inverse fermion propagator that we
computed up to 1-loop and up to ${\cal O}(a^2)$. We make the following
definitions for convenience: $p^2\equiv\sum_\rho p^2_\rho$,
$p4\equiv\sum_\rho p^4_\rho$, $\pslash=\sum_{\rho}\gamma_\rho p_\rho$
and $\pslash\,^3\equiv\sum_{\rho}\gamma_\rho p^3_\rho$.

A very important issue is that the ${\cal O}(a^2)$ terms in $Z_q$ depend not
only on $|p|$, but also on the direction
of the renormalization scale, $p_\rho$, as manifested by the presence of $\sum_\rho
p^4_\rho$:
\be
{\cal V}_q^{\rm pert.}(p) = -\frac{i}{4}{\rm Tr}\Bigg{[} \frac{\pslash
    -\frac{a^2}{6}\pslash\,^3}{p^2}\Bigg(1+\frac{a^2}{3}\frac{\sum_{\rho}
    p_\rho^4}{p^2} \Bigg)
  \,\cdot\, S^{-1}_{\rm pert.}(p)\Bigg{]} + {\cal O}(a^4\,g^2,g^4).
\ee
As a consequence, alternative renormalization prescriptions, involving
different directions of the renormalization scale $\mu_\rho=p_\rho$,
treat lattice artifacts differently.

By implementing the perturbative expressions of $S^{-1}(p)$ and
$\Lambda_\Gamma(p)$ in Eqs.~(\ref{eq:rimom}) and (\ref{eq:zqri}), we obtain the
corresponding RCs. 
For the following special choices (independent): tree-level Symanzik gauge action,
$\csw=0$, Landau gauge, and general mass $m$ the results of the
RCs under study are (for $\Zq$ we have also kept $\mu$ and ${{
 M}=\sqrt{m^2+\mu ^2}}$ as free parameters): 

\vspace*{0.5cm}
\begin{minipage}{17.5cm}
\bdg[style={\normalsize}, spread={8pt}]
\bdm[label={Zqpert}]
Z_q^{\rm pert.} \hiderel{=} 1
\edm
\bdms
+ 
\gttwo \Green{\left\{}-13.0232725(2) +\Red{a}\, m \Red{\Bigg{[}}0.3393996(2)
                          +\frac{3 \ln[a^2 M^2+a^2 p2]}{2}
                          +\frac{3 M^2}{2 p2}
                          -\frac{3 M^4 \ln[1+\frac{p2}{M^2}]}{2 p2^2}
                    \Red{\Bigg{]}}
             +\Red{a^2} \Red{\Big{[}} 1.1471643(7)\,p2
                           -0.2145514(2)\,m^2
                           +1.15904388(6)\,M^2
                           +\frac{2.1064977(2)\,p4}{p2}
                           -\frac{9 m^2 M^2}{2 p2}
                           -\frac{209 M^4}{360 p2}
                           -\frac{M^6}{240 p2^2}
                           +\frac{7 M^8}{40 p2\,^3}
                           -\left(\frac{73 p2}{360}+\frac{3 m^2}{2}+\frac{2 M^2}{3}+\frac{157 p4}{180 p2}\right)\,\ln[a^2 M^2+a^2 p2]
                           +\left(\frac{1}{24}+\frac{9 m^2}{2 p2}+\frac{43 M^2}{72 p2}-\frac{M^4}{12 p2^2}-\frac{7 M^6}{40 p2^3}\right)\,\frac{M^4 \ln[1+\frac{p2}{M^2}]}{p2}
                           +\frac{p4}{p2} \left( -\frac{43 M^2}{80 p2}
                                      +\frac{169 M^4}{180 p2^2}
                                      +\frac{M^6}{120 p2^3}
                                      -\frac{7 M^8}{20 p2^4}
                                      +\left(\frac{1}{12}-\frac{35 M^2}{36 p2}+\frac{M^4}{6 p2^2}+\frac{7 M^6}{20 p2^3}\right)\,\frac{M^4 \ln[1+\frac{p2}{M^2}]}{p2^2}
                               \right)
                    \Red{\Bigg{]}}   
        \Green{\right\}}\Bigg{|}_{p_{\rho}=\mu_{\rho}}
+ {\cal O}(\Red{a^3}\,\Green{g^2},\Green{g^4}) 
\edms
\edg
\end{minipage}
\vspace*{0.5cm}

\vspace*{0.5cm}

\noindent
The RCs of the bilinear operators have lengthy expressions and
are not shown in the main text; they are presented in
Appendix~\ref{appB}. We also include Appendix~\ref{appC} which is
related to the perturbative results appearing in our publication for RCs of
one-derivative operators~\cite{Alexandrou:2010me}.

\section{Conversion to the continuum $\rm \overline{MS}$ scheme at a
reference renormalization scale}
\label{conversion}

\subsection{Conversion factors}

In this section we provide the expressions for the conversion factors
to the $\rm \overline{MS}$ scheme, as adapted from Ref.~\cite{Gracey:2003yr}. In
our analysis we use 2-loop formulae; 3-loop corrections
for the particular expressions are at the one per cent level. We
use the following definitions for the conversion factors:
$Z_q^{\rm \overline{MS}} = C_q\,Z_q^{\rm RI'-MOM}$ 
(note we use $C_q$ in contrast to Ref.~\cite{Gracey:2003yr} where the same quantity was denoted with $C_q^{-1}$ ), and 
$Z_{\cal O}^{\rm \overline{MS}} = C_{\cal O}^{-1}\,Z_{\cal O}^{\rm RI'-MOM}$.  

\bea
\label{Cq}
C_q &=& 1 + \lambda \frac{g^2\,C_F}{16\,\pi^2}  
- \left[\left( 8 \lambda^2 + 5 \right) C_F \right. \nonumber \\
&& \left. - \left( 9 \lambda^2
- 24 \zeta(3) \lambda + 52 \lambda - 24 \zeta(3) + 82 \right) N_c
+ 14 N_F -\lambda^2 C_F^2\right] \frac{C_F}{8}\left(\frac{g^2}{16\,\pi^2}\right)^2 \\[2ex]
C_{S,P} &=& 1 - (\lambda + 4) C_F \frac{g^2}{16\,\pi^2} 
+ \left[ \left( 
24 \lambda^2 + 96 \lambda - 288 \zeta(3) + 57 \right) C_F \right. \nonumber \\
&& \left. + 166 N_F - \left( 18 \lambda^2 + 84\lambda - 432 \zeta(3) 
+ 1285 \right) N_c \right] \frac{C_F}{24}\left(\frac{g^2}{16\,\pi^2}\right)^2 \\[2ex]
C_{A,V} &=& 1 + {\cal O}(g^8)\\[2ex]
C_{T,T'} &=& 1 + \lambda C_F \frac{g^2}{16\,\pi^2} +  
\left[ \left( 216 \lambda^2 + 4320 \zeta(3) - 4815 \right) C_F - 626 N_F 
\right. \nonumber \\
&& \left. + \left( 162 \lambda^2 + 756 \lambda 
- 3024 \zeta(3) + 5987 \right) N_c \right] \frac{C_F}{216}\left(\frac{g^2}{16\,\pi^2}\right)^2
\label{CT}
\eea
The variables $g$, $\lambda$ correspond to the RI$'$
scheme coupling constant and covariant gauge parameter (defined in
Ref.~\cite{Gracey:2003yr}); in the Landau gauge, $\lambda=0$.
$\zeta(n)$ is the Riemann zeta function.
The coupling constant, $g$ is related to the bare coupling, $g_0$,
and up to ${\cal O}(g^6)$ the relation takes the form
\be
\frac{g^2}{4\pi} = \frac{g_0^2}{4\pi} + d_1(a\mu) \left(\frac{g_0^2}{4\pi}\right)^2 + d_2(a\mu) \left(\frac{g_0^2}{4\pi}\right)^3 \,.
\ee
The coefficients $d_1$ and $d_2$ depend on the renormalization scale
$a\mu$ and  are given by~\cite{Christou:1998ws}:
\bea
d_1(a\mu)&=&-\frac{1}{2\pi}
\left(\frac{11}{3}N_c-\frac{2}{3}N_F\right)\ln(a\mu)
-\frac{\pi}{2N_c} + 2.13573007 N_c - 0.08414443(8) N_F\,,\nonumber\\
d_2(a\mu) &=& d_1(a\mu)^2 -
\frac{1}{24 \pi^2} \left[ 34 N_c^2 - N_F \left(
13 N_c- \frac{3}{N_c}\right)\right]\ln(a\mu)  \nonumber \\
&&+ \frac{3\pi^2}{8N_c^2} - 2.8626216 + 1.2491158 N_c^2 +
N_F \left[ 0.18898(22) \frac{1}{N_c} - 0.15789(26) N_c\right].\nonumber
\eea

\subsection{Evolution to a reference scale}

All our $Z$-factors have been evaluated for a range of renormalization
scales. In this subsection we use 3-loop perturbative expressions to
extrapolate to a scale $\mu = 2$ GeV. Thus, each result is extrapolated
to $2$ GeV, maintaining information of the initial renormalization
scale at which it was computed. 

The scale dependence is predicted by the renormalization group equation
(at fixed bare parameters), that is~\cite{Gockeler:1998ye} 
\be
Z^{\rm \overline{\rm MS}}_{\cal O}(\mu) = R_{\cal O}(\mu,\mu_0)\,Z^{\rm \overline{\rm MS}}_{\cal O}(\mu_0)
\label{ZRI}
\ee
where
\begin{equation} 
 R_{\cal O} (\mu,\mu_0) = 
  \frac{\exp F \left(\frac{\bar{g}^2(\mu^2)}{16 \pi^2}\right)}
       {\exp F \left(\frac{\bar{g}^2(\mu_0^2)}{16 \pi^2}\right)}
\label{R}
\end{equation}
with
\begin{eqnarray} 
  F(x) & = & \frac{\gamma_0}{2 \beta_0} \ln(x)
     + \frac{\beta_0 \gamma_2 - \beta_2 \gamma_0}{4 \beta_0 \beta_2}
        \ln(\left(\beta_0 + \beta_1 x + \beta_2 x^2 \right))
   \nonumber \\ 
  {} & {} & {}  + \frac{2 \beta_0 \beta_2 \gamma_1 
        - \beta_1 \beta_2 \gamma_0 - \beta_0 \beta_1 \gamma_2 }    
            {2 \beta_0 \beta_2 \sqrt{4 \beta_0 \beta_2 - \beta_1^2}}
       \arctan \left( \frac{\beta_1 + 2 \beta_2 x}
                        {\sqrt{4 \beta_0 \beta_2 - \beta_1^2}} \right) \,.
\end{eqnarray}

To 3 loops, the running coupling, $\beta$-function and
anomalous dimension $\gamma$ are as follows~\cite{Chetyrkin:1999pq,Gockeler:1998ye,Vermaseren:1997fq,Gracey:2000am,Gracey:2003mr}, for $N_c=3$:

\bea
  \frac{\bar{g}^2 (\mu^2)}{16 \pi^2} & = & 
       \frac{1}{\beta_0 \ln(\mu^2/\Lambda^2)} -
        \frac{\beta_1}{\beta_0^3} \frac{\ln(\ln(\mu^2/\Lambda^2)) }
                              {\ln^2(\mu^2/\Lambda^2)}  
    \nonumber \\ [3ex]
 {} & {} & {} + \frac{1}{\beta_0^5 \ln^3(\mu^2/\Lambda^2)}
    \left( \beta_1^2 \ln^2(\ln(\mu^2/\Lambda^2))
           - \beta_1^2 \ln(\ln(\mu^2/\Lambda^2)) + \beta_2 \beta_0 - \beta_1^2 \right) + \cdots \\[3ex]
\label{b0}
 \beta_0 &=& 11 - \frac{2}{3} \,N_F  \\[2ex]
 \beta_1 &=& 102 - \frac{38}{3}\,N_F   \\[2ex]
 \beta_2 &=& \frac{2857}{2}-\frac{5033\,N_F}{18}
 +\frac{325\,N_F^2}{54} \\ [3ex]
%
\label{gq}
      \gamma^q_0 &=& 0 \\[2ex]
      \gamma^q_1 &=& -2\,\left(\frac{67}{3} - \frac{4}{3} \,N_F\right) \\[2ex]
      \gamma^q_2 &=& -2\,\left( \frac{20729}{36} -\frac{79}{2}\,\zeta(3) -\frac{550}{9} \,N_F + \frac{20}{27} \,N_F^2\right) \\[2ex]
%
      \gamma^{S,P}_0 &=& -2\,\frac{3\,C_F}{4}\\[2ex]
      \gamma^{S,P}_1 &=& -2\,\left(\frac{202}{3} - \frac{20}{9}\,N_F\right) \\[2ex]
      \gamma^{S,P}_2 &=& -2498 + \left(\frac{4432}{27} + \frac{320}{3}\,\zeta(3)\right)\,N_F   + \frac{280}{81}\,N_F^2 \\[2ex]
%
      \gamma^{V,A}_0 &=& \gamma^{V,A}_1 =\gamma^{V,A}_2 =0\\ [2ex]
%
      \gamma^{T,T'}_0 &=& \frac{8}{3}\\[2ex]
      \gamma^{T,T'}_1 &=& -\frac{4}{27}\,\left(26\,N_F-543\right) \\ [2ex]
      \gamma^{T,T'}_2 &=& -\frac{2}{81}\,\left(36\,N_F^2 +
      1440\,\zeta(3)\,N_F + 5240\,N_F + 2784\,\zeta(3) - 52555\right)
\label{gTTp}
\eea

Eqs.~(\ref{gq}) - (\ref{gTTp}) differ by numerical factors compared to
Refs.~\cite{Chetyrkin:1999pq,Gockeler:1998ye,Vermaseren:1997fq,Gracey:2000am,Gracey:2003mr}
due to alternative definitions of the factor $R_{\cal O} (\mu,\mu_0)$.

\section{Non-perturbative calculation}
\label{sec4}

In the literature there are two main approaches
that have been employed for the non-perturbative evaluation of the renormalization 
constants. They both start by considering that 
the operators can all be written in the form
\begin{equation}
   \Op(z) = \sum_{z'} \overline u(z) \J(z,z') d(z')\, ,
\end{equation}
where $u$ and $d$ denote quark fields in the physical basis and $\J$ denotes the operator we are interested in,
e.g. $\J(z,z') = \delta_{z,z'} \gamma_\mu$ would correspond to the local vector current.
For each operator we define a bare vertex function given by
\begin{equation}\label{vfun}  
   G(p) = \frac{a^{12}}{V}\sum_{x,y,z,z'} e^{-ip(x-y)} \langle u(x) \overline u(z) \J(z,z') d(z') \overline d(y) \rangle \, ,
\end{equation}
where $p$ is a momentum allowed by the boundary conditions, $V$ is the lattice volume, and the gauge average, denoted by the brackets, is
performed over gauge-fixed configurations. We have suppressed the Dirac and color indices 
of $G(p)$.
The first approach relies on translation invariance to shift the coordinates
of the correlators in
Eq.~(\ref{vfun}) to position $z=0$ \cite{Constantinou:2010gr, Dimopoulos:2007fn, Blossier:2010vt}. 
Having shifted to $z=0$, one can calculate the 
amputated vertex function for a given operator $\J$ for {\it any} momentum
with one inversion per quark flavor.

In this work we explore the second approach, 
introduced in Ref.~\cite{Gockeler:1998ye},
which uses directly Eq.~(\ref{vfun}) without employing translation
invariance. One must now use a source that is momentum dependent but
can couple to any operator. For twisted mass fermions, we use the
symmetry $S^u(x,y)=\gamma_5S^{d\dagger}(y,x)\gamma_5$ between the $u-$
and $d-$quark propagators. Therefore with a single inversion one can extract the vertex function for
 a {\em single} momentum.
The advantage of this approach is a high statistical
accuracy and the evaluation of the vertex 
for any operator including extended operators at
no significant additional computational cost. Since
 we are interested in a number
of operators with their associated renormalization constants 
we use the second approach. 
We fix to Landau gauge using a stochastic over-relaxation
algorithm~\cite{deForcrand:1989im}, converging to a gauge
transformation which minimizes the functional
\begin{equation}
   F = \sum_{x,\mu} {\rm Re}\ {\rm tr} \left[ U_\mu(x) + U^\dagger_\mu(x-\hat\mu)\right] \, .
\end{equation}
Questions related to the Gribov ambiguity will not be addressed in this work.
The propagator in momentum space, in the physical basis, is defined by
\begin{equation}\label{pprop}
   S^u(p) = \frac{a^8}{V}\sum_{x,y} e^{-ip(x-y)} \left\langle u(x) \overline u(y) \right\rangle\, , \qquad
   S^d(p) = \frac{a^8}{V}\sum_{x,y} e^{-ip(x-y)} \left\langle d(x) \overline d(y) \right\rangle \, .
\end{equation}
An amputated vertex function is given by
\begin{equation}
   \Gamma(p) = (S^u(p))^{-1}\, G(p)\, (S^d(p))^{-1} \, .
\end{equation}
and the corresponding renormalized quantities are assigned the values
\begin{equation}
   S_R(p)      = \Zq S(p) \, , \qquad \qquad
   \Gamma_R(p) = \Zq^{-1} Z_\Op \Gamma(p) \quad.
\end{equation}
In the twisted basis at maximal twist, Eq.~(\ref{vfun}) takes the form
\begin{equation}\label{vfun_tm}
   G(p) = \frac{a^{12}}{4V}\sum_{x,y,z,z'} e^{-ip(x-y)} \left\langle(\eins+i\gamma_5) u(x) \overline u(z)(\eins+i\gamma_5) \J(z,z') (\eins-i\gamma_5) d(z')\
 \overline d(y)(\eins-i\gamma_5) \right\rangle \, .
\end{equation}
After integration over the fermion fields, and using $\s^u(x,z)=\gamma_5 {\s^d}^\dagger(z,x)\gamma_5$ this becomes
\begin{equation}
   G(p) = -\frac{a^{12}}{4V} \sum_{z,\,z'} \left\langle (\eins-i\gamma_5){\breve {\s^d}}^\dagger(z,p)(\eins-i\gamma_5) \J(z,z')
   (\eins-i\gamma_5)\breve \s^d(z',p)(\eins-i\gamma_5) \right\rangle^G \, ,
\end{equation}
where $\langle ... \rangle^G$ denotes the integration over gluon fields,
and $\breve \s(z,p) = \sum_y e^{ipy} \s(z,y)$ is the Fourier
transformed propagator on one of its argument on a particular gauge
background. It can be obtained by inversion using the Fourier source 
\begin{equation}
   b_\alpha^a(x) = e^{ipx} \delta_{\alpha \beta}\delta_{a b} \, ,
\end{equation}
for all Dirac $\alpha$ and color $a$ indices.
The propagators in the physical basis given in Eq.~(\ref{pprop}) can be obtained from
\begin{eqnarray}\label{pprop_tm}
   S^d(p) &=& \phantom{-}\frac{1}{4} \sum_z e^{-ipz} \langle (\eins-i\gamma_5)\breve \s^d(z,p)(\eins-i\gamma_5) \rangle^G \nonumber \\
   S^u(p) &=&           -\frac{1}{4} \sum_z e^{+ipz} \langle (\eins-i\gamma_5){\breve {\s^d}}^\dagger(z,p)(\eins-i\gamma_5) \rangle^G \, ,
\end{eqnarray}
which evidently only need 12 inversions despite the occurrence of both $u$ and $d$ quarks in the original
expression.

We evaluate Eq.~(\ref{vfun_tm}) and Eq.~(\ref{pprop_tm}) for each momentum
separately employing Fourier sources over a range of $a^2p^2$ for which
perturbative results can be trusted and finite $a$ corrections
are reasonably small.

\section{Non-perturbative results}
\label{sec5}

We perform the non-perturbative calculation of renormalization
constants for three values of the lattice spacing, corresponding to
$\beta=3.9,\,4.05$ and $4.20$~\cite{Boucaud:2008xu}. In this work we 
use the lattice spacing as determined from
the nucleon mass. The values we obtained are 0.089(1)(5)~fm, 0.070(1)(4)~fm and 0.056(2)(3)~fm for $\beta=3.9$, 4.05 and 4.2,
respectively~\cite{Alexandrou:2010hf} and they are in agreement with the ones
determined from the pion sector.
To extract the renormalization constants reliably one needs to
consider momenta in the range $\Lambda_{QCD}<p<1/a$. We relax the
upper bound to be $\sim 2/a$ to $5/a$, which is justified by the
linear dependence of our results on $a^2$. Therefore, we consider
momenta spanning the range $0.5<a^2p^2<5$ for which perturbation
theory is trustworthy and lattice artifacts are still small enough. It
is important to note that the extrapolation to the continuum limit,
$(a\,p)^2\to 0$, is performed for a fixed momentum range in physical
units. In Table~\ref{tab2} we summarize the various
parameters of the action, that we used in our simulations, and in
Table \ref{tab3} we present the values we used for the momenta
$(p_t,p_x,p_y,p_z)=(2\pi/L_t\,n_t,2\pi/L_x\,n_x,2\pi/L_y\,n_y,2\pi/L_z\,n_z))$.
\begin{table}[h]
\begin{center}
\begin{minipage}{15cm}
\begin{tabular}{ccccc}
\hline
\hline\\[-2.25ex]
$\beta$ & a (fm)& $a\mu_0$ & $m_\pi$~(GeV) & $L^3\times T$ \\
\hline
\hline\\[-2.25ex]
3.9  & 0.089  & 0.0040 &  0.3021(14)  &$24^3\times48$  \\
3.9  & 0.089  & 0.0064 &  0.37553(80) &$24^3\times48$  \\
3.9  & 0.089  & 0.0085 &  0.4302(11)  &$24^3\times48$  \\
3.9  & 0.089  & 0.01   &  0.4675(12)  &$24^3\times48$  \\
4.05 & 0.070  & 0.003  &  0.2925(18)  &$32^3\times64$  \\
4.05 & 0.070  & 0.006  &  0.4082(31)  &$24^3\times48$  \\
4.05 & 0.070  & 0.006  &  0.404(2)    &$32^3\times64$  \\
4.05 & 0.070  & 0.008  &  0.465(1)    &$32^3\times64$  \\
4.20 & 0.056  & 0.002  &  0.2622(11)  &$24^3\times48$  \\
4.20 & 0.056  & 0.0065 &  0.476(2)    &$32^3\times64$  \\
\hline
\hline
\end{tabular}
\end{minipage}
\end{center}
\caption{$\beta$-values and lattice size used in the simulations are
  given in the first and last columns respectively. The lattice
  spacing $a$ in fm is determined from the nucleon mass. We also give
  the bare light quark mass $a\mu_0$ and pion mass.}
\label{tab2}
\end{table}

\noindent
The number of configuration in each ensemble varies between
10 to 100. 
 Using even 10 configurations leads to results with very high
statistical accuracy, easily below 0.5$\%$. Thus, in the plots presented
here the statistical errors are too small to be visible. We mostly use in
our computation democratic momenta, in the sense that they have the
same $p_x,\,p_y,\,p_z$. We have also tested a few non-democratic momentum,
which turn out to behave similarly to democratic ones 
(e.g. $(n_t,n_x,n_y,n_z)$=(3,3,3,2) is similar to (3,3,3,3)). 
We would like to point out that the non-perturbative results have a
significant dependence on the value of the momentum in the spatial
direction, indicating large lattice artifacts in some cases. Such a study 
appeared in Ref.~\cite{Alexandrou:2010me}.

\begin{table}[h]
\begin{center}
\begin{minipage}{17cm}
\begin{tabular}{lr@{}lr@{}lr@{}l}
\hline
\hline\\[-2.25ex]
\multicolumn{1}{c}{$\,\,\beta=3.9\,\,$} &
\multicolumn{2}{c}{$\,\,\beta=4.05\,\,$}&
\multicolumn{2}{c}{$\,\,\beta=4.20\,\,$} \\
\hline
\hline\\[-2.25ex]
($n_t$,2,2,2), $n_t:4-8,\,10-14$  &&\qquad($n_t$,2,2,2), $n_t:4-8,\,10,\,13-14$   &&\qquad($n_t$,2,2,2), $n_t:4-8,\,10,\,13-14$  \\
($n_t$,3,3,3), $n_t:2-6,\,8-9$    &&\qquad($n_t$,3,3,3), $n_t:2-6,\,8-11,\,13$    &&\qquad($n_t$,3,3,3), $n_t:2-6,\,8-11,\,13$   \\
($n_t$,4,4,4), $n_t:4-9$          &&\qquad($n_t$,4,4,4), $n_t:8-10$               &&\qquad($n_t$,4,4,4), $n_t:7-11$              \\
(3,3,3,2)                         &&\qquad(3,3,3,2)                               &&\qquad($n_t$,5,5,5), $n_t:2-4$               \\
$\quad$                           &&\qquad$\quad$                                 &&\qquad(3,3,3,2)                              \\
\hline
\hline
\end{tabular}
\end{minipage}
\end{center}
\caption{Values of momentum used for the various ensembles at $\beta=3.9,\,4.05,\,4.20$.}
\label{tab3}
\end{table}

\subsection{Pion mass dependence}
\label{PionMass}

In Table~\ref{tab2} we give the number
of pion mass that we studied for each of the three $\beta$ values.
These ensembles have been produced by the ETM Collaboration~\cite{Boucaud:2007uk,Boucaud:2008xu,Urbach:2007rt,Baron:2009wt}. 
The pion mass dependence of $Z_{\rm q},\, Z_{\rm
V},\,Z_{\rm A},\,\ZT$ is displayed in Fig.~\ref{fig3} and is
not significant. A linear extrapolation to the data shown in Fig.~\ref{fig3} yields a
slope consistent with zero. This behavior is also observed at the
other $\beta$ values. Thus, it would be sufficient to obtain the
results on the aforementioned RCs at one pion mass value, although
we perform the chiral extrapolation with all available data on
different pion masses. The need of having simulations at a number of
pion masses comes from the fact that one has to perform the
subtraction of the pion pole contribution. This is discussed in Subsection~\ref{Sub_pole}.
\begin{center}
\begin{figure}[!h]
\centerline{\psfig{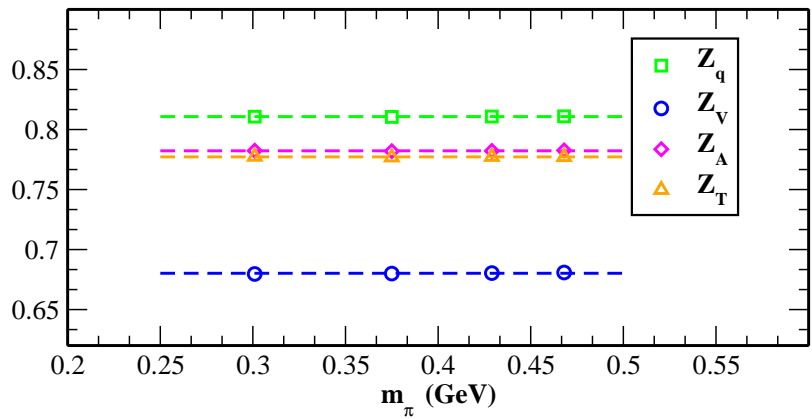}}
\begin{minipage}{12cm}
\caption{$Z_{\rm q},\,Z_{\rm V},\,Z_{\rm A},\,\ZT$ at $\beta=3.9$, as a
  function of the pion mass. Computations were performed at pion masses of
 $m_\pi=0.302$~GeV ($a\mu_0=0.004$),
  $m_\pi=0.375$~GeV ($a\mu_0=0.0064$), $m_\pi=0.429$~GeV
  ($a\mu_0=0.0085$) and $m_\pi=0.468$~GeV ($a\mu_0=0.01$).}
\label{fig3}
\end{minipage}
\end{figure}
\end{center}

\subsection{Volume dependence}

We perform the evaluation of the RCs at $\beta=4.05$ and $\mu=0.006$
for two volumes, $24^3\times48$ and $32^3\times64$ in order to check
for finite volume effects. For this comparison we used momenta that
correspond to the same renormalization scale. For the small lattice we use
$(a\,p)=2\pi(3/48,3/24,3/24,3/24)$, in lattice units, whereas for the larger
one we employ $(a\,p)=2\pi(4/64,4/32,4/32,4/32)$. The volume effects
appear to be in the worst case $\sim$ 0.1$\%$, as can be seen from
Table \ref{tab4}. We would like to point out 
that the $Z_P$ estimator shows the largest volume dependence, which
however tends to decrease after the pion pole subtraction
(Subsection~\ref{Sub_pole}, Fig.~\ref{fig4}).
\begin{table}[h!]
\begin{center}
\begin{minipage}{15cm}
\begin{tabular}{lr@{}lr@{}lr@{}lr@{}lr@{}lr@{}l}
\hline
\hline\\[-2.25ex]
\multicolumn{1}{c}{lattice}&
\multicolumn{2}{c}{$\,\,Z_{\rm q}\,\,$} &
\multicolumn{2}{c}{$\,\,Z_{\rm S}\,\,$} &
\multicolumn{2}{c}{$\,\,Z_{\rm P}\,\,$} &
\multicolumn{2}{c}{$\,\,Z_{\rm V}\,\,$} &
\multicolumn{2}{c}{$\,\,Z_{\rm A}\,\,$} &
\multicolumn{2}{c}{$\,\,Z_{\rm T}\,\,$}   \\
\hline
\hline\\[-2.25ex]
$24^3$x$48$  &$\,\,\,0$.&82315(7)   &$\,\,\,0$.&743(2)  &$\,\,\,0$.&512(2)    &$\,\,\,0$.&7068(1)    &$\,\,\,0$.&7935(2)   &$\,\,\,0$.&7759(1)\\
$32^3$x$64$  &$\,\,\,0$.&82303(3)   &$\,\,\,0$.&744(1)  &$\,\,\,0$.&508(1)    &$\,\,\,0$.&7069(1)    &$\,\,\,0$.&7935(1)    &$\,\,\,0$.&7759(1)\\
\hline
\hline
\end{tabular}
\end{minipage}
\end{center}
\caption{Renormalization constants at $\beta=4.05, \mu_0=0.006$
 using two lattice sizes, namely $32^3\times64$ and momentum (4,4,4,4) and
  $24^3$x$48$ with momentum (3,3,3,3), and at a scale of $(a\,p)^2\sim 2$.}
\label{tab4}
\end{table}

\subsection{Pion-pole subtraction}
\label{Sub_pole}

The correlation functions of the pseudoscalar operator
have pion-pole contamination and therefore need to be treated carefully.
In order to subtract the pole contribution we use the following Ansatz
for the pseudoscalar amputated vertex function, $\Lambda_P$,
\be
\Lambda_{P} = a_{P} + b_{P}\,m_\pi^2 + \frac{c_{P}}{m_\pi^2}\,,
\label{PionPole}
\ee
which we apply to data produced at a given value of $\beta$.
Once we have the fitting
parameters we subtract the pion-pole using the value of $c_{P}$, determined from the fitting, i.e. we take
\be
\Lambda_{P}^{\rm sub} = \Lambda_{P} - \frac{c_{P}}{m_\pi^2}\,.
\label{PionPoleSub}
\ee
To reliably obtain the three fitting parameters of Eq.~(\ref{PionPole}) we
need the RC of the pseudoscalar operator for at least 4
pion masses; this is feasible for $\beta=3.9$. On the contrary, for $\beta=4.05$
we have data for three pion masses, and for $\beta=4.20$ only for two pion masses. 
At $\beta=3.9$ we determine the parameters using results at three of the four pion masses 
$\beta=3.9$ and compare them with the fit resulting when using all available
data. The conclusion is that the values obtained are compatible. 
Therefore at $\beta=4.05$ we determine the 
parameters using results at the three pion masses that
are available. One may observe the
effectiveness of the pion-pole subtraction in Fig.~\ref{fig4}, where
we show results before and after the pion pole subtraction.
After the subtraction results obtained at different pion mass
fall on each other.
While the pion-pole term has an appreciable contribution,
the quadratic term with the $b_{P}$ coefficient
is expected to be small.
The values extracted for $b_{P}$ at $\beta=3.9$ and $\beta=4.05$ 
are indeed small showing a very weak pion mass dependence of the
$b_{P}$ coefficient. This is consistent with the weak 
pion mass dependence observed for the vector and axial-vector RCs (see
Subsection~\ref{PionMass} for the other RCs). It is also verified by
our data: after subtracting the pion-pole term determined from fitting
to the data, the remaining pion mass dependence ($b_{P}\,m_p^2$) is
negligible for all the ensembles. This allows us to perform a two
parameter fit at $\beta=4.2$ of the form: 
\be
\Lambda_{P} = a_{P} + \frac{c_{P}}{m_\pi^2}\,,
\label{PionPole2}
\ee
using data on the two pion masses $m_\pi= 476$ and 262 MeV.
The two sets correspond to different lattice size, $24^3$x$48$ 
($32^3\times64$) for the lower (higher) pion mass. As a result,
momenta with the same values for $(n_t,n_x,n_y,n_z)$ correspond to
different $(a\,p)^2$. Thus, in order to perform the fit using the Ansatz of Eq.~(\ref{PionPole2})
we carefully choose the momenta in the two ensembles
 to have almost the same $(a\,p)^2$.
In general, this could lead to additional uncertainties, but we have already
checked that volume effects are negligible. Indeed the fit using the Ansatz of
Eq.~(\ref{PionPole2}) yields a value for $c_{P(S)}$ that accurately removes the
pion pole as demonstrated in Fig.~\ref{fig11}. 

\begin{center}
\begin{figure}[t]
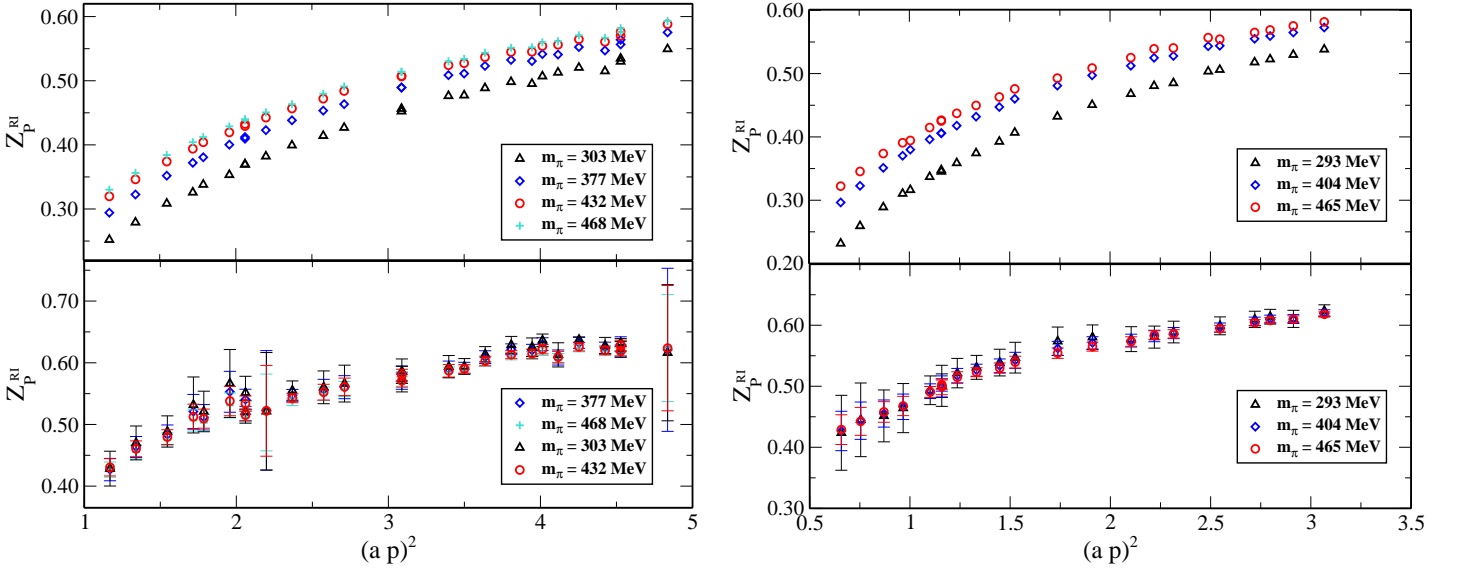

\centerline{\psfig{figure=plots/Z_p_with_pole_b3.9.eps,height=7.5truecm}
$\quad$\psfig{figure=plots/Z_p_with_pole_b4.05.eps,height=7.5truecm}}
\caption{$\ZP$ at $\beta=3.9$ (left panel) and $\beta=4.05$
(right panel) for various masses. The upper plot
shows the results before the pion pole subtraction
as described by Eq.~(\ref{PionPole}), while the lower figure the results upon the
appropriate subtraction given in Eq.~(\ref{PionPoleSub}).}
\label{fig4}
\end{figure}
\end{center}

\begin{center}
\begin{figure}[h]
\centerline{\psfig{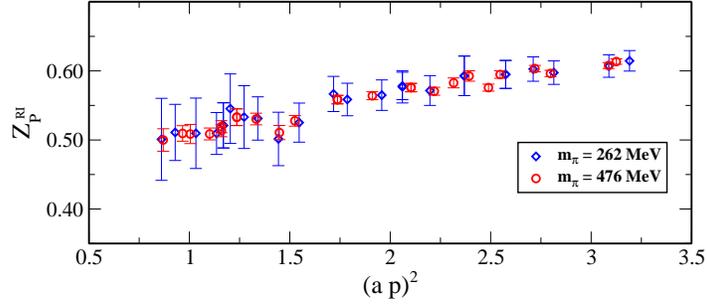}}
\caption{$\ZP$ for $\beta=4.20$
for the two pion masses. Results are shown upon the pion pole subtraction
as described in Eq.~(\ref{PionPoleSub}).}
\label{fig11}
\end{figure}
\end{center}
The errors shown in Figs.~\ref{fig4} - \ref{fig11} are computed in two ways:
using super jackknife error
analysis~\cite{AliKhan:2001tx, Bratt:2010jn} and requiring
that a correlated change of the fit parameters
increases the minimum value of $\chi$ by one.
We find that both methods lead to similar errors. 

Our data for the ratio $\ZP/\ZS$ also show dependence on the pion mass, 
and thus to form this ratio we used the subtracted data of Fig.~\ref{fig4}
which we compute in the chiral limit.
This procedure leads to the values shown in Fig.~\ref{fig5}. With black circles
we show the non-perturbative results after subtracting the pion pole
from $\ZP$ using Eq.~(\ref{PionPoleSub}). 
If one further subtracts from $\ZS$ and $\ZP$ the perturbative ${\cal
O}(a^2)$ contributions, presented in Sections~\ref{prop} - \ref{oper},
one obtains the values shown with the magenta diamonds in Fig.~\ref{fig5}.
The ratio $\ZP/\ZS$ is renormalization scale independent and therefore
one can directly use the ${\cal O}(a^2)$-perturbatively subtracted non-perturbative 
results to extrapolate to the continuum limit, eliminating any remaining cut-off effects.

\begin{center}
\begin{figure}
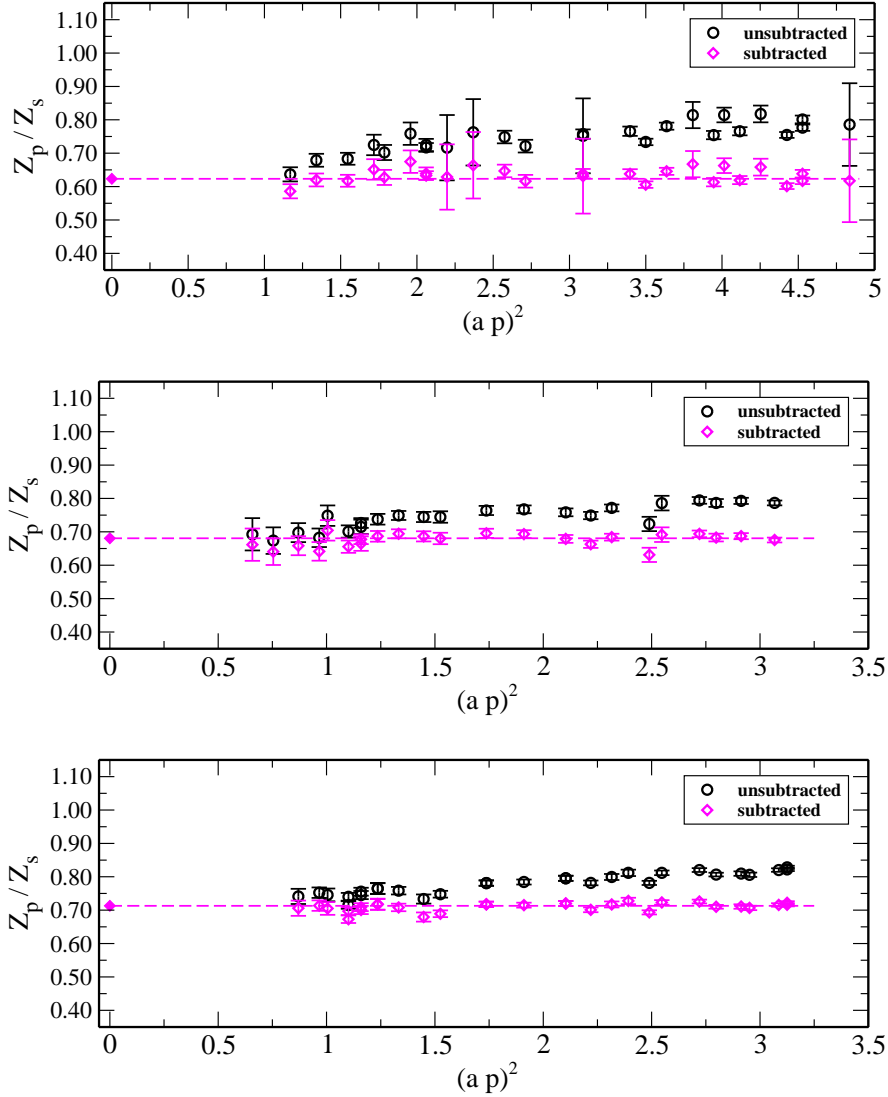

\centerline{\psfig{figure=plots/Zp_over_Zs_b3.9.eps,height=4.5truecm}}
\vspace{0.5cm}
\centerline{\psfig{figure=plots/Zp_over_Zs_b4.05.eps,height=4.5truecm}}
\vspace{0.5cm}
\centerline{\psfig{figure=plots/Zp_over_Zs_b4.20.eps,height=4.5truecm}}
\caption{$Z_{\rm P}/Z_{\rm S}$ at $\beta=3.9$ (upper plot), $\beta=4.05$
(middle plot) and $\beta=4.20$ (lower plot) as a function of $(a\,p)^2$. 
In each plot we demonstrate the effect of subtracting the ${\cal O}(a^2)$-terms by
plotting the non-perturbative results before (black circles) and
after (magenta diamonds) subtraction. The pion-pole 
term has been removed from all the data that we show here.} 
\label{fig5}
\end{figure}
\end{center}

\subsection{Results in RI$'$-MOM scheme}

In this section we present results in the RI$'$-MOM scheme for
$\Zq,\,\ZS,\,\ZP$, as well as for the scale-independent RCs $\ZV$ and $\ZA$.
We have also performed a computation of $\ZT$ and its results are presented 
in the next section. In all cases we subtract
the leading discretization effects of ${\cal O}(a^2)$ computed
to one-loop in perturbation theory from the non-perturbative
results. All renormalization constants are evaluated at the three
$\beta$-values, where the 
simulations were carried out. 
For all $\beta$ values we perform a chiral
extrapolation using results at different pion masses; the results have
negligible dependence on the quark mass
 as demonstrated in Fig.~\ref{fig3} and therefore we use
a constant fit to extrapolate to the chiral limit.

The renormalization constant of the fermion
field is needed as an input in
various expressions, and the results obtained are displayed in Fig.~\ref{figZq}.
\begin{center}
\begin{figure}[!h]
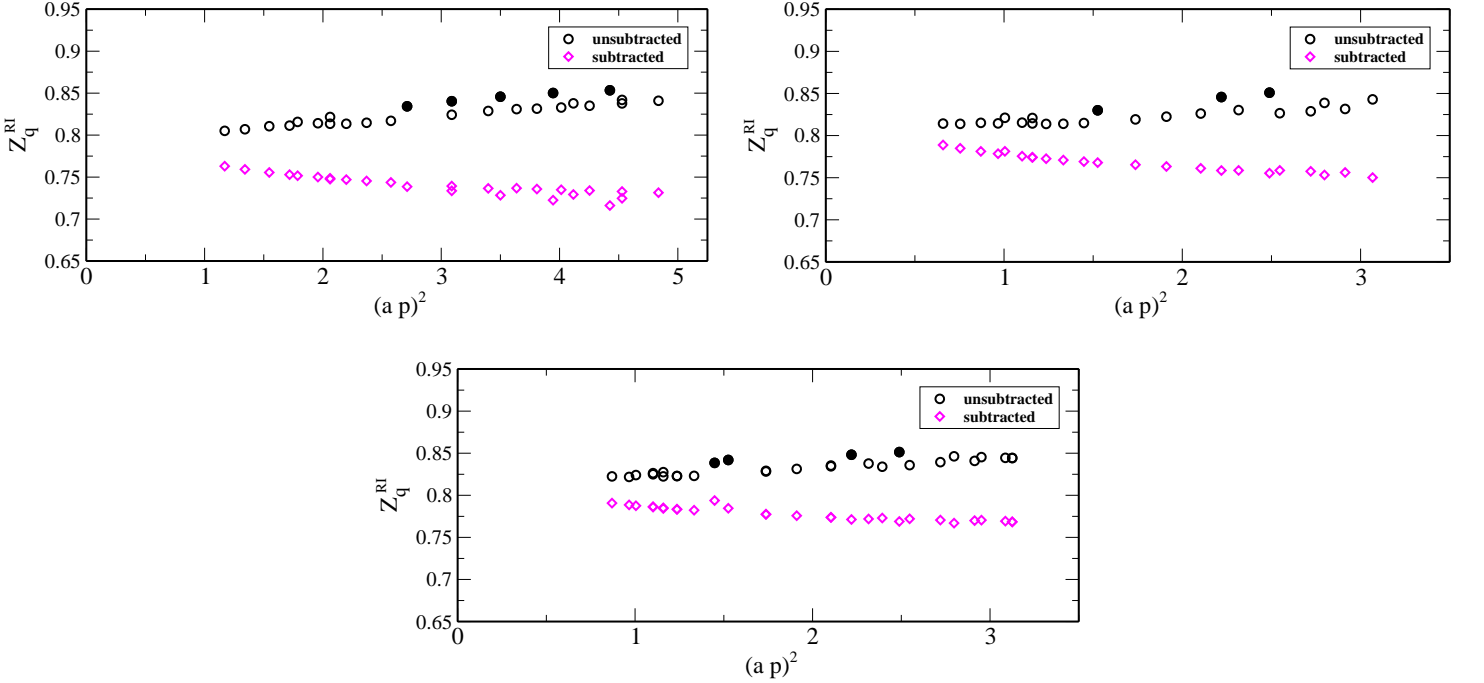

\centerline{\psfig{figure=plots/Zq_RI_b3.9.eps,height=4.25truecm}
\quad
\psfig{figure=plots/Zq_RI_b4.05.eps,height=4.25truecm}}
\vspace{0.5cm}
\centerline{\psfig{figure=plots/Zq_RI_b4.20.eps,height=4.25truecm}}
\begin{minipage}{14cm}
\caption{Non-perturbative results on $\Zq$ for $\beta=3.9$ (upper left
plot), $\beta=4.05$ (upper right plot) and $\beta=4.20$ (lower
plot). In all plots we show chirally extrapolated results. 
Black circles (magenta diamonds) represent the non-perturbative
data before (after) subtracting the ${\cal O}(a^2)$-terms.
The unsubtracted data that suffer
from large non-Lorentz invariant contributions are show with filled
black circles.}
\label{figZq}
\end{minipage}
\end{figure}
\end{center}
The non-perturbative values of $\Zq$ are obtained
for all available momenta and they reveal a non-smooth behavior as
a function of the momentum (see Fig.~\ref{figZq}), which becomes smoother once we subtract the ${\cal O}(a^2)$
perturbative terms. We would like to point out that in all our
non-perturbative results before subtraction
we have some data that fluctuate several
standard deviations around the
mean value and these correspond to momenta that lead to large
non-Lorentz invariant contributions in the perturbative expressions of
Sections~\ref{prop}-\ref{oper}. These terms are of the form $(\sum_\rho
p_\rho^4)/(\sum_\rho p_\rho^2)$. After subtraction these
non-Lorentz invariant contributions are removed (to ${\cal O}(a^2\,g^2)$), resulting in
the much smoother behavior of the subtracted data. The unsubtracted data of $\Zq$ that suffer
from large non-Lorentz invariant contributions are show in Fig.~\ref{figZq} with filled
black circles. Note that as the lattice spacing decreases, the discrepancy between unsubtracted
and subtracted data becomes smaller.

The RCs $\ZV$ and $\ZA$ are scale-independent and therefore
there is no need to evolve them. In Fig.~\ref{fig6}
we show results on $\ZV$ and $\ZA$ before and after
subtraction of the perturbatively determined ${\cal O}(a^2)$-terms. As can be seen, the
subtraction weakens the dependence on $(a\,p)^2$. In fact, fitting
the subtracted data to a straight line of the form $z+s(ap)^2$
results in a value of the slope $s$ consistent with zero.
This shows that
leading cut-off effects are effectively removed by the
subtraction of perturbatively determined ${\cal O}(a^2)$-terms. 
The small remaining lattice
artifacts are removed by extrapolating linearly to the continuum line.
 The unsubtracted data can also be extrapolated linearly but,
in this case, the slope is generally non-zero as can be seen in Fig.~\ref{fig6}.
As the lattice spacing decreases, the deviation between subtracted and 
unsubtracted data decreases. We 
note that, although the value found at the continuum limit
 for the unsubtracted data approaches that extracted for the subtracted data,
small differences still remain.
This is an indication that the systematic error due to cut-off effects
is larger than the statistical error and therefore the subtraction 
of ${\cal O}(a^2)$-terms helps in diminishing the uncertainty in the
choice of the fit range.

In order to perform
 the continuum extrapolation we choose the same momentum range in
physical units for all $\beta$ values and we thus extract all
renormalization constants using the same physical momentum range,
$p^2\sim15\,-\,32$ (GeV)$^2$. This momentum range is in line with
what has been chosen in our previous work on the RCs for one-derivative 
bilinear operators~\cite{Alexandrou:2010me}, ensuring
 that we use data in a region where an approximate plateau exists. 
The momentum range in lattice units at each $\beta$-value
is as follows: $\beta=3.9$: $(a\,p)^2=3\, - \,5$, $\beta=4.05$:
$(a\,p)^2=1.8\,-\,3.1$, $\beta=4.20$: $(a\,p)^2=1.2\,-\,2.5$. The choice
for the momentum range is not so relevant for $\ZV$ and $\ZA$ as it
was for the case of the RCs for one-derivative operators, since the
subtracted data are almost constant. However, for consistency we
use the same range as in Ref.~\cite{Alexandrou:2010me}.

\begin{center}
\begin{figure}[!h]
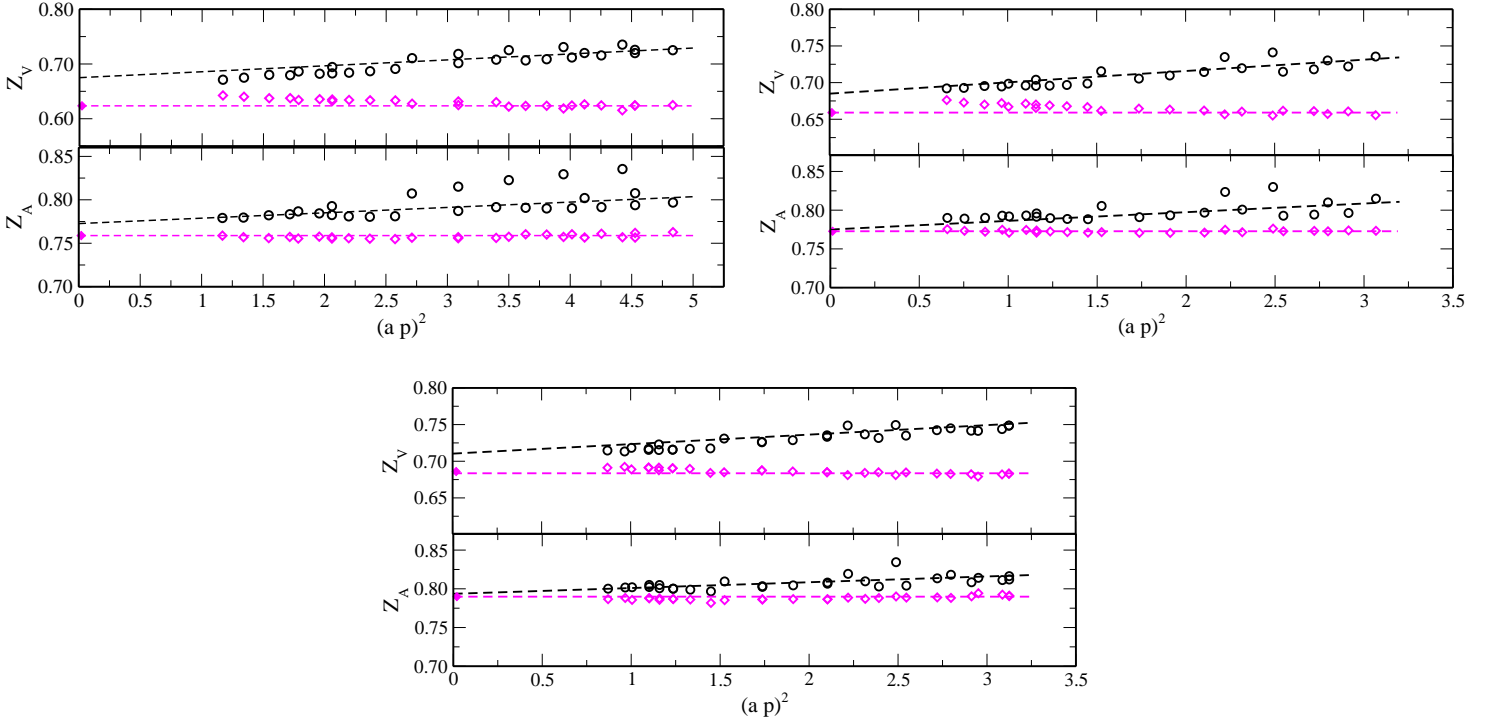

\centerline{\psfig{figure=plots/Z_v_a_b3.9.eps,height=4.5truecm}
\quad
\psfig{figure=plots/Z_v_a_b4.05.eps,height=4.5truecm}}
\vspace{0.5cm}
\centerline{\psfig{figure=plots/Z_v_a_b4.20.eps,height=4.5truecm}}
\begin{minipage}{14cm}
\caption{The renormalization constants for
the vector and axial-vector operators, $\ZV$ and $\ZA$,
 for $\beta=3.9$ (upper left
plot), $\beta=4.05$ (upper right plot) and $\beta=4.20$ (lower
plot). In all plots we show chirally extrapolated results.
Black circles (magenta diamonds) represent the non-perturbative
  data before after subtracting the ${\cal O}(a^2)$-terms.}
\label{fig6}
\end{minipage}
\end{figure}
\end{center}

\subsection{${\overline{\rm MS}}$ scheme}

In this Section we present our results on $\Zq,\,\ZS,\,\ZP$ and $\ZT$
converted to the continuum ${\overline{\rm MS}}$ scheme and at a
reference scale of $\mu = 2$ GeV. For the conversion from RI$'$-MOM to
${\overline{\rm MS}}$ we use the formulae given in Eqs.~(\ref{Cq}) - (\ref{CT}).
We use the 3-loop formulae of Eqs.~(\ref{ZRI}) - (\ref{R}) to evolve the scale from
$\mu$ to 2 GeV.

As already discussed, a ``renormalization window'' should exist for $\Lambda_{QCD}^2 <<
\mu^2 << 1/a^2$ where perturbation theory holds and finite $a$
artifacts are small, leading to scale-independent results
(plateau). In practice such a condition is hard to satisfy. The right
inequality is extended to $(2-5)/a^2$ leading to lattice artifacts in
our results that are of ${\cal O}(a^2p^2)$. Fortunately our
perturbative calculations allow us to subtract the leading
perturbative ${\cal O}(a^2)$ lattice artifacts which alleviates the
problem. To remove the remaining ${\cal O}(a^2p^2)$ artifacts we extrapolate
linearly to $(ap)^2=0$ as demonstrated in Figs. \ref{fig6} -\ref{fig9}.
The statistical errors are negligible, however an estimate of the
systematic errors is important. The largest systematic error comes
from the choice of the momentum range to
use for the extrapolation to $(ap)^2=0$. One way to estimate
this systematic error is to vary the momentum range where we perform
the fit. Another approach is to fix a range and then eliminate a given
momentum in the fit range and refit. The spread of the results about
the mean gives an estimate of the systematic error. In the final
results we give as systematic error the largest of the two, which is the 
one obtained by modifying the fit
range. As already mentioned we choose the same momentum range in
physical units for the three $\beta$-values and extract all renormalization
constants using the same physical momentum range, $p^2\sim 15-32$
(GeV)$^2$; within this range the data fall on a straight line of a
small slope. We note that the ${\cal O}(a^2)$
perturbative terms which we subtract, tend to decrease with increasing $\beta$,
as expected. The error bars in Fig.~\ref{fig8} are due
to the fit uncertainties in performing the pion
pole subtraction.

\begin{center}
\begin{figure}[!h]
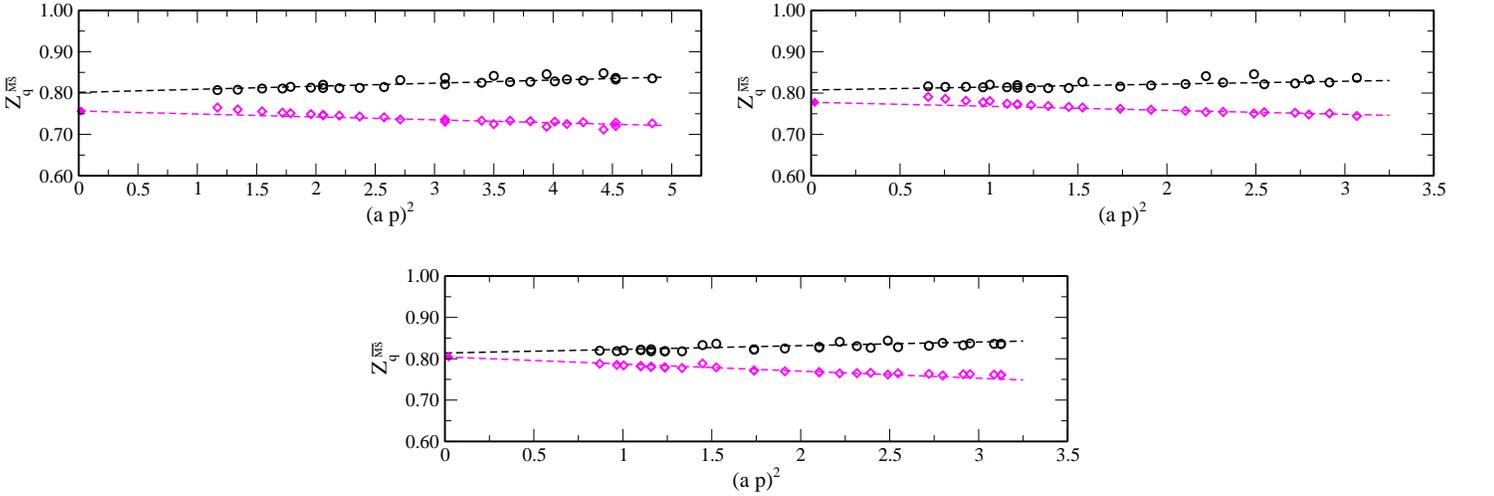

\centerline{\psfig{figure=plots/Zq_MS_2GeV_b3.9.eps,height=3truecm}
\quad
\psfig{figure=plots/Zq_MS_2GeV_b4.05.eps,height=3truecm}}
\vspace{0.5cm}
\centerline{\psfig{figure=plots/Zq_MS_2GeV_b4.20.eps,height=3truecm}}
\begin{minipage}{14cm}
\caption{Results on $\Zq$ for $\beta=3.9$ (upper left plot),
  $\beta=4.05$ (upper right plot) and $\beta=4.20$ (lower plot). 
 In all plots we show chirally extrapolated results.
 Black circles (magenta diamonds) represent the non-perturbative
  data before (after) subtracting the ${\cal O}(a^2)$-terms.The corresponding dashed lines show the extrapolation
to the continuum limit and the filled diamond shows the final
value in the continuum. Statistical errors are smaller than the size of the
symbols.\\}
\label{fig7}
\end{minipage}
\end{figure}
\end{center}

\begin{center}
\begin{figure}[!h]
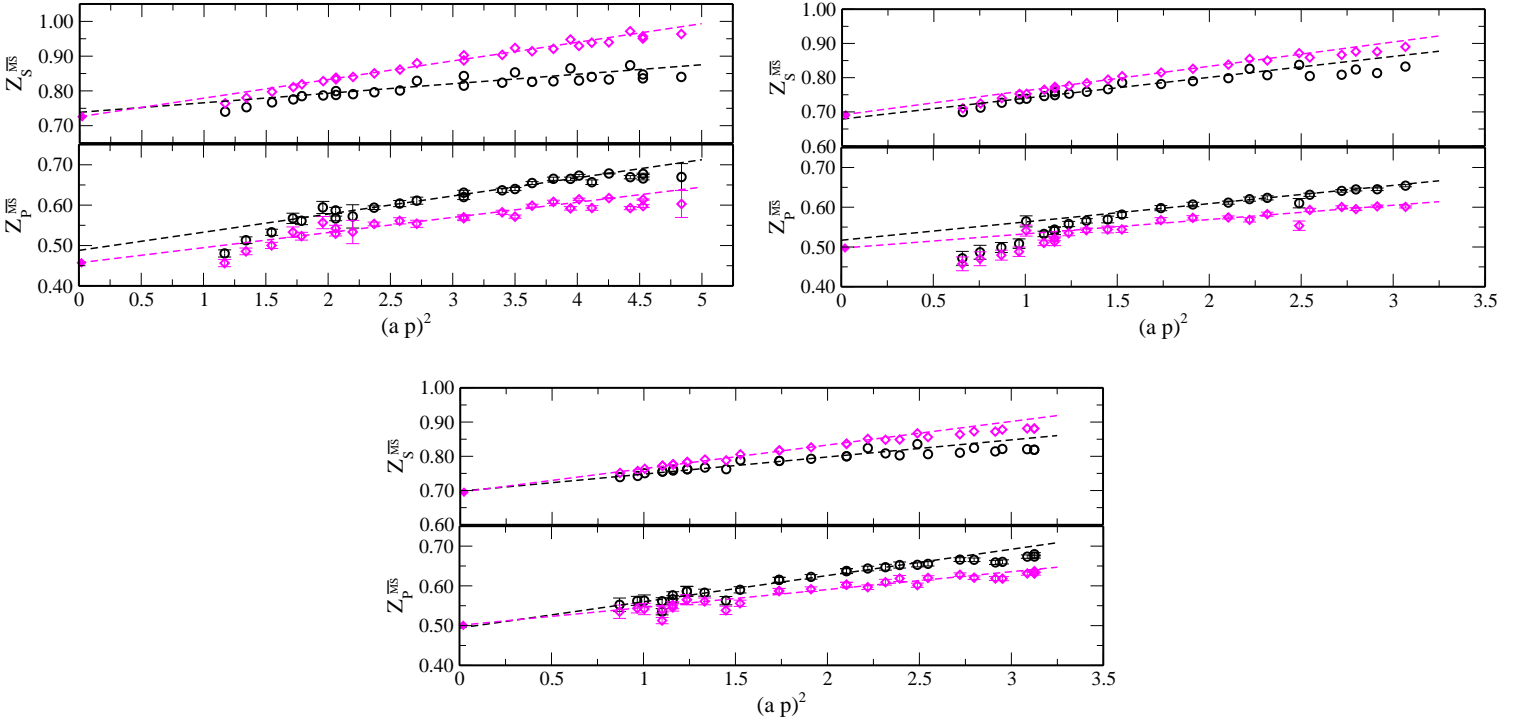

\centerline{\psfig{figure=plots/Z_s_p_MS_2GeV_b3.9.eps,height=4.5truecm}
\quad
\psfig{figure=plots/Z_s_p_MS_2GeV_b4.05.eps,height=4.5truecm}}
\vspace{0.5cm}
\centerline{\psfig{figure=plots/Z_s_p_MS_2GeV_b4.20.eps,height=4.5truecm}}
\begin{minipage}{14cm}
\caption{Results on $\ZS$ and $\ZP$ at the chiral limit for $\beta=3.9$
  (upper left panel), $\beta=4.05$ (upper right panel) and $\beta=4.20$
(lower plot). }
\label{fig8}
\end{minipage}
\end{figure}
\end{center}

\begin{center}
\begin{figure}[!h]
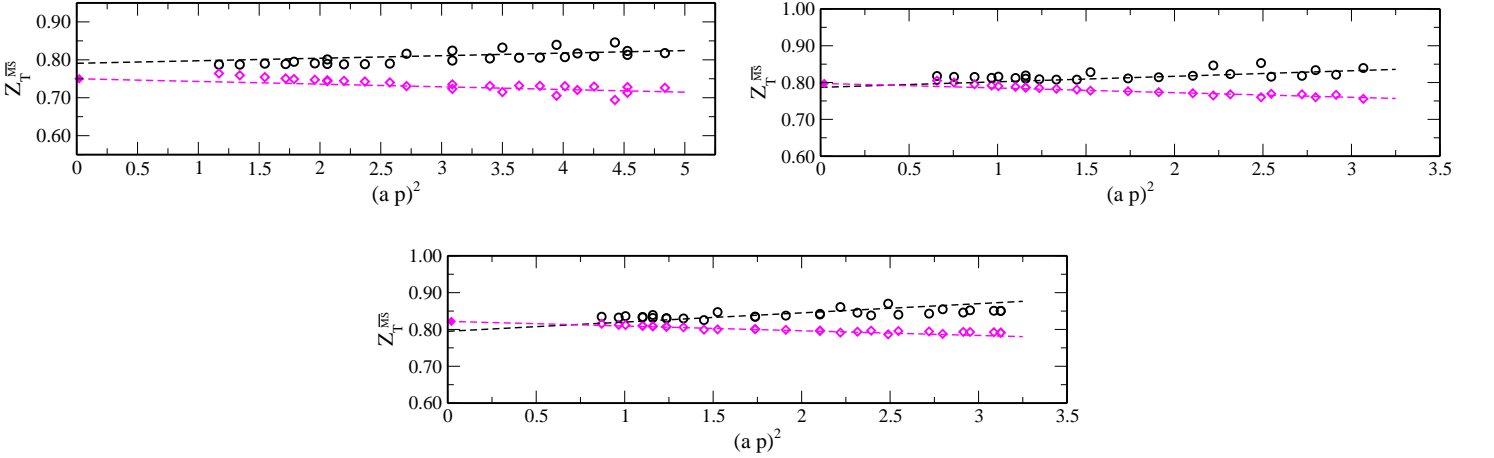

\centerline{\psfig{figure=plots/Zt_MS_2GeV_b3.9.eps,height=2.75truecm}
\quad
\psfig{figure=plots/Zt_MS_2GeV_b4.05.eps,height=2.75truecm}}
\vspace{0.5cm}
\centerline{\psfig{figure=plots/Zt_MS_2GeV_b4.20.eps,height=2.75truecm}}
\begin{minipage}{14cm}
\caption{Results on $\ZT$ for $\beta=3.9$ (upper left plot),
  $\beta=4.05$ (upper right plot) and $\beta=4.20$ (lower plot). 
 In all plots we show chirally extrapolated results. Black circles represent the non-perturbative
  data before and magenta diamonds after subtracting the ${\cal O}(a^2)$ -terms.}
\label{fig9}
\end{minipage}
\end{figure}
\end{center}

Our final results for the $Z$-factors in the $\overline{\rm MS}$-scheme at
$\mu= 2$~GeV are given in Table~\ref{tab5}. As pointed out, these are
obtained in the continuum limit by extrapolating linearly in $(ap)^2$
using data in a fixed momentum range $p^2\approx 15-32$ (GeV)$^2$. The
continuum extrapolation was carried out at the chiral limit.
The systematic error due to the continuum extrapolation,
is estimated from the difference between results using the fit range
$p^2\approx 15-32$ (GeV)$^2$ and the range $p^2\approx 17-24$ (GeV)$^2$. 
The
results at $\beta=3.9$ and $\beta=4.05$ agree within error bars with the results of Ref.~\cite{Constantinou:2010gr}.
Since their evaluation procedure differs, this agreement provides a nice confirmation
of the values obtained. 
In Ref.~\cite{Constantinou:2010gr} the vertex computation 
 employs translation invariance to evaluate the
correlation functions for all values of the momentum, whereas
we calculate the vertex for a given momentum dependent source, leading to
smaller statistical errors.
More importantly, the two procedures differ in the analysis of the lattice
data, both in the way the chiral extrapolation of the renormalization constants at 
fixed $p^2$ is carried out as well as in the way the systematic
errors associated with the extrapolation $p^2 \to 0$ are estimated. 
Due to the different approaches used, the statistical and systematic errors
 between the two 
computations is somewhat different.
In this work, we additionally compute the renormalization constants at
 $\beta=4.2$ but not at $\beta=3.8$, which
were included in Ref.~\cite{Constantinou:2010gr}.
Given the consistency between our values of the renormalization constant
and those of Ref.~\cite{Constantinou:2010gr}  
at $\beta=3.9$ and $\beta=4.05$ consolidates our value
at $\beta=4.20$ \footnote{We note that the preliminary value of $Z_P=0.50(2)$ in
$\overline{\rm MS}$ at 2 GeV was used in quark mass
evaluation~\cite{Blossier:2010cr} and
$b$-physics~\cite{Dimopoulos:2011gx} ETMC papers is totally consistent with the
result of this analysis.}.

\begin{table}[h]
\begin{center}
\begin{minipage}{17cm}
\begin{tabular}{lr@{}lr@{}lr@{}lr@{}lr@{}lr@{}lr@{}l}
\hline
\hline\\[-2.25ex]
\multicolumn{1}{c}{$\beta$}&
\multicolumn{2}{c}{$\,\,\Zq\,\,$} &
\multicolumn{2}{c}{$\,\,\ZS\,\,$} &
\multicolumn{2}{c}{$\,\,\ZP\,\,$} &
\multicolumn{2}{c}{$\,\,\ZP/\ZS\,\,$} &
\multicolumn{2}{c}{$\,\,\ZV\,\,$} &
\multicolumn{2}{c}{$\,\,\ZA\,\,$} &
\multicolumn{2}{c}{$\,\,\ZT\,\,$}  \\
\hline
\hline\\[-2.25ex]
3.90  &$\,\,$&0.754(9)(9)   &$\,\,$&0.726(5)(11)  &$\,\,$&0.457(10)(16)  &$\,\,$&0.639(3)(1)  &$\,\,$&0.627(1)(3)  &$\,\,$&0.758(1)(1)  &$\,\,$&0.750(9)(10) \\
4.05  &$\,\,$&0.775(4)(5)   &$\,\,$&0.691(9)(16)  &$\,\,$&0.497(8)(15)   &$\,\,$&0.682(2)(1)  &$\,\,$&0.662(1)(3)  &$\,\,$&0.773(1)(1)  &$\,\,$&0.798(7)(8)  \\
4.20  &$\,\,$&0.798(4)(9)   &$\,\,$&0.695(10)(13) &$\,\,$&0.501(8)(10)   &$\,\,$&0.713(2)(2)  &$\,\,$&0.686(1)(1)  &$\,\,$&0.789(1)(2)  &$\,\,$&0.822(4)(6)  \\
\hline
\hline
\end{tabular}
\end{minipage}
\end{center}
\caption{Final results of the renormalization constants
  $\Zq,\,\ZS,\,\ZP,\,\ZT$ in the ${\overline{\rm MS}}$ scheme, as well
  as for the scale-independent $\ZP/\ZS,\,\ZV$ and $\ZA$. 
Statistical errors
  are shown in the first parenthesis. The number in the second
  parenthesis is the systematic error due to the continuum
  extrapolation, taken as the difference between results using the fit range
  $p^2\approx 15-32$ (GeV)$^2$ and the range $p^2\approx 17-24$
  (GeV)$^2$.}
\label{tab5}
\end{table}

\section{Conclusions}
\label{sec6}
The values of the renormalization factor for the fermion field $\Zq,$ and for the
scalar, pseudoscalar, vector, axial-vector and tensor 
 local operators, $\ZS,\,\ZP,\,\ZV,\,\ZA,\,\ZT$, have been
calculated non-perturbatively. The method of choice is to use a
momentum dependent source and extract the renormalization factors
for all the relevant operators. This leads to a very accurate
evaluation of these factors using a small ensemble of
gauge configurations. The precision of the results allows us to reliably 
investigate the
light quark mass dependence. For most of the renormalization
constants studied in this work we do not find any light quark mass
dependence within our small statistical errors.
For all $\beta$ values we obtain the value at the chiral
limit by fitting the data to a constant.
For the RC of the pseudoscalar operator, $\ZP$, we
find a quark mass dependence due to the pion-pole,
which we subtract. Once the pole is subtracted, the behavior of the
data show a weak dependence on the light quark mass and therefore we
again compute the value at the chiral limit by fitting the pion pole
subtracted data to a constant. We also show that, despite using a
lattice spacing smaller than 0.1~fm, ${\cal O}(a^2\,p^2)$ cut-off
effects are visible given the high precision with which the RCs are
calculated. Thus we perform a perturbative subtraction of ${\cal
  O}(a^2\,g^2)$ terms. This leads to a milder dependence of the
renormalization constants on $(ap)^2$. Residual ${\cal O}(a^2p^2)$
effects are removed by extrapolating to zero. In this way we can
accurately determine the renormalization constants in the RI$'$-MOM
scheme. In order to compare with experiment we convert our values to
the $\rm \overline{MS}$ scheme at a scale of 2~GeV. The statistical
errors are in general smaller than the systematic ones. The latter are
estimated by changing the window of values of the momentum used to
extrapolate to $a^2p^2=0$. Our final values are given in Table~\ref{tab5}.

\vspace*{1cm}
{\bf{Acknowledgments:}}
This work was partly supported by funding received from the
Cyprus Research Promotion Foundation under contracts EPYAN/0506/08,
and TECHNOLOGY/$\Theta$E$\Pi$I$\Sigma$/0308(BE)/17.

\newpage
\appendix
\section{Strong IR divergent integrals}
\label{appA}

The integrals, with strong IR divergences (convergent only beyond
$D>6$), encountered in the present calculation are listed below
with their results. For completeness we also include the integrals
that appeared in our related publication~\cite{Alexandrou:2010me} for the matrix
elements of twist-2 operators. All these integrals can be found in
electronic form in the Mathematica file Zfactors.m, with the names:
IntegralPropagator1 - IntegralPropagator3, IntegralBilinears1 -
IntegralBilinears6, and IntegralExtendedBilinears1 -
IntegralExtendedBilinears2. To avoid heavy notation we define:
\bea
M^2 &=& (m_0)^2 +\mu^2  \,,\quad
M_j^2 = (m_0^f)^2 +\mu_j^2  ,\nonumber \\
p2 &=& \sum_\rho p_\rho^2  \,,\quad
p4 = \sum_\rho p_\rho^4  ,\nonumber \\
\hat{q}_\nu &=& 2\,\sin(\frac{q_\nu}{2})  \,, \quad
\hat{q}^2 = 4\sum_\rho \sin^2(\frac{q_\rho}{2})  , \nonumber
\eea
where $q$ stands for $k$ or $k+a\,p$, while $k$ is the loop
momentum and $p$ is the external momentum. No summation over the
indices $\nu_i$ is implied.

\vspace*{0.5cm}
{\bf{\underline{Propagator}}}
\vspace*{0.35cm}

\begin{minipage}{16cm}
\bdg[style={\small}, spread={8pt}]
\bdm[label={intp1}]
\bullet\int_{-\pi}^\pi \frac{d^4k}{\left(2\pi\right)^4}\frac{1}{\hat k^2\, \left(\widehat{k+a\,p}^2 + a^2 M^2\right)}
\hiderel{=} {\rm IntegralPropagator1} + {\cal O}\left(\Red{a^4}\right)
\edm
\bdms
=
0.03667832907475711(1) 
- \frac{\ln[a^2 M^2 + a^2 p2]}{16 \pi^2} 
- \frac{M^2 \ln[1 + \frac{p2}{M^2}]}{16 \pi^2 p2} 
+ \Red{a^2} \Red{\left(} 0.00007524033(9) p2 
        -0.00396328514(4) M^2 
        + \frac{M^2 \ln[a^2 M^2 + a^2 p2]}{128 \pi^2} 
        - \frac{M^4}{128 \pi^2 p2} 
        + \left(\frac{1}{64 \pi^2} + \frac{M^2}{128 \pi^2 p2}\right) \frac{M^4 \ln[1 + \frac{p2}{M^2}]}{p2} 
        + \frac{p4}{p2} \left( \frac{1}{384 \pi^2} 
              + \frac{M^2}{128 \pi^2 p2}
              + \frac{M^4}{64 \pi^2 p2^2}  
              - \left(\frac{1}{192 \pi^2} + \frac{M^2}{64 \pi^2 p2} + \frac{M^4}{64 \pi^2 p2^2}\right) \frac{M^2 \ln[1 + \frac{p2}{M^2}]}{p2}
             \right)
       \Red{\right)}
+ {\cal O}\left(\Red{a^4}\right)
\edms
\edg
\end{minipage}
\vspace*{0.5cm}

\begin{minipage}{16cm}
\bdg[style={\small}, spread={8pt}]
\bdm[label={intp2}]
\bullet\int_{-\pi}^\pi \frac{d^4k}{\left(2\pi\right)^4}\frac{\sin{k_{\nu_1}}}{\hat k^2\, \left(\widehat{k+a\,p}^2 + a^2 M^2\right)} 
\hiderel{=} {\rm IntegralPropagator2} + {\cal O}\left(\Red{a^5}\right) 
\edm
\bdms
= 
\Red{a}\,p_{\nu_1} \Red{\left(}-0.008655827647937295(1) 
                  + \frac{\ln[a^2 M^2 + a^2 p2]}{32 \pi^2} 
                  - \frac{M^2}{32 \pi^2 p2} 
                  + \left(\frac{1}{16 \pi^2} + \frac{M^2}{32 \pi^2 p2}\right) \frac{M^2 \ln[1 + \frac{p2}{M^2}]}{p2}
            \Red{\right)}
+ \Red{a^3} \Red{\left(}\frac{p_{\nu_1} p4}{p2} \left(- \frac{1}{768 \pi^2} 
                              - \frac{M^2}{96 \pi^2 p2}
                              - \frac{3 M^4}{128 \pi^2 p2^2} 
                              - \frac{M^6}{64 \pi^2 p2^3}  
                              + \left(\frac{1}{192 \pi^2} + \frac{M^2}{48 \pi^2 p2} + \frac{M^4}{32 \pi^2 p2^2} + \frac{M^6}{64 \pi^2 p2^3}\right) \frac{M^2 \ln[1 + \frac{p2}{M^2}]}{p2}
                        \right)
       + p_{\nu_1} \left(-0.0005107794(2) p2 
                       + \frac{p2 \ln[a^2 M^2 + a^2 p2]}{768 \pi^2} 
                       -0.00028240872(9) M^2
                       + \frac{13 M^4}{1536 \pi^2 p2}  
                       + \frac{M^6}{256 \pi^2 p2^2}
                       - \left(\frac{1}{128 \pi^2} + \frac{M^2}{96 \pi^2 p2} + \frac{M^4}{256 \pi^2 p2^2}\right) \frac{M^4 \ln[1 + \frac{p2}{M^2}]}{p2}
                 \right) 
       + p_{\nu_1}^3 \left(0.0011713297148098348(1) 
                        - \frac{\ln[a^2 M^2 + a^2 p2]}{384 \pi^2} 
                        + \frac{5 M^2}{384 \pi^2 p2}  
                        + \frac{13 M^4}{768 \pi^2 p2^2} 
                        + \frac{M^6}{128 \pi^2 p2^3}
                        - \left(\frac{1}{96 \pi^2} + \frac{M^2}{48 \pi^2 p2} + \frac{M^4}{48 \pi^2 p2^2} + \frac{M^6}{128 \pi^2 p2^3}\right) \frac{M^2 \ln[1 + \frac{p2}{M^2}]}{p2}
                  \right)
     \Red{\right)}
+ {\cal O}\left(\Red{a^5}\right)
\edms
\edg
\end{minipage}
\vspace*{0.75cm}

\newpage

\begin{minipage}{16cm}
\bdg[style={\small}, spread={8pt}]
\bdm[label={intp3}]
\bullet\int_{-\pi}^\pi \frac{d^4k}{\left(2\pi\right)^4}\frac{\sin{k_{\nu_1}} \sin{k_{\nu_2}}}{\left(\hat k^2\right)^2\, \left(\widehat{k+a\,p}^2 + a^2 M^2\right)} 
\hiderel{=} {\rm IntegralPropagator3} + {\cal O}\left(\Red{a^4}\right) 
\edm
\bdms
=
\delta_{\nu_1\, \nu_2} \left(0.004327913823968648(1) 
                     - \frac{\ln[a^2 M^2 + a^2 p2]}{64 \pi^2} 
                     - \frac{M^2}{64 \pi^2 p2} 
                     + \frac{M^4 \ln[1 + \frac{p2}{M^2}]}{64 \pi^2 p2^2}
                    \right) 
+ \frac{p_{\nu_1} p_{\nu_2}}{p2} \left(\frac{1}{32 \pi^2} 
                        + \frac{M^2}{16 \pi^2 p2} 
                        - \left(\frac{1}{16 \pi^2} + \frac{M^2}{16 \pi^2 p2}\right) \frac{M^2 \ln[1 + \frac{p2}{M^2}]}{p2}
                 \right)
+ \Red{a^2} \Red{\left(}\delta_{\nu_1\, \nu_2} \left(0.00025539124(4) p2 
                                     - \frac{p2 \ln[a^2 M^2 + a^2 p2]}{1536 \pi^2}  
                                     + 0.00010358434(2) M^2
                                     + \frac{5 M^4}{3072 \pi^2 p2} 
                                     + \frac{M^6}{512 \pi^2 p2^2}
                                     - \left(\frac{1}{384 \pi^2} + \frac{M^2}{512 \pi^2 p2}\right) \frac{M^6 \ln[1 + \frac{p2}{M^2}]}{p2^2}
                            \right) 
       + p_{\nu_1} p_{\nu_2} \left(-0.00037885376(9) 
                               + \frac{\ln[a^2 M^2 + a^2 p2]}{768 \pi^2} 
                               - \frac{M^2}{768 \pi^2 p2}
                               - \frac{23 M^4}{1536 \pi^2 p2^2} 
                               - \frac{3 M^6}{256 \pi^2 p2^3} 
                               + \left(\frac{1}{128 \pi^2} + \frac{M^2}{48 \pi^2 p2} + \frac{3 M^4}{256 \pi^2 p2^2}\right) \frac{M^4 \ln[1 + \frac{p2}{M^2}]}{p2^2}
                         \right)
               + \delta_{\nu_1\, \nu_2} p_{\nu_1}^2 \left(-0.00013565411323668763(1) 
                                                   + \frac{\ln[a^2 M^2 + a^2 p2]}{768 \pi^2} 
                                                   + \frac{5 M^2}{768 \pi^2 p2}
                                                   + \frac{31 M^4}{1536 \pi^2 p2^2} 
                                                   + \frac{3 M^6}{256 \pi^2 p2^3} 
                                                   - \left(\frac{1}{64 \pi^2} + \frac{5 M^2}{192 \pi^2 p2} + \frac{3 M^4}{256 \pi^2 p2^2}\right) \frac{M^4 \ln[1 + \frac{p2}{M^2}]}{p2^2}
                                             \right)
       + \frac{p_{\nu_1} p_{\nu_2} \left(p_{\nu_1}^2 + p_{\nu_2}^2\right)}{p2} \left(- \frac{1}{384 \pi^2} 
                                                                 - \frac{M^2}{48 \pi^2 p2} 
                                                                 - \frac{3 M^4}{64 \pi^2 p2^2}
                                                                 - \frac{M^6}{32 \pi^2 p2^3} 
                                                                 + \left(\frac{1}{96 \pi^2} + \frac{M^2}{24 \pi^2 p2} + \frac{M^4}{16 \pi^2 p2^2} + \frac{M^6}{32 \pi^2 p2^3}\right) \frac{M^2 \ln[1 + \frac{p2}{M^2}]}{p2}
                                                          \right)
       + \frac{\delta_{\nu_1\, \nu_2} p4}{p2} \left(\frac{1}{1536 \pi^2}  
                                          - \frac{M^2}{768 \pi^2 p2} 
                                          - \frac{M^4}{256 \pi^2 p2^2} 
                                          - \frac{M^6}{128 \pi^2 p2^3} 
                                          + \left(\frac{1}{384 \pi^2} + \frac{M^2}{128 \pi^2 p2} + \frac{M^4}{128 \pi^2 p2^2}\right) \frac{M^4 \ln[1 + \frac{p2}{M^2}]}{p2^2}
                                 \right) 
             + \frac{p_{\nu_1} p_{\nu_2} p4}{p2^2} \left(\frac{1}{768 \pi^2}
                                    + \frac{5 M^2}{192 \pi^2 p2}  
                                    + \frac{11 M^4}{128 \pi^2 p2^2}
                                    + \frac{5 M^6}{64 \pi^2 p2^3} 
                                    - \left(\frac{1}{96 \pi^2} + \frac{M^2}{16 \pi^2 p2} + \frac{M^4}{8 \pi^2 p2^2} + \frac{5 M^6}{64 \pi^2 p2^3}\right) \frac{M^2 \ln[1 + \frac{p2}{M^2}]}{p2}
                              \right) 
       \Red{\right)}
+ {\cal O}\left(\Red{a^4}\right)
\edms
\edg
\end{minipage}

\vspace*{0.5cm}
{\bf{\underline{Bilinears}}}
\vspace*{0.5cm}

\begin{minipage}{16cm}
\bdg[style={\small}, spread={8pt}]
\bdm[label={intb1}]
\bullet\int_{-\pi}^\pi \frac{d^4k}{\left(2\pi\right)^4}\frac{1}{\hat k^2\, \left(\widehat{k+a\,p}^2 + a^2 M^2\right)\, \left(\widehat{k+a\,p}^2 + a^2 M_2^2\right)} 
\hiderel{=} {\rm IntegralBilinears1} + {\cal O}\left(\Red{a^2}\right) 
\edm
\bdms
=
0.0039632853(1) 
- \frac{p4}{128 \pi^2 p2^2}
+\sum^{2}_{j=1}\MB{\left\{}
\frac{(-1)^{j+1}}{\Red{a^2}} \Red{\left(}\frac{\ln[a^2 M_j^2 + a^2 p2]}{16 \pi^2 \left(M^2 - M_2^2\right)} 
                + \frac{M_j^2 \ln[1 + \frac{p2}{M_j^2}]}{16 \pi^2 p2 \left(M^2 - M_2^2\right)} 
                \Red{\right)}
+ \frac{M_j^2}{128 \pi^2 p2} 
+ (-1)^{j} \frac{M_j^2 \ln[a^2 M_j^2 + a^2 p2]}{128 \pi^2 \left(M^2 - M_2^2\right)} 
+ (-1)^{j} \left(\frac{1}{64 \pi^2} + \frac{M_j^2}{128 \pi^2 p2}\right) \frac{M_j^4 \ln[1 + \frac{p2}{M_j^2}]}{p2 \left(M^2 - M_2^2\right)}
+ \frac{p4}{p2^2} \left(-\frac{M_j^2}{64 \pi^2 p2}  
      + (-1)^{j+1} \left(\frac{1}{192 \pi^2} + \frac{M_j^2}{64 \pi^2 p2} + \frac{M_j^4}{64 \pi^2 p2^2}\right) \frac{M_j^2 \ln[1 + \frac{p2}{M_j^2}]}{\left(M^2 - M_2^2\right)}
     \right)
\MB{\right\}}
+ {\cal O}\left(\Red{a^2}\right)
\edms
\edg
\end{minipage}
\vspace*{0.5cm}
\newpage
\begin{minipage}{16cm}
\bdg[style={\small}, spread={8pt}]
\bdm[label={intb2}]
\bullet\int_{-\pi}^\pi \frac{d^4k}{\left(2\pi\right)^4}\frac{\sin\left(k_{\nu_1}+a\,p_{\nu_1}\right)}{\hat k^2\, \left(\widehat{k+a\,p}^2 + a^2 M^2\right)\, \left(\widehat{k+a\,p}^2 + a^2 M_2^2\right)} 
\hiderel{=} {\rm IntegralBilinears2} + {\cal O}\left(\Red{a^3}\right) 
\edm
\bdms
=
\frac{1}{\Red{a}} \frac{p_{\nu_1}}{32 \pi^2 p2}
+ \Red{a} \Red{\left(} \frac{p_{\nu_1} p4}{p2^2} \left(\frac{1}{384 \pi^2} 
                             + \frac{ M^2 M_2^2 }{64 \pi^2 p2^2}
                       \right)
            - p_{\nu_1} \left(0.0002071688(1) 
                            + \frac{M^2 M_2^2}{256 \pi^2 p2^2}
                     \right)
            - \frac{p_{\nu_1}^3}{p2} \left(\frac{1}{384 \pi^2}  
                             + \frac{M^2 M_2^2}{128 \pi^2 p2^2}
                       \right)  
      \Red{\right)}
+\sum^{2}_{j=1} \MB{\left\{}
(-1)^{j+1} \frac{p_{\nu_1}}{\Red{a}} \Red{\left(} \frac{\ln[a^2 M_j^2 + a^2 p2]}{32 \pi^2 \left(M^2 - M_2^2\right)} 
                                   - \frac{M_j^4 \ln[1 + \frac{p2}{M_j^2}]}{32 \pi^2 p2^2 \left(M^2 - M_2^2\right)} 
                            \Red{\right)} 
+ \Red{a} \Red{\left(} \frac{p_{\nu_1} p4}{p2^2} \Bigg{(} \frac{M_j^2}{128 \pi^2 p2}
                             + \frac{M_j^4}{64 \pi^2 p2^2} 
                             +(-1)^{j} \left(\frac{1}{192 \pi^2} + \frac{M_j^2}{64 \pi^2 p2} + \frac{M_j^4}{64 \pi^2 p2^2}\right) \frac{M_j^4 \ln[1 + \frac{p2}{M_j^2}]}{p2 \left(M^2 - M_2^2\right)} 
                                                 \Bigg{)}
       + p_{\nu_1} \Bigg{(}- \frac{5 M_j^2}{1536 \pi^2 p2}
                      - \frac{M_j^4}{256 \pi^2 p2^2}
                      + (-1)^{j+1} \frac{p2 \ln[a^2 M_j^2 + a^2 p2]}{768 \pi^2 \left(M^2 - M_2^2\right)} 
                      + (-1)^{j+1} \left( \frac{1}{192 \pi^2} + \frac{M_j^2}{256 \pi^2 p2}\right) \frac{M_j^6 \ln[1 + \frac{p2}{M_j^2}]}{p2^2 \left(M^2 - M_2^2\right)}
               \Bigg{)}
       + \frac{p_{\nu_1}^3}{p2} \Bigg{(}- \frac{5 M_j^2}{768 \pi^2 p2} 
                       - \frac{M_j^4}{128 \pi^2 p2^2} 
                       + (-1)^{j} \frac{p2 \ln[a^2 M_j^2 + a^2 p2]}{384 \pi^2 \left(M^2 - M_2^2\right)} 
                       + (-1)^{j+1} \left( \frac{1}{192 \pi^2} + \frac{M_j^2}{96 \pi^2 p2} + \frac{M_j^4}{128 \pi^2 p2^2}\right) \frac{M_j^4 \ln[1 + \frac{p2}{M_j^2}]}{p2 \left(M^2 - M_2^2\right)}
                              \Bigg{)}
    \Red{\right)}
\MB{\right\}}
+ {\cal O}\left(\Red{a^3}\right)
\edms
\edg
\end{minipage}
\vspace*{0.5cm}

\bdm[style={\small}, spread={8pt}, label={intb3}]
\hspace*{-4cm}\bullet\int_{-\pi}^\pi \frac{d^4k}{\left(2\pi\right)^4}\frac{\sin\left(k_{\nu_1}+a\,p_{\nu_1}\right) \sin\left(k_{\nu_2}+a\,p_{\nu_2}\right)}{\hat k^2\, \left(\widehat{k+a\,p}^2 + a^2 M^2\right)\, \left(\widehat{k+a\,p}^2 + a^2 M_2^2\right)} 
= {\rm IntegralBilinears3} + {\cal O}\left(\Red{a^4}\right)
\edm
\vspace*{0.5cm}

\bdm[style={\small}, spread={8pt}, label={intb4}]
\hspace*{-4cm}\bullet\int_{-\pi}^\pi \frac{d^4k}{\left(2\pi\right)^4}\frac{\sin{k_{\nu_1}} \sin{k_{\nu_2}}}{\left(\hat k^2\right)^2\, \left(\widehat{k+a\,p}^2 + a^2 M^2\right)\, \left(\widehat{k+a\,p}^2 + a^2 M_2^2\right)} 
= {\rm IntegralBilinears4} + {\cal O}\left(\Red{a^2}\right)
\edm
\vspace*{0.5cm}

\bdm[style={\small}, spread={8pt}, label={intb5}]
\hspace*{-4cm}\bullet\int_{-\pi}^\pi \frac{d^4k}{\left(2\pi\right)^4}\frac{\sin{k_{\nu_1}} \sin{k_{\nu_2}} \sin\left(k_{\nu_3}+a\,p_{\nu_3}\right)}{\left(\hat k^2\right)^2\, \left(\widehat{k+a\,p}^2 + a^2 M^2\right)\, \left(\widehat{k+a\,p}^2 + a^2 M_2^2\right)} 
= {\rm IntegralBilinears5} + {\cal O}\left(\Red{a^3}\right)
\edm
\vspace*{0.5cm}

\bdm[style={\small}, spread={8pt}, label={intb6}]
\hspace*{-4cm}\bullet\int_{-\pi}^\pi \frac{d^4k}{\left(2\pi\right)^4}\frac{\sin{k_{\nu_1}} \sin{k_{\nu_2}} \sin\left(k_{\nu_3}+a\,p_{\nu_3}\right) \sin\left(k_{\nu_4}+a\,p_{\nu_4}\right)}{\left(\hat k^2\right)^2\, \left(\widehat{k+a\,p}^2 + a^2 M^2\right)\, \left(\widehat{k+a\,p}^2 + a^2 M_2^2\right)} 
= {\rm IntegralBilinears6} + {\cal O}\left(\Red{a^4}\right)
\edm
\vspace*{1cm}

{\bf{\underline{Extended Bilinears}}}
\vspace{0.5cm}
\bdm[style={\small}, spread={8pt}, label={inteb1}]
\hspace*{-2.5cm}\bullet\int_{-\pi}^\pi \frac{d^4k}{\left(2\pi\right)^4}\frac{\sin\left(k_{\nu_1}+a\,p_{\nu_1}\right) \sin\left(k_{\nu_2}+a\,p_{\nu_2}\right) \sin\left(k_{\nu_3}+a\,p_{\nu_3}\right)}{\hat k^2\, \left(\widehat{k+a\,p}^2 + a^2 M^2\right)\, \left(\widehat{k+a\,p}^2 + a^2 M_2^2\right)} 
= {\rm IntegralExtendedBilinears1} + {\cal O}\left(\Red{a^5}\right)
\edm

\bdm[style={\small}, spread={8pt}, label={inteb2}]
\hspace*{-2.5cm}\bullet\int_{-\pi}^\pi \frac{d^4k}{\left(2\pi\right)^4}\frac{\sin\left(k_{\nu_1}+a\,p_{\nu_1}\right) \sin\left(k_{\nu_2}+a\,p_{\nu_2}\right) \sin\left(k_{\nu_3}+a\,p_{\nu_3}\right) \kc_{\nu_4} \kc_{\nu_5} }{\left(\hat k^2\right)^2\, \left(\widehat{k+a\,p}^2 + a^2 M^2\right)\, \left(\widehat{k+a\,p}^2 + a^2 M_2^2\right)} 
= {\rm IntegralExtendedBilinears2} + {\cal O}\left(\Red{a^5}\right)
\edm

\newpage
\section{Analytic expressions for RCs of bilinear operators}
\label{appB}

In this Appendix we provide the analytic expressions for the RCs of the ultra-local bilinears,
as defined in Eq.~(\ref{eq:rimom}):


\begin{minipage}{17.5cm}
\bdg[style={\normalsize}, spread={8pt}]
\bdm[label={Zs}]
Z_S^{\rm pert.} \hiderel{=} 1 
\edm
\bdms
 +\gttwo \Green{\left\{}-13.606731(1)
                +3 \ln[a^2 m^2+a^2 p2]
                +\frac{9 m^2 \ln[1+\frac{p2}{m^2}]}{p2}
                +\Red{a}\, m \Red{\left[}2.7312983(2)
                            -\frac{15}{2} \ln[a^2 m^2+a^2 p2]
                            +\frac{3 m^2}{2 p2}
                            +\frac{6 m^2}{m^2+p2}
                            -\left(24 + \frac{3 m^2}{2 p2} \right) \frac{m^2 \ln[1+\frac{p2}{m^2}]}{p2}
                     \Red{\right]}
                +\Red{a^2} \Red{\left[}-1.207563(2)\,p2
                           -10.853390(2)\,m^2
                           -\frac{1289 m^4}{360 p2}
                           -\frac{721 m^6}{240 p2^2}
                           +\frac{7 m^8}{40 p2^3}
                           -\frac{18 m^4}{m^2+p2}
                           +\frac{3 m^6}{\left(m^2+p2\right)^2}
                           +\left(\frac{107 m^2}{6}+\frac{17 p2}{360}\right) \ln[a^2 m^2+a^2 p2]
                           +\left(-1 +\frac{1321 m^2}{24 p2}+\frac{367 m^4}{72 p2^2}+\frac{35 m^6}{12 p2^3}-\frac{7 m^8}{40 p2^4}\right)\,m^2 \ln[1+\frac{p2}{m^2}]\qquad
                           +\frac{p4}{p2} \Bigg{(}-0.3935023(2)
                                    -\frac{157 \ln[a^2 m^2+a^2 p2]}{180}
                                    +\frac{117 m^2}{80 p2}
                                    -\frac{371 m^4}{180 p2^2}
                                    +\frac{721 m^6}{120 p2^3}
                                    -\frac{7 m^8}{20 p2^4}
                                    +\frac{2 p2}{m^2+p2}
                                    +\left(1 +\frac{m^2}{12 p2} -\frac{35 m^4}{36 p2^2} -\frac{35 m^6}{6 p2^3} +\frac{7 m^8}{20 p2^4}\right) \frac{m^2 \ln[1+\frac{p2}{m^2}]}{p2}
                              \Bigg{)}
                     \Red{\right]}
          \Green{\right\}}\Bigg{|}_{p_{\rho}=\mu_{\rho}}
+ {\cal O}(\Red{a^3},\Green{g^4})
\edms
\edg
\end{minipage}
\vspace*{0.5cm}

\begin{minipage}{17.5cm}
\bdg[style={\normalsize}, spread={8pt}]
\bdm[label={Zp}]
Z_P^{\rm pert.} \hiderel{=} 1
\edm
\bdms
+ \gttwo \Green{\left\{}-21.733356(1)
               +3 \ln[a^2 m^2+a^2 p2]
               +\frac{3 m^2 \ln[1+\frac{p2}{m^2}]}{p2}
               +\Red{a}\, m \Red{\left[}7.0252230(2)
                          -\frac{3}{2} \ln[a^2 m^2+a^2 p2]
                          +\frac{3 m^2}{2 p2}
                          -\left(6 +\frac{3 m^2}{2 p2}\right) \frac{m^2 \ln[1+\frac{p2}{m^2}]}{p2}
                   \Red{\right]}
               +\Red{a^2} \Red{\Bigg{[}}0.440762(2)\,p2
                            -5.520750(2)\,m^2
                            -\frac{1769 m^4}{360 p2}
                            -\frac{27 m^6}{80 p2^2}
                            +\frac{7 m^8}{40 p2^3}
                            -\frac{3 m^4}{2(m^2+p2)}
                            +\left(\frac{3 m^2}{2}+\frac{17 p2}{360}\right) \ln[a^2 m^2+a^2 p2]
                            +\left(-\frac{1}{3}+\frac{229 m^2}{24 p2}+\frac{367 m^4}{72 p2^2}+\frac{m^6}{4 p2^3}-\frac{7 m^8}{40 p2^4}\right)\,m^2 \ln[1+\frac{p2}{m^2}]
                            +\frac{p4}{p2} \left(1.6064977(2)
                                      -\frac{157 \ln[a^2 m^2+a^2 p2]}{180}
                                      -\frac{227 m^2}{720 p2}
                                      +\frac{109 m^4}{180 p2^2}
                                      +\frac{27 m^6}{40 p2^3}
                                      -\frac{7 m^8}{20 p2^4}
                                      +\left(\frac{1}{3} +\frac{m^2}{12 p2} -\frac{35 m^4}{36 p2^2} -\frac{m^6}{2 p2^3} +\frac{7 m^8}{20 p2^4}\right) \frac{m^2 \ln[1+\frac{p2}{m^2}]}{p2}
                               \right)
                      \Red{\Bigg{]}}
         \Green{\right\}}\Bigg{|}_{p_{\rho}=\mu_{\rho}}
+ {\cal O}(\Red{a^3},\Green{g^4})
\edms
\edg
\end{minipage}
\vspace*{0.5cm}

\newpage
\begin{minipage}{17.5cm}
\bdg[style={\normalsize}, spread={8pt}]
\bdm[label={Zv}]
Z_V^{\rm pert.} \hiderel{=} 1
\edm
\bdms
 + \gttwo \Green{\Bigg{\{}}-16.6028865(8)
                                        +\Red{a}\, m \Red{\left[}2.2261230(2)
                                                  +3 \ln[a^2 m^2+a^2 p2]
                                                  +\frac{p_{\nu_1}^2}{p2} \left(-3 +\frac{6 m^2}{p2} \right)
                                                  +\left(3 -\frac{6 m^2 p_{\nu_1}^2}{p2^2}\right) \frac{m^2 \ln[1+\frac{p2}{m^2}]}{p2}
                                            \Red{\right]}
                                         +\Red{a^2} \Red{\Bigg{[}}1.125750(1)\,p2
                                                     +1.102770(2)\,m^2
                                                      +\frac{65 m^4}{48 p2}
                                                      +\frac{m^6}{8 p2^2}
                                                      +\left(-\frac{25 m^2}{4}+\frac{76 p_{\nu_1}^2}{45}-\frac{7 p2}{24}\right) \ln[a^2 m^2+a^2 p2]
                                                      +p_{\nu_1}^2 \left(-2.714031(1)
                                                                        +\frac{5017 m^2}{360 p2}
                                                                        -\frac{9401 m^4}{360 p2^2}
                                                                        -\frac{m^6}{10 p2^3}
                                                                        +\frac{7 m^8}{10 p2^4}
                                                                        -\frac{6 m^2}{m^2+p2}
                                                                  \right)
                                                      +\frac{p_{\nu_1}^4}{p2} \left(\frac{323}{180}
                                                                       -\frac{59 m^2}{18 p2}
                                                                       -\frac{35 m^4}{18 p2^2}
                                                                       -\frac{14 m^6}{3 p2^3}
                                                                       +\frac{14 m^8}{3 p2^4}
                                                                 \right)
                                                      +\left(-\frac{41}{4}
                                                             -\frac{17 m^2}{12 p2}
                                                             -\frac{m^4}{8 p2^2}
                                                             +\frac{p_{\nu_1}^2}{p2} \left(-\frac{11}{12}+\frac{236 m^2}{9 p2}-\frac{m^4}{4 p2^2}-\frac{7 m^6}{10 p2^3}\right) 
                                                             +\frac{p_{\nu_1}^4}{p2^2} \left(\frac{11}{3}+\frac{14 m^2}{3 p2}+\frac{7 m^4}{3 p2^2}-\frac{14 m^6}{3 p2^3}\right)
                                                       \right) \frac{m^4 \ln[1+\frac{p2}{m^2}]}{p2}
                                                     +\frac{p4}{p2} \left(2.0773310(2)
                                                               -\frac{157 \ln[a^2 m^2+a^2 p2]}{180}
                                                               -\frac{67 m^2}{120 p2}
                                                               +\frac{m^4}{120 p2^2}
                                                               -\frac{8 m^6}{15 p2^3}
                                                               +\frac{7 m^8}{30 p2^4}
                                                               +\frac{p_{\nu_1}^2}{p2} \left(\frac{7}{120}
                                                                                            +\frac{m^2}{12 p2}
                                                                                            +\frac{77 m^4}{12 p2^2}
                                                                                            +\frac{6 m^6}{p2^3}
                                                                                            -\frac{7 m^8}{p2^4}
                                                                          \right)
                                                               +\left(+\frac{1}{2}
                                                                      +\frac{5 m^2}{18 p2}
                                                                      +\frac{5 m^4}{12 p2^2}
                                                                      -\frac{7 m^6}{30 p2^3}
                                                                      +\frac{p_{\nu_1}^2}{p2} \left(-\frac{5}{2}
                                                                                                 -\frac{10 m^2}{p2}
                                                                                                 -\frac{5 m^4}{2 p2^2}
                                                                                                 +\frac{7 m^6}{p2^3}
                                                                                          \right)
                                                               \right) \frac{m^4 \ln[1+\frac{p2}{m^2}]}{p2^2}
                                                        \right)
                                                \Red{\Bigg{]}}
                                  \Green{\Bigg{\}}}\Bigg{|}_{p_{\rho}=\mu_{\rho}}
+ {\cal O}(\Red{a^3},\Green{g^4})
\edms
\edg
\end{minipage}
\vspace*{0.5cm}

\newpage

\begin{minipage}{17.5cm}
\bdg[style={\normalsize}, spread={8pt}]
\bdm[label={Za}]
Z_A^{\rm pert.} \hiderel{=} 1 
\edm
\bdms
1+ \gttwo \Green{\Bigg{\{}}-12.5395741(8)
                                       +\frac{2 m^2}{p2}
                                       -\frac{8 p_{\nu_1}^2 m^2}{p2^2}
                                       +\left(2 -\frac{2 m^2}{p2} 
                                                + \frac{p_{\nu_1}^2}{p2}\left(4 +\frac{8 m^2}{p2} \right) 
                                       \right) \frac{m^2 \ln[1+\frac{p2}{m^2}]}{p2} 
                                       +\Red{a} m \Red{\left[}-1.4208394(2)
                                                  -\frac{5 m^2}{p2}
                                                  +\frac{4 m^2}{m^2+p2}
                                                  +\frac{p_{\nu_1}^2}{p2} \left(-1
                                                                  +\frac{26 m^2}{p2}
                                                                  +\frac{4 p2}{m^2+p2}
                                                            \right) 
                                                  +\left(-3 m
                                                         +\frac{5 m^2}{p2}
                                                         - \frac{p_{\nu_1}^2}{p2} \left(12 +\frac{26 m^2}{p2}\right)
                                                  \right)\frac{m^2 \ln[1+\frac{p2}{m^2}]}{p2} 
                                           \Red{\right]}
                                       +\Red{a^2} \Red{\Bigg{[}}-0.153718(1)\,p2 
                                                   +1.290617(2)\,m^2
                                                   +\frac{557 m^4}{48 p2}
                                                   -\frac{13 m^6}{12 p2^2}
                                                   +\frac{3 m^8}{4 p2^3}
                                                   -\frac{10 m^4}{m^2+p2}
                                                   +\frac{2 m^6}{\left(m^2+p2\right)^2} 
                                                   +\left(\frac{23 m^2}{12}-\frac{14 p_{\nu_1}^2}{45}+\frac{5 p2}{24}\right) \ln[a^2 m^2+a^2 p2] 
                                                   +p_{\nu_1}^2 \left(-0.892808(1)
                                                                   +\frac{2707 m^2}{360 p2}
                                                                   -\frac{23201 m^4}{360 p2^2}
                                                                   -\frac{3 m^6}{5 p2^3}
                                                                   -\frac{23 m^8}{10 p2^4}
                                                                   -\frac{11 m^2}{m^2+p2}
                                                                   +\frac{2 m^4}{\left(m^2+p2\right)^2}
                                                             \right)
                                                   +\frac{p_{\nu_1}^4}{p2} \left(+\frac{323}{180}
                                                                               +\frac{5 m^2}{18 p2}
                                                                               +\frac{145 m^4}{18 p2^2}
                                                                               +\frac{8 m^6}{p2^3}
                                                                               -\frac{46 m^8}{3 p2^4}
                                                                        \right) 
                                                   +\left(-\frac{2}{3}
                                                          +\frac{73 m^2}{12 p2}
                                                          -\frac{11 m^4}{p2^2}
                                                          +\frac{17 m^6}{24 p2^3}
                                                          -\frac{3 m^8}{4 p2^4}
                                                          +\frac{p_{\nu_1}^2}{p2} \left(\frac{4}{3} +\frac{99 m^2}{4 p2} +\frac{581 m^4}{9 p2^2} +\frac{7 m^6}{4 p2^3} +\frac{23 m^8}{10 p2^4}\right)
                                                          +\frac{p_{\nu_1}^4}{p2^2} \left(-2 -\frac{3 m^2}{p2}-\frac{40 m^4}{3 p2^2}-\frac{m^6}{3 p2^3}+\frac{46 m^8}{3 p2^4}\right) 
                                                    \right)\,m^2 \ln[1+\frac{p2}{m^2}] 
                                                   +\frac{p4}{p2} \left(0.7439977(2)
                                                             -\frac{157 \ln[a^2 m^2+a^2 p2]}{180}
                                                             +\frac{19 m^2}{60 p2}
                                                             -\frac{13 m^4}{40 p2^2}
                                                             +\frac{163 m^6}{60 p2^3}
                                                             -\frac{34 m^8}{15 p2^4}
                                                             +\frac{4 p2}{3 \left(m^2+p2\right)}
                                                             +\Bigg{(}\frac{2}{3}
                                                                    +\frac{m^2}{6 p2}
                                                                    -\frac{11 m^4}{9 p2^2}
                                                                    -\frac{19 m^6}{12 p2^3}
                                                                    +\frac{34 m^8}{15 p2^4}
                                                                    +\frac{p_{\nu_1}^2}{p2} \left(\frac{2}{3}
                                                                                     +\frac{11 m^2}{2 p2}
                                                                                     +\frac{14 m^4}{p2^2}
                                                                                     -\frac{5 m^6}{2 p2^3}
                                                                                     -\frac{23 m^8}{p2^4}
                                                                              \right)
                                                              \Bigg{)} \frac{m^2 \ln[1+\frac{p2}{m^2}]}{p2}
                                                             +\frac{p_{\nu_1}^2}{p2} \left(+\frac{167}{120}
                                                                             -\frac{4 p2}{3 m^2}
                                                                             -\frac{41 m^2}{12 p2}
                                                                             -\frac{91 m^4}{12 p2^2}
                                                                             -\frac{9 m^6}{p2^3}
                                                                             +\frac{23 m^8}{p2^4}
                                                                             +\frac{4 p2^2 }{3 \left(m^2+p2\right)\,m^2}
                                                                       \right)
                                                       \right)
                                             \Red{\Bigg{]}}
                                 \Green{\Bigg{\}}}\Bigg{|}_{p_{\rho}=\mu_{\rho}}
+ {\cal O}(\Red{a^3},\Green{g^4})
\edms
\edg
\end{minipage}
\vspace*{0.5cm}

\newpage
\begin{minipage}{17.5cm}
\bdg[style={\normalsize}, spread={8pt}]
\bdm[label={Zt}]
\left(Z_T^{\rm pert.}\right)^{\nu_1 \hiderel{\neq}\, \nu_2} \hiderel{=} 1
\edm
\bdms
1+ \gttwo \Green{\Bigg{\{}}-13.5382926(8)
                                                        -\ln[a^2 m^2+a^2 p2]
                                                        -\frac{2 m^2}{p2}
                                                        +\left(1+ \frac{2 m^2}{p2} \right) \frac{m^2 \ln[1+\frac{p2}{m^2}]}{p2}
                                                        +\frac{\left(p_{\nu_1}^2+p_{\nu_2}^2\right)}{p2} \left(\frac{4 m^2}{p2}
                                                                                                         -\left(2 +\frac{4 m^2}{p2}\right) \frac{m^2 \ln[1+\frac{p2}{m^2}]}{p2}
                                                                                         \right) 
                                                        +\Red{a}\, m \Red{\left[}0.4107689(2)
                                                                   +\frac{7}{2} \ln[a^2 m^2+a^2 p2]
                                                                   +\frac{13 m^2}{2 p2}
                                                                   -\frac{13 m^4 \ln[1+\frac{p2}{m^2}]}{2 p2^2}
                                                                   +\frac{\left(p_{\nu_1}^2+p_{\nu_2}^2\right)}{p2} \left(-1
                                                                                                          -\frac{10 m^2}{p2}
                                                                                                          -\frac{2 p2}{m^2+p2}
                                                                                                          +\left(6 +\frac{10 m^2}{p2} \right) \frac{m^2 \ln[1+\frac{p2}{m^2}]}{p2} 
                                                                                                    \right) 
                                                            \Red{\right]}
                                                        +\Red{a^2} \Red{\left[}1.000358(2)\,p2
                                                                    +1.509337(2)\,m^2
                                                                    -\frac{628 m^4}{45 p2}
                                                                    +\frac{407 m^6}{720 p2^2}
                                                                    -\frac{23 m^8}{40 p2^3}
                                                                    +\frac{3 m^4}{2 (m^2+p2)}
                                                                    -\left(\frac{55 m^2}{9}+\frac{41 p2}{120}\right) \ln[a^2 m^2+a^2 p2] 
                                                                    +\frac{p_{\nu_1}^2 p_{\nu_2}^2}{p2} \left(\frac{20}{9}
                                                                                                         -\frac{38 m^2}{9 p2}
                                                                                                         +\frac{4 m^4}{p2^2}
                                                                                                         +\frac{16 m^6}{3 p2^3}
                                                                                                   \right)  
                                                                    +\left(\frac{1}{3}
                                                                           -\frac{65 m^2}{24 p2}
                                                                           +\frac{109 m^4}{8 p2^2}
                                                                           -\frac{5 m^6}{18 p2^3}
                                                                           +\frac{23 m^8}{40 p2^4}
                                                                           + \frac{p_{\nu_1}^2 p_{\nu_2}^2}{p2^2}\left(\frac{8 m^2}{3 p2}
                                                                                                                  -\frac{20 m^4}{3 p2^2}
                                                                                                                  -\frac{16 m^6}{3 p2^3}
                                                                                                           \right)
                                                                    \right)\,m^2 \ln[1+\frac{p2}{m^2}]
                                                                    +\left(p_{\nu_1}^2+p_{\nu_2}^2\right) \Bigg{(}-2.0217225(2)
                                                                                                           +\ln[a^2 m^2+a^2 p2]
                                                                                                           +\frac{383 m^2}{72 p2}
                                                                                                           +\frac{103 m^4}{6 p2^2}
                                                                                                           -\frac{29 m^6}{12 p2^3}
                                                                                                           +\frac{3 m^8}{2 p2^4}
                                                                                                           +\frac{5 m^2}{2 \left(m^2+p2\right)}
                                                                                                           -\frac{m^4}{\left(m^2+p2\right)^2}
                                                                                                           +\left(-\frac{2}{3}
                                                                                                                  -\frac{85 m^2}{6 p2}
                                                                                                                  -\frac{95 m^4}{6 p2^2}
                                                                                                                  +\frac{5 m^6}{3 p2^3}
                                                                                                                  -\frac{3 m^8}{2 p2^4}
                                                                                                            \right) \frac{m^2 \ln[1+\frac{p2}{m^2}]}{p2}
                                                                                                     \Bigg{)}
                                                                    +\frac{\left(p_{\nu_1}^4+p_{\nu_2}^4\right)}{p2} \Bigg{(}\frac{10}{9}
                                                                                                           -\frac{35 m^2}{9 p2}
                                                                                                           -\frac{3 m^4}{p2^2}
                                                                                                           -\frac{11 m^6}{3 p2^3}
                                                                                                           +\frac{10 m^8}{p2^4}
                                                                                                           +\left(1
                                                                                                                  +\frac{14 m^2}{3 p2}
                                                                                                                  +\frac{17 m^4}{3 p2^2}
                                                                                                                  -\frac{4 m^6}{3 p2^3}
                                                                                                                  -\frac{10 m^8}{p2^4}
                                                                                                            \right) \frac{m^2 \ln[1+\frac{p2}{m^2}]}{p2}
                                                                                                     \Bigg{)} 
                                                                    +\frac{p4}{p2} \Bigg{(}2.4814977(2)
                                                                              -\frac{157 \ln[a^2 m^2+a^2 p2]}{180}
                                                                              -\frac{497 m^2}{720 p2}
                                                                              -\frac{73 m^4}{90 p2^2}
                                                                              -\frac{43 m^6}{40 p2^3}
                                                                              +\frac{43 m^8}{20 p2^4}
                                                                              +\left(-\frac{1}{3}
                                                                                     +\frac{11 m^2}{12 p2}
                                                                                     +\frac{55 m^4}{36 p2^2}
                                                                                     -\frac{43 m^8}{20 p2^4}
                                                                               \right) \frac{m^2 \ln[1+\frac{p2}{m^2}]}{p2} 
                                                                              +\frac{\left(p_{\nu_1}^2+p_{\nu_2}^2\right)}{p2} \Bigg{(}-\frac{2}{3}
                                                                                                                     +\frac{2 p2}{3 m^2}
                                                                                                                     +\frac{7 m^2}{4 p2}
                                                                                                                     +\frac{7 m^4}{p2^2}
                                                                                                                     +\frac{15 m^6}{2 p2^3}
                                                                                                                     -\frac{15 m^8}{p2^4}
                                                                                                                     -\frac{2 p2}{3 \left(m^2+p2\right)}
                                                                                                                     +\bigg{(}-\frac{1}{3}
                                                                                                                              -\frac{4 m^2}{p2}
                                                                                                                              -\frac{12 m^4}{p2^2}
                                                                                                                              +\frac{15 m^8}{p2^4}
                                                                                                                       \bigg{)} \frac{m^2 \ln[1+\frac{p2}{m^2}]}{p2}
                                                                                                               \Bigg{)}
                                                                        \Bigg{)}
                                                              \Red{\right]} 
                                                 \Green{\Bigg{\}}}\Bigg{|}_{p_{\rho}=\mu_{\rho}}
+ {\cal O}(\Red{a^3},\Green{g^4})
\edms
\edg
\end{minipage}
\vspace*{0.5cm}

\newpage

\begin{minipage}{17.5cm}
\bdg[style={\normalsize}, spread={8pt}]
\bdm[label={Ztp}]
\left(Z_{T'}^{\rm pert.}\right)^{\nu_1 \hiderel{\neq}\, \nu_2} \hiderel{=} 1 
\edm
\bdms
1+ \gttwo \Green{\Bigg{\{}}-13.5382926(8)
                                                         -\ln[a^2 m^2+a^2 p2]
                                                         +\frac{2 m^2}{p2}
                                                         -\left(1 + \frac{2 m^2}{p2}\right) \frac{m^2 \ln[1+\frac{p2}{m^2}]}{p2}
                                                         +\frac{\left(p_{\nu_1}^2+p_{\nu_2}^2\right)}{p2} \left(-\frac{4 m^2}{p2}
                                                                                                            +\left(2 + \frac{4 m^2}{p2}\right) \frac{m^2 \ln[1+\frac{p2}{m^2}]}{p2}
                                                                                          \right)
                                                         +\Red{a}\, m \Red{\left[}-2.5892311(2)
                                                                    +\frac{7}{2} \ln[a^2 m^2+a^2 p2]
                                                                    -\frac{7 m^2}{2 p2}
                                                                    +\frac{2 m^2}{m^2+p2}
                                                                    +\left(6 + \frac{7 m^2}{2 p2} \right) \frac{m^2 \ln[1+\frac{p2}{m^2}]}{p2} 
                                                                    +\frac{\left(p_{\nu_1}^2+p_{\nu_2}^2\right)}{p2} \left(1
                                                                                                            +\frac{10 m^2}{p2}
                                                                                                            +\frac{2 p2}{m^2+p2}
                                                                                                            -\left(6 + \frac{10 m^2}{p2} \right) \frac{m^2 \ln[1+\frac{p2}{m^2}]}{p2}                                                                                                      
                                                                                                      \right)
                                                             \Red{\right]}
                                                         +\Red{a^2} \Red{\left[}0.089747(2)\,p2
                                                                     +7.217670(2)\,m^2
                                                                     +\frac{469 m^4}{90 p2}
                                                                     +\frac{587 m^6}{720 p2^2}
                                                                     +\frac{37 m^8}{40 p2^3}
                                                                     -\frac{2 m^4}{m^2+p2}
                                                                     +\frac{m^6}{\left(m^2+p2\right)^2}
                                                                     +\left(\frac{79 p2}{120}-\frac{55 m^2}{9}\right) \ln[a^2 m^2+a^2 p2]
                                                                     +\frac{p_{\nu_1}^2 p_{\nu_2}^2}{p2} \left(\frac{20}{9}
                                                                                               -\frac{38 m^2}{9 p2}
                                                                                               +\frac{4 m^4}{p2^2}
                                                                                               +\frac{16 m^6}{3 p2^3}
                                                                                         \right) 
                                                                     +\left(-\frac{1}{3}
                                                                            -\frac{373 m^2}{24 p2}
                                                                            -\frac{133 m^4}{24 p2^2}
                                                                            -\frac{23 m^6}{18 p2^3}
                                                                            -\frac{37 m^8}{40 p2^4}
                                                                            +\frac{p_{\nu_1}^2 p_{\nu_2}^2}{p2^2} \left(\frac{8 m^2}{3 p2}
                                                                                                      -\frac{20 m^4}{3 p2^2}
                                                                                                      -\frac{16 m^6}{3 p2^3}
                                                                                                \right) 
                                                                      \right)\,m^2 \ln[1+\frac{p2}{m^2}]
                                                                     +\left(p_{\nu_1}^2+p_{\nu_2}^2\right) \Bigg{(}-0.2004998(2)
                                                                                                            -\ln[a^2 m^2+a^2 p2]
                                                                                                            -\frac{79 m^2}{72 p2}
                                                                                                            -\frac{127 m^4}{6 p2^2}
                                                                                                            -\frac{35 m^6}{12 p2^3}
                                                                                                            -\frac{3 m^8}{2 p2^4}
                                                                                                            -\frac{5 m^2}{2 \left(m^2+p2\right)}
                                                                                                            +\frac{m^4}{\left(m^2+p2\right)^2}
                                                                                                            +\left(\frac{2}{3}
                                                                                                                   +\frac{23 m^2}{2 p2}
                                                                                                                   +\frac{45 m^4}{2 p2^2}
                                                                                                                   +\frac{11 m^6}{3 p2^3}
                                                                                                                   +\frac{3 m^8}{2 p2^4}
                                                                                                            \right) \frac{m^2 \ln[1+\frac{p2}{m^2}]}{p2}
                                                                                                      \Bigg{)} 
                                                                     +\frac{\left(p_{\nu_1}^4+p_{\nu_2}^4\right)}{p2} \Bigg{(}\frac{10}{9}
                                                                                                            -\frac{m^2}{3 p2}
                                                                                                            +\frac{7 m^4}{p2^2}
                                                                                                            +\frac{9 m^6}{p2^3}
                                                                                                            -\frac{10 m^8}{p2^4}
                                                                                                            +\left(-1
                                                                                                                   -\frac{2 m^2}{p2}
                                                                                                                   -\frac{37 m^4}{3 p2^2}
                                                                                                                   -\frac{4 m^6}{p2^3}
                                                                                                                   +\frac{10 m^8}{p2^4}
                                                                                                            \right) \frac{m^2 \ln[1+\frac{p2}{m^2}]}{p2}
                                                                                                      \Bigg{)} 
                                                                     +\frac{p4}{p2} \Bigg{(}1.8148310(2)
                                                                               -\frac{157 \ln[a^2 m^2+a^2 p2]}{180}
                                                                               -\frac{517 m^2}{720 p2}
                                                                               +\frac{107 m^4}{90 p2^2}
                                                                               +\frac{11 m^6}{120 p2^3}
                                                                               -\frac{57 m^8}{20 p2^4}
                                                                               +\frac{2 p2}{3 \left(m^2+p2\right)}
                                                                               +\left(\frac{1}{3}
                                                                                      +\frac{m^2}{4 p2}
                                                                                      -\frac{53 m^4}{36 p2^2}
                                                                                      +\frac{4 m^6}{3 p2^3}
                                                                                      +\frac{57 m^8}{20 p2^4}
                                                                                \right) \frac{m^2 \ln[1+\frac{p2}{m^2}]}{p2}
                                                                               +\frac{\left(p_{\nu_1}^2+p_{\nu_2}^2\right)}{p2} \Bigg{(}\frac{2}{3}
                                                                                                                      -\frac{2 p2}{3 m^2}
                                                                                                                      -\frac{7 m^2}{4 p2}
                                                                                                                      -\frac{7 m^4}{p2^2}
                                                                                                                      -\frac{15 m^6}{2 p2^3}
                                                                                                                      +\frac{15 m^8}{p2^4}
                                                                                                                      +\frac{2 p2^2}{3 \left(m^2+p2\right)\,m^2}
                                                                                                                      +\left(\frac{1}{3}
                                                                                                                             +\frac{4 m^2}{p2}
                                                                                                                             +\frac{12 m^4}{p2^2}
                                                                                                                             -\frac{15 m^8}{p2^4}
                                                                                                                       \right) \frac{m^2 \ln[1+\frac{p2}{m^2}]}{p2}
                                                                                                                \Bigg{)} 
                                                                         \Bigg{)}
                                                               \Red{\right]}
                                                   \Green{\Bigg{\}}}\Bigg{|}_{p_{\rho}=\mu_{\rho}}
 + {\cal O}(\Red{a^3},\Green{g^4})
\edms
\edg
\end{minipage}
\vspace*{0.5cm}

where
\bea
M^2 &=& m_0^2 +\mu_0^2\,, \\
\nu_1 &=& \mu\,, \\
\nu_2 &=& \nu\,.
\eea

\newpage
\section{Analytic expressions for one-derivative operators}
\label{appC}

Here we present the $Z$-factors for the one-derivative vector, axial and
tensor operators published separately in Ref.~\cite{Alexandrou:2010me}, defined as follows
\begin{eqnarray}
\Op_{\rm DV}^{\{\mu\,\nu\}} &= \overline \chi \gamma_{\{\mu}\overleftrightarrow D_{\nu\}}\tau^a \chi 
                                              &= \begin{cases} \overline \psi  \gamma_5\gamma_{\{\mu}\overleftrightarrow D_{\nu\}} \tau^2 \psi   & a=1 \\
                                                              -\overline \psi  \gamma_5\gamma_{\{\mu}\overleftrightarrow D_{\nu\}} \tau^1 \psi   & a=2 \\
                                                               \overline \psi  \gamma_{\{\mu}\overleftrightarrow D_{\nu\}}         \tau^3 \psi   & a=3 \end{cases} \\[3ex]
\Op_{\rm DA}^{\{\mu\,\nu\}} &= \overline \chi \gamma_5\gamma_{\{\mu}\overleftrightarrow D_{\nu\}}\tau^a \chi 
                                              &= \begin{cases} \overline \psi  \gamma_{\{\mu}\overleftrightarrow D_{\nu\}} \tau^2 \psi   &\quad a=1 \\
                                                              -\overline \psi  \gamma_{\{\mu}\overleftrightarrow D_{\nu\}} \tau^1 \psi   & \quad a=2 \\
                                                               \overline \psi  \gamma_5\gamma_{\{\mu}\overleftrightarrow D_{\nu\}} \tau^3 \psi   & \quad a=3 \end{cases}\\[3ex]
\Op_{\rm DT}^{\mu\,\{\nu\,\rho\}} &= \overline \chi \gamma_5\sigma_{\mu\{\nu}\overleftrightarrow D_{\rho\}}\tau^a \chi 
                                              &= \begin{cases} \overline \psi  \gamma_5\sigma_{\mu\{\nu}\overleftrightarrow D_{\rho\}}\tau^a \psi   & a=1,2 \\
                         -i\,\overline \psi  \sigma_{\mu\{\nu}\overleftrightarrow D_{\rho\}}\eins \psi            & a=3 \end{cases}\,.
\end{eqnarray}
The above operators are symmetrized over two Lorentz indices and are made traceless 
\be
\Op^{\{\sigma\,\tau\}} \equiv \frac{1}{2}\Big(\Op^{\sigma\,\tau}+\Op^{\tau\,\sigma}
\Big) - \frac{1}{4}\delta^{\sigma\,\tau} \sum_\lambda \Op^{\lambda\,\lambda}\,.\nonumber
\ee
The one derivative operators fall into different irreducible
representations of the hypercubic group, depending on the choice of
indices:
\begin{eqnarray}
   \Op_{\rm DV1} &=& \Op_{\rm DV} \ {\rm with} \ \mu=\nu \nonumber\\
\Op_{\rm DV2} &=& \Op_{\rm DV} \ {\rm with} \ \mu\neq\nu \nonumber\\
   \Op_{\rm DA1} &=& \Op_{\rm DA} \ {\rm with} \ \mu=\nu \nonumber\\
   \Op_{\rm DA2} &=& \Op_{\rm DA} \ {\rm with} \ \mu\neq\nu \nonumber\\
   \Op_{\rm DT1} &=& \Op_{\rm DT} \ {\rm with} \ \mu\neq\nu=\rho\nonumber\\
   \Op_{\rm DT2} &=& \Op_{\rm DT} \ {\rm with} \ \mu\neq\nu\neq\rho\neq\mu\,.\nonumber
\end{eqnarray}
Thus, $Z_{\rm DV1},\,Z_{\rm DA1}$ will be different from $Z_{\rm
  DV2},\,Z_{\rm DA2}$, respectively. More details on the
one-derivative renormalization factors can be found in
Ref.~\cite{Alexandrou:2010me}.

We have computed, to ${\cal O}(a^2)$, the forward matrix
elements of these operators for general external indices $\mu,\,\nu$
(and $\rho$ for the tensor operator), external momentum
$p,\,m,\,g,\,N_c,\,a,\,\csw$ and gauge fixing. Our final results were
obtained for the 10 sets of Symanzik coefficients given in
Table~\ref{tab1}. 

The amputated Green†¢s functions of the $\Op_{D\Gamma}$ operator appear
in the Mathematica file Zfactors.m as below:
\bea
\L_{DV}^{\rm pert.} &=& {\rm LDV[Action,csw,beta,g2tilde,a,m]} + {\cal O}(\Red{a^3},\MB{g^4}) ,\nonumber \\
\L_{DA}^{\rm pert.} &=& {\rm LDA[Action,csw,beta,g2tilde,a,m]} + {\cal O}(\Red{a^3},\MB{g^4}) , \nonumber\\
\L_{DT}^{\rm pert.} &=& {\rm LDT[Action,csw,beta,g2tilde,a,m]} + {\cal O}(\Red{a^3},\MB{g^4}) \,. \nonumber
\eea

In order to define $Z_{\cal O}$, we have used a renormalization
prescription which is most amenable to non-perturbative treatment:
\be
\Zq^{-1} Z_{\cal O}{\rm Tr}\Big{[} L^{O}(p)\,\cdot\,L^{O}_{\rm tree}(p)\Big{]}_{p_{\lambda}=\mu_{\lambda}} = 
{\rm Tr}\Big{[}L^{O}_{\rm tree}(p) \,\cdot\,L^{O}_{\rm tree}(p)\Big{]}_{p_{\lambda}=\mu_{\lambda}}
\label{ZO}
\ee
where $L^{O}$ denotes the amputated 2-point Green's function of the
operators up to 1-loop and up to ${\cal O}(a^2)$. These $Z$-factors
appear in electronic form with the name:
\bea
\left(Z_{\rm DV1}^{\rm pert.}\right)^{\nu_1 = \nu_2} &=& {\rm ZDV1[Action,csw,beta,g2tilde,a,m]} + {\cal O}(\Red{a^3},\MB{g^4}),\nonumber \\
\left(Z_{\rm DV2}^{\rm pert.}\right)^{\nu_1 \neq\, \nu_2} &=& {\rm ZDV2[Action,csw,beta,g2tilde,a,m]} + {\cal O}(\Red{a^3},\MB{g^4}), \nonumber\\
\left(Z_{\rm DA1}^{\rm pert.}\right)^{\nu_1 = \nu_2} &=& {\rm ZDA1[Action,csw,beta,g2tilde,a,m]} + {\cal O}(\Red{a^3},\MB{g^4}), \nonumber\\
\left(Z_{\rm DA2}^{\rm pert.}\right)^{\nu_1 \neq\, \nu_2} &=& {\rm ZDA2[Action,csw,beta,g2tilde,a,m]} + {\cal O}(\Red{a^3},\MB{g^4}), \nonumber\\
\left(Z_{\rm DT1}^{\rm pert.}\right)^{\nu_1 = \nu_3 \neq\, \nu_2} &=& {\rm ZDT1[Action,csw,beta,g2tilde,a,m]} + {\cal O}(\Red{a^3},\MB{g^4}), \nonumber\\
\left(Z_{\rm DT2}^{\rm pert.}\right)^{\nu_1 \neq\, \nu_2 \neq\, \nu_3 \neq\, \nu_1} &=& {\rm ZDT2[Action,csw,beta,g2tilde,a,m]} + {\cal O}(\Red{a^3},\MB{g^4})\,.\nonumber
\eea

Due to very lengthy expressions we only show the
results for specific choices of the action parameters, that is Landau
gauge, tree-level Symanzik gluons, $\csw=0$, $m=0$:

\vspace*{1cm}

\begin{minipage}{16cm}
\bdg[style={\footnotesize}, spread={8pt}]
\bdm[label={ZDV1}]
\left(Z_{DV1}^{\rm pert.}\right)^{\nu_1 \hiderel{=}\, \nu_2} \hiderel{=} {\rm ZDV1[2,0,1,g2tilde,a,0]} + {\cal O}(\Red{a^4},\MB{g^4})
\edm
\bdms
=
\Green{\delta_{\nu_1\,\nu_1} \left(}1+ \gttwo \MB{\Bigg{(}}1.41698(1)
                                      -\frac{8}{3} \ln[a^2 p2] 
                                      +\frac{2 p_{\nu_1}^2}{3 p2}
                                      -\frac{6 p_{\nu_1}^2}{8 p_{\nu_1}^2+p2}
                                      +\Red{a^2} \Red{\Bigg{(}}1.62067(6) p2
                                                  -6.4175(7) p_{\nu_1}^2
                                                  +\frac{21 p_{\nu_1}^4}{10 p2}
                                                  +\frac{23.328(6) p_{\nu_1}^4}{8 p_{\nu_1}^2+p2}
                                                  -\frac{16 p_{\nu_1}^6}{\left(8 p_{\nu_1}^2+p2\right)^2}
                                                  +\left(-\frac{19 p2}{180}
                                                        -\frac{334 p_{\nu_1}^2}{45}
                                                        +\frac{232 p_{\nu_1}^4}{3 \left(8 p_{\nu_1}^2+p2\right)}
                                                  \right) \ln[a^2 p2]
                                                  +p4 \Bigg{(}\frac{2.0544143(2)}{p2}
                                                           +\frac{29 p_{\nu_1}^2}{180 p2^2}
                                                           -\frac{2.0100(1)}{8 p_{\nu_1}^2+p2}
                                                           -\frac{2 p_{\nu_1}^2}{\left(8 p_{\nu_1}^2+p2\right)^2} 
                                                           +\left(-\frac{157}{180 p2}
                                                                  +\frac{79}{60 \left(8 p_{\nu_1}^2+p2\right)}
                                                           \right) \ln[a^2 p2]
                                                     \Bigg{)}
                                            \Red{\Bigg{)}}
                         \MB{\Bigg{)}}
            \Green{\right)}
+ {\cal O}(\Red{a^4},\MB{g^4})
\edms
\edg
\end{minipage}
\vspace*{0.75cm}

\begin{minipage}{16cm}
\bdg[style={\footnotesize}, spread={8pt}]
\bdm[label={ZDV2}]
\left(Z_{DV2}^{\rm pert.}\right)^{\nu_1 \hiderel{\neq}\, \nu_2} \hiderel{=} {\rm ZDV2[2,0,1,g2tilde,a,0]} + {\cal O}(\Red{a^4},\MB{g^4})
\edm
\bdms
= 
\Green{\delta_{\nu_1\,\nu_1} \delta_{\nu_2\,\nu_2}\left(} 1+ \gttwo \MB{\Bigg{(}}2.02248(1)
                                                            -\frac{8}{3} \ln[a^2 p2]
                                                            +\frac{4 p_{\nu_1}^2 p_{\nu_2}^2}{3 p2 \left(p_{\nu_1}^2+p_{\nu_2}^2\right)}
                                                            +\Red{a^2} \Red{\Bigg{(}}1.01505(3) p2
                                                                          +\frac{209 p_{\nu_1}^2 p_{\nu_2}^2}{90 p2}
                                                                          -2.1276(1) \left(p_{\nu_1}^2+p_{\nu_2}^2\right)
                                                                          +\frac{0.2456(3) p_{\nu_1}^2 p_{\nu_2}^2}{p_{\nu_1}^2+p_{\nu_2}^2}
                                                                          -\frac{8 p_{\nu_1}^4 p_{\nu_2}^4}{9 p2 \left(p_{\nu_1}^2+p_{\nu_2}^2\right)^2}
                                                                          +\left(-\frac{22 p2}{45}
                                                                                 +\frac{19}{40} \left(p_{\nu_1}^2+p_{\nu_2}^2\right)
                                                                                 +\frac{421 p_{\nu_1}^2 p_{\nu_2}^2}{90 \left(p_{\nu_1}^2+p_{\nu_2}^2\right)}
                                                                           \right) \ln[a^2 p2]
                                                                          +p4 \left(\frac{2.5967755(2)}{p2}
                                                                                    -\frac{157 \ln[a^2 p2]}{180 p2}
                                                                                    +\frac{29 p_{\nu_1}^2 p_{\nu_2}^2}{90 p2^2 \left(p_{\nu_1}^2+p_{\nu_2}^2\right)}
                                                                              \right)
                                                                 \Red{\Bigg{)}}
                                                 \MB{\Bigg{)}}
                       \Green{\right)}
+ {\cal O}(\Red{a^4},\MB{g^4})
\edms
\edg
\end{minipage}
\vspace*{0.75cm}

\begin{minipage}{16cm}
\bdg[style={\footnotesize}, spread={8pt}]
\bdm[label={ZDA1}]
\left(Z_{DA1}^{\rm pert.}\right)^{\nu_1 \hiderel{=}\, \nu_2} \hiderel{=} {\rm ZDA1[2,0,1,g2tilde,a,0]} + {\cal O}(\Red{a^4},\MB{g^4})
\edm
\bdms
= 
\Green{\delta_{\nu_1\,\nu_1} \left(}1+ \gttwo \MB{\Bigg{(}}3.48606(1)
                                        -\frac{8}{3} \ln[a^2 p2]
                                        +\frac{2 p_{\nu_1}^2}{3 p2}
                                        -\frac{6 p_{\nu_1}^2}{8 p_{\nu_1}^2+p2}
                                        +\Red{a^2} \Red{\Bigg{(}} 0.46577(6) p2
                                                   +3.8584(7) p_{\nu_1}^2
                                                   +\frac{21 p_{\nu_1}^4}{10 p2}
                                                   -\frac{62.787(6) p_{\nu_1}^4}{8 p_{\nu_1}^2+p2}
                                                   -\frac{16 p_{\nu_1}^6}{\left(8 p_{\nu_1}^2+p2\right)^2}
                                                   +\left(-\frac{199 p2}{180}
                                                          +\frac{866 p_{\nu_1}^2}{45}
                                                          -\frac{488 p_{\nu_1}^4}{3 \left(8 p_{\nu_1}^2+p2\right)}
                                                   \right) \ln[a^2 p2]
                                                   +p4 \Bigg{(}\frac{2.0544143(2)}{p2}
                                                             +\frac{29 p_{\nu_1}^2}{180 p2^2}
                                                             -\frac{1.2140(1)}{8 p_{\nu_1}^2+p2}
                                                             -\frac{2 p_{\nu_1}^2}{\left(8 p_{\nu_1}^2+p2\right)^2}
                                                             +\left(-\frac{157}{180 p2}
                                                                   +\frac{79}{60 \left(8 p_{\nu_1}^2+p2\right)}
                                                             \right) \ln[a^2 p2]
                                                      \Bigg{)}
                                            \Red{\Bigg{)}}  
                             \MB{\Bigg{)}}
            \Green{\right)}
+ {\cal O}(\Red{a^4},\MB{g^4})
\edms
\edg
\end{minipage}
\vspace*{0.75cm}

\newpage

\begin{minipage}{16cm}
\bdg[style={\footnotesize}, spread={8pt}]
\bdm[label={ZDA2}]
\left(Z_{DA2}^{\rm pert.}\right)^{\nu_1 \hiderel{\neq}\, \nu_2} \hiderel{=} {\rm ZDA2[2,0,1,g2tilde,a,0]} + {\cal O}(\Red{a^4},\MB{g^4})
\edm
\bdms
= 
\Green{\delta_{\nu_1\,\nu_1} \delta_{\nu_2\,\nu_2}\left(} 1+ \gttwo \MB{\Bigg{(}}3.07868(1)
                                                            -\frac{8}{3} \ln[a^2 p2]
                                                            +\frac{4 p_{\nu_1}^2 p_{\nu_2}^2}{3 p2 \left(p_{\nu_1}^2+p_{\nu_2}^2\right)}
                                                            +\Red{a^2} \Red{\Bigg{(}}0.38848(3) p2
                                                                        +\frac{209 p_{\nu_1}^2 p_{\nu_2}^2}{90 p2}
                                                                        -1.8613(1) \left(p_{\nu_1}^2+p_{\nu_2}^2\right)
                                                                        +\frac{-1.1283(3) p_{\nu_1}^2 p_{\nu_2}^2}{p_{\nu_1}^2+p_{\nu_2}^2}
                                                                        -\frac{8 p_{\nu_1}^4 p_{\nu_2}^4}{9 p2 \left(p_{\nu_1}^2+p_{\nu_2}^2\right)^2}
                                                                        +\left(\frac{8 p2}{45}
                                                                               +\frac{19}{40} \left(p_{\nu_1}^2+p_{\nu_2}^2\right)
                                                                               -\frac{179 p_{\nu_1}^2 p_{\nu_2}^2}{90 \left(p_{\nu_1}^2+p_{\nu_2}^2\right)}
                                                                         \right) \ln[a^2 p2]
                                                                        +p4 \left(\frac{2.5967755(2)}{p2}
                                                                                  -\frac{157 \ln[a^2 p2]}{180 p2}
                                                                                  +\frac{29 p_{\nu_1}^2 p_{\nu_2}^2}{90 p2^2 \left(p_{\nu_1}^2+p_{\nu_2}^2\right)}
                                                                            \right)
                                                                  \Red{\Bigg{)}}
                                                 \MB{\Bigg{)}}
                       \Green{\right)}
+ {\cal O}(\Red{a^4},\MB{g^4})
\edms
\edg
\end{minipage}
\vspace*{0.75cm}

\begin{minipage}{16cm}
\bdg[style={\footnotesize}, spread={8pt}]
\bdm[label={ZDT1}]
\left(Z_{DT1}^{\rm pert.}\right)^{\nu_1 \hiderel{=}\, \nu_3 \hiderel{\neq}\, \nu_2} \hiderel{=} {\rm ZDT1[2,0,1,g2tilde,a,0]} + {\cal O}(\Red{a^4},\MB{g^4})
\edm
\bdms
= 
\Green{\delta_{\nu_1\,\nu_1} \delta_{\nu_2\,\nu_2}\left(} 1+ \gttwo \MB{\Bigg{(}}3.88296(1)
                                                            -\frac{18.7832(1) p_{\nu_1}^2}{8 p_{\nu_1}^2-p_{\nu_2}^2+p2}
                                                            +\left(-4 
                                                                   +\frac{12 p_{\nu_1}^2}{8 p_{\nu_1}^2-p_{\nu_2}^2+p2}
                                                            \right) \ln[a^2 p2]
                                                            +\Red{a^2} \Red{\left(}-0.7730(3) p2
                                                                       -26.269(8) p_{\nu_1}^2
                                                                       -3.3610(1) p_{\nu_2}^2
                                                                       -\frac{78 p_{\nu_1}^2 p_{\nu_2}^2}{5 p2}
                                                                       -\frac{7058 p_{\nu_1}^4}{45 p2}
                                                                       +\frac{28.163(3) p_{\nu_1}^2 p2}{8 p_{\nu_1}^2-p_{\nu_2}^2+p2}
                                                                       +\frac{2.4703(2) p2^2}{8 p_{\nu_1}^2-p_{\nu_2}^2+p2}
                                                                       +\frac{302.33(1) p_{\nu_1}^4}{8 p_{\nu_1}^2-p_{\nu_2}^2+p2}
                                                                       +\frac{3800 p_{\nu_1}^6}{3 p2 \left(8 p_{\nu_1}^2-p_{\nu_2}^2+p2\right)}
                                                                       +\frac{6.2611(1) p_{\nu_1}^2 p2^2}{\left(8 p_{\nu_1}^2-p_{\nu_2}^2+p2\right)^2}
                                                                       +\frac{100.1772(9) p_{\nu_1}^4 p2}{\left(8 p_{\nu_1}^2-p_{\nu_2}^2+p2\right)^2}
                                                                       +\frac{350.620(3) p_{\nu_1}^6}{\left(8 p_{\nu_1}^2-p_{\nu_2}^2+p2\right)^2}
                                                                       +\Bigg{(}\frac{80 p2}{90}
                                                                              +\frac{82 p_{\nu_2}^2}{45}
                                                                              -\frac{739 p_{\nu_1}^2}{180}
                                                                              -\frac{293 p_{\nu_1}^2 p2}{45 \left(8 p_{\nu_1}^2-p_{\nu_2}^2+p2\right)}
                                                                              -\frac{193 p2^2}{180 \left(8 p_{\nu_1}^2-p_{\nu_2}^2+p2\right)}
                                                                              +\frac{658 p_{\nu_1}^4}{9 \left(8 p_{\nu_1}^2-p_{\nu_2}^2+p2\right)}
                                                                              -\frac{4 p_{\nu_1}^2 p2^2}{\left(8 p_{\nu_1}^2-p_{\nu_2}^2+p2\right)^2}
                                                                              -\frac{64 p_{\nu_1}^4 p2}{\left(8 p_{\nu_1}^2-p_{\nu_2}^2+p2\right)^2}
                                                                              -\frac{224 p_{\nu_1}^6}{\left(8 p_{\nu_1}^2-p_{\nu_2}^2+p2\right)^2}
                                                                        \Bigg{)} \ln[a^2 p2]
                                                                       +p4 \Bigg{(}\frac{2.4606643(2)}{p2}
                                                                                 -\frac{2.4703(1)}{8 p_{\nu_1}^2-p_{\nu_2}^2+p2}
                                                                                 -\frac{131 p_{\nu_1}^2}{180 p2 \left(8 p_{\nu_1}^2-p_{\nu_2}^2+p2\right)}
                                                                                 -\frac{6.26108(4) p_{\nu_1}^2}{\left(8 p_{\nu_1}^2-p_{\nu_2}^2+p2\right)^2}
                                                                                 +\left(-\frac{157}{180 p2}
                                                                                       +\frac{193}{180 \left(8 p_{\nu_1}^2-p_{\nu_2}^2+p2\right)}
                                                                                       +\frac{4 p_{\nu_1}^2}{\left(8 p_{\nu_1}^2-p_{\nu_2}^2+p2\right)^2}
                                                                                 \right) \ln[a^2 p2]
                                                                           \Bigg{)}
                                                                 \Red{\right)}
                                          \MB{\Bigg{)}}
                       \Green{\right)}
+ {\cal O}(\Red{a^4},\MB{g^4})
\edms
\edg
\end{minipage}
\vspace*{0.75cm}

\begin{minipage}{16cm}
\bdg[style={\footnotesize}, spread={8pt}]
\bdm[label={ZDT2}]
\left(Z_{DT2}^{\rm pert.}\right)^{\nu_1 \hiderel{\neq}\, \nu_2 \hiderel{\neq}\,\nu_3 \hiderel{\neq}\, \nu_1} \hiderel{=} {\rm ZDT2[2,0,1,g2tilde,a,0]} + {\cal O}(\Red{a^4},\MB{g^4})
\edm
\bdms
= 
\Green{\delta_{\nu_1\,\nu_1} \delta_{\nu_2\,\nu_2} \delta_{\nu_3\,\nu_3}\left(} 1+ \gttwo \MB{\Bigg{(}} 2.82413(1)
                                                                                -3 \ln[a^2 p2]
                                                                                +\Red{a^2} \Red{\Bigg{(}}0.92582(3) p2
                                                                                            -0.73604(2) p_{\nu_2}^2
                                                                                            +\frac{67 p_{\nu_1}^2 p_{\nu_3}^2}{45 p2}
                                                                                            -2.1124(1) \left(p_{\nu_1}^2+p_{\nu_3}^2\right)
                                                                                            +\frac{67 p_{\nu_2}^4}{90 p2}
                                                                                            +\frac{1.2403(3) p_{\nu_1}^2 p_{\nu_3}^2}{p_{\nu_1}^2+p_{\nu_3}^2}
                                                                                            +\frac{67 p_{\nu_1}^2 p_{\nu_2}^2 p_{\nu_3}^2}{15 p2 \left(p_{\nu_1}^2+p_{\nu_3}^2\right)}
                                                                                            +\left(-\frac{p2}{2}
                                                                                                   +\frac{301 p_{\nu_2}^2}{360}
                                                                                                   +\frac{331}{720} \left(p_{\nu_1}^2+p_{\nu_3}^2\right)
                                                                                                   +\frac{71 p_{\nu_1}^2 p_{\nu_3}^2}{20 \left(p_{\nu_1}^2+p_{\nu_3}^2\right)}
                                                                                             \right) \ln[a^2 p2]
                                                                                            +p4 \left(\frac{2.1064977(2)}{p2}
                                                                                                      +\frac{41}{60 p2}
                                                                                                      -\frac{157 \ln[a^2 p2]}{180 p2}
                                                                                                \right)
                                                                                      \Red{\Bigg{)}}
                                                                      \MB{\Bigg{)}}
                       \Green{\right)}
+ {\cal O}(\Red{a^4},\MB{g^4})
\edms
\edg
\end{minipage}
\vspace*{0.75cm}

where for ZDV1, ZDV2, ZDA1, ZDA2:
\bea
\nu_1 &=& \mu , \nonumber\\
\nu_2 &=& \nu ,\nonumber
\eea
and for ZDT1, ZDT2:
\bea
\nu_1 &=& \rho , \nonumber\\
\nu_2 &=& \mu , \nonumber\\
\nu_3 &=& \nu .\nonumber
\eea

\newpage

\section{Notation in Mathematica file: Zfactors.m }
\label{appD}

The full body of our results can be accessed online through the
 Mathematica file Zfactors.m, which is a Mathematica input file. It includes
the expressions for the amputated Green's functions of the inverse propagator:
\vspace{-0.25cm}
\be
S^{-1}_{\rm pert.} = \delta^{f' \, g'} \bigg( {\rm propagator[Action,csw,beta,g2tilde,a,m,mu]} + {\cal O}(\Red{a^3},\MB{g^4}) \bigg)\,,
\ee
from which one can construct the fermion field renormalization constant for any renormalization scheme.
This expression depends on the variables:
\vspace{-0.25cm}
\begin{itemize}
\item action: Selection of improved gauge action as follows, 1 $\rightarrow$
  Plaquette, 2 $\rightarrow$ Tree Level Symanzik, 3 $\rightarrow$ TILW
  ($\beta\,c_0=8.60$), 4 $\rightarrow$ TILW ($\beta\,c_0=8.45$), 5
  $\rightarrow$ TILW ($\beta\,c_0=8.30$), 6 $\rightarrow$ TILW
  ($\beta\,c_0=8.20$), 7 $\rightarrow$ TILW ($\beta\,c_0=8.10$), 8
  $\rightarrow$ TILW ($\beta\,c_0=8.00$), 9 $\rightarrow$ Iwasaki, 10
  $\rightarrow$ DBW2
\vspace{-0.25cm}\item csw: clover parameter
\vspace{-0.25cm}\item beta: gauge parameter (Landau/Feynman/Generic correspond to 1/0/beta)
\vspace{-0.25cm}\item g2tilde=$\ggcf$ g: coupling constant
\vspace{-0.25cm}\item a: lattice spacing
\vspace{-0.25cm}\item m: Lagrangian mass
\vspace{-0.25cm}\item mu: twisted mass parameter
\end{itemize}

The expression for the critical mass is defined in the variable mcritical:
$m_{\rm cr} = {\rm mcritical[Action,csw,g2tilde,aL]} + \frac{1}{a} {\cal O}(g^4)$.
The reader may also find the amputated Green's functions relevant to the ultra-local operators:
\bea
\label{LS}
\Lambda_S^{\rm pert.} =& {\rm scalar}&[{\rm Action,csw,beta,g2tilde,a,m,mu1,mu2}] + {\cal O}(\Red{a^3},\MB{g^4}) , \\
\Lambda_P^{\rm pert.} =& {\rm pseudoscalar}&[{\rm Action,csw,beta,g2tilde,a,m,mu1,mu2}] + {\cal O}(\Red{a^3},\MB{g^4}) , \\
\Lambda_V^{\rm pert.} =& {\rm vector}&[{\rm Action,csw,beta,g2tilde,a,m,mu1,mu2}] + {\cal O}(\Red{a^3},\MB{g^4}) , \\
\Lambda_A^{\rm pert.} =& {\rm axial}&[{\rm Action,csw,beta,g2tilde,a,m,mu1,mu2}] + {\cal O}(\Red{a^3},\MB{g^4}) , \\
\Lambda_T^{\rm pert.} =& {\rm tensor}&[{\rm Action,csw,beta,g2tilde,a,m,mu1,mu2}] + {\cal O}(\Red{a^3},\MB{g^4}) , \\
\Lambda_{Tp}^{\rm pert.} =& {\rm tensorprime}&[{\rm Action,csw,beta,g2tilde,a,m,mu1,mu2}] + {\cal O}(\Red{a^3},\MB{g^4}) ,\label{LT}
\eea
as well as the Green's functions of the one-derivative vector, axial and tensor operators:
\bea
\label{LDV}
\Lambda_{DV}^{\rm pert.} =& {\rm LDV}&[{\rm Action,csw,beta,g2tilde,a,m}] + {\cal O}(\Red{a^3},\MB{g^4}) , \\
\Lambda_{DA}^{\rm pert.} =& {\rm LDA}&[{\rm Action,csw,beta,g2tilde,a,m}] + {\cal O}(\Red{a^3},\MB{g^4}) , \\
\Lambda_{DT}^{\rm pert.} =& {\rm LDT}&[{\rm Action,csw,beta,g2tilde,a,m}] + {\cal O}(\Red{a^3},\MB{g^4}) \label{LDT}
\eea
We note that Eqs. (\ref{LS}) - (\ref{LT}) hold for quarks with the same Lagrangian mass and $\mu_1 = \pm \mu_2$,
while Eqs. (\ref{LDV}) - (\ref{LDT}) correspond to zero twisted mass parameters, $\mu_1,\,\mu_2$, and
non-zero Lagrangian mass.

The RCs are interesting quantities for other studies and are also
provided in the Mathematica file Zfactors.m
at the RI$'$-MOM scheme by employing Eq.~(\ref{eq:zqri}) and Eq.~(\ref{eq:rimom}), for the
fermion field and fermion operator RCs, respectively.

\be
Z_q^{\rm pert.} \hiderel{=} {\rm zq[2,0,1,g2tilde,a,m,mu,p2,p4]} + {\cal O}(\Red{a^3}\,\MB{g^2},\MB{g^4})
\ee

\bea
Z_S^{\rm pert.} \hiderel{=} {\rm zs[2,0,1,g2tilde,a,m,mu1,mu2]} + {\cal O}(\Red{a^3},\MB{g^4})\\
Z_P^{\rm pert.} \hiderel{=} {\rm zp[2,0,1,g2tilde,a,m,mu1,mu2]} + {\cal O}(\Red{a^3},\MB{g^4})\\
Z_V^{\rm pert.} \hiderel{=} {\rm zv[2,0,1,g2tilde,a,m,mu1,mu2]} + {\cal O}(\Red{a^3},\MB{g^4})\\
Z_A^{\rm pert.} \hiderel{=} {\rm za[2,0,1,g2tilde,a,m,mu1,mu2]} + {\cal O}(\Red{a^3},\MB{g^4})\\
\left(Z_T^{\rm pert.}\right)^{\nu_1 \hiderel{\neq}\, \nu_2} \hiderel{=} {\rm zt[2,0,1,g2tilde,a,m,mu1,mu2]} + {\cal O}(\Red{a^3},\MB{g^4})\\
\left(Z_{Tp}^{\rm pert.}\right)^{\nu_1 \hiderel{\neq}\, \nu_2} \hiderel{=} {\rm ztp[2,0,1,g2tilde,a,m,mu1,mu2]} + {\cal O}(\Red{a^3},\MB{g^4})
\eea

The additional variables are
\vspace{-0.25cm}
\begin{itemize}
\item p2: $\sum_{i=1}^4 p_i^2$
\item p4: $\sum_{i=1}^4 p_i^4$
\end{itemize}

For completeness we include in the Mathematica file Zfactors.m the conversion factors to
the $\rm \overline{MS}$ scheme for the one-derivative RCs, studied in
Ref.~\cite{Alexandrou:2010me}.
\bea
C_{DV1,DA1} &=& {\rm CDV1[alphaRIprime, lambdaRIprime, CA, Cf, Nf, Tf]} + {\cal O}(q^8)  \\
C_{DV2,DA2} &=& {\rm CDV2[alphaRIprime, lambdaRIprime, CA, Cf, Nf, Tf]} + {\cal O}(g^8)  
\eea
where
\bea
{\rm alphaRIprime} &\equiv& g^2/(16 \pi^2)  \nonumber \\
{\rm lambdaRIprime} &\equiv& \lambda \equiv 1- {\rm beta} \nonumber
\eea

\newpage


\bibliographystyle{apsrev}                     
\bibliography{Zlocal}

\end{document}